\documentclass[12pt,oneside]{book}
\usepackage[T1]{fontenc}
\usepackage{times}
\usepackage{geometry}
\usepackage{fancyhdr}
\usepackage{makeidx}
\makeindex
\usepackage{graphics}

\makeatletter

\newcommand{\LyX}{L\kern-.1667em\lower.25em\hbox{Y}\kern-.125emX\@}
\newcommand{\noun}[1]{\textsc{#1}}

\usepackage[]{fontenc}
\usepackage[dvips]{epsfig}
\usepackage{a4wide}

\makeatletter
\def\LyX{L\kern-.1667em\lower.25em\hbox{Y}\kern-.125emX\spacefactor1000}%
\newcommand{\lyxtitle}[1] {\thispagestyle{empty}
\global\@topnum\z@
\section*{\LARGE \centering \sffamily \bfseries \protect#1 }
}

\makeatother

\makeatother

\begin{document}

\title{{\huge Quantum Fluctuations of Coherent Light in Nonlinear Media}\huge }

\maketitle
{\par\centering {\Large This work is dedicated to Dr. Arani Chakravarti \& Dr.
Samita Sil without whose encouragement it would be impossible for me to complete
the work.}\Large \par}
\newpage

~

\newpage
{\par\centering \textbf{\textsc{\underbar{\Large Acknowledgement}}}\Large \par}
\vspace{1cm}

\textsf{\textit{\large For a Ph.D work the essential condition is that there
will be a student and a supervisor. In the present case the condition is fulfilled
by me (student) and Dr. Swapan Mandal (my supervisor). I am thankful to my supervisor
for his help and guidance. The essential condition stated above is not sufficient
for a PhD work and one needs a lot of journals, preprints, reprints, software
and scientific discussions in order to achieve the sufficiency condition. In
last four years many people have helped me to achieve the sufficiency condition.
I am grateful to all of them. It is not possible to acknowledge everyone by
name. Still I would like to express my sincere thanks to Professor E C G Sudarshan,
Professor B K Talukdar, Professor F M Fern\'{a}ndez, Dr. Sreekantha Sil and
Dr. Arani Chakravarti for some valuable technical discussions. I also wish to
thank Chiranjib Sur, Madhumita Gupta, Anuj Kumar Saw, Susanta Madhab Das, Pika
Jha, Purnendu Chakraborty, Abhijit Sen, Tarun Kanti Ghosh, Tanya Bhattacharya,
Sutapa Datta, Moumita Maiti, Bipul Sarkar and Biswajit Sen for their cooperation
and technical support. }}{\large \par}

\textsf{\textit{\large I am also indebted to Kishore Mandal and other nonteaching
staff of our department for their constant cooperation.}}{\large \par}

\textsf{\textit{\large I am grateful to the Council of Scientific and Industrial
Research, India for their financial support during my research period.}}{\large \par}

\textsf{\textit{\large I would like to take this opportunity to express my deep
sense of gratitude to my grandmother, parents, uncles, aunts, sisters, brothers
and all other family members who were always there with me with their constant
encouragement and support. Lastly, but certainly not the least, I wish to thank
my friend Papia Chowdhury for her cooperation and interest in the present work
without which it would not be possible for me to devote all my time to the present
study.}}{\large \par}
\vspace{2cm}

{\large Date: \( 6^{th} \) May, 2002~~~~~~~~~~~~~~~~~~~~~~~~~~~~~~~~~~(Anirban
Pathak)}{\large \par}

\newpage
{\par\centering \underbar{\large ABSTRACT}\large \par}
\vspace{1cm}

There are substantial quantum fluctuations even in a pure coherent state. For
example, the quadrature fluctuation in coherent state \( |\alpha > \) is \( \frac{1}{2} \)
and the fluctuation in photon number is \( |\alpha |^{2} \). Further, randomization
in the case of general classical states which are random superposition of coherent
states can only increase these fluctuations. But there are some phenomena in
which quantum fluctuations reduce below the coherent state level. For example,
an electromagnetic field is said to be electrically squeezed field if uncertainties
in the quadrature phase observable \( X \) is less than the coherent state
level (i.e. \( (\triangle X)^{2}<\frac{1}{2} \)). Correspondingly a magnetically
squeezed field is one for which \( (\triangle \dot{X})^{2}<\frac{1}{2}. \)
\textbf{}On the other hand, antibunching is a phenomenon in which the fluctuation
in photon number is reduced below the Poisson level (i.e \( (\triangle N)^{2}<\, <N> \)).
All phenomena in which quantum fluctuations reduce below those of the coherent
state level are called nonclassical. It can be noted that the interaction of
dielectric medium with an intense electromagnetic field can gives rise to these
nonclassical nonlinear optical effects. These effects which help us to understand
the nature of quantum world in further detail are also important from the application
point of view. Keeping that in mind, the present thesis is devoted to the study
of quantum fluctuations of coherent light in non-linear media. 

The interaction of coherent light with a nonlinear medium is modeled here by
a general quantum anharmonic oscillator. The model of a quartic anharmonic oscillator
emerges if the lowest order of nonlinearity (i.e third order nonlinear susceptibility)
is assumed. The quartic oscillator model has considerable importance in the
study of non-linear and quantum optical effects present in a nonlinear medium
of inversion symmetry. Silica crystals constitute an inversion symmetric third
order nonlinear medium and these crystals are used to construct optical fibers.
In optical communications electromagnetic beams pass through these fibers and
the interaction of the electromagnetic field (single mode) with the fiber can
be described by the one dimensional quartic anharmonic oscillator Hamiltonian.
Depending on the nature of nonlinearity in physical problems the treatment of
higher anharmonic oscillators assumes significance.

Anharmonic oscillator models are not exactly solvable in a closed analytical
form. But we need operator solutions of the equations of motion corresponding
to these models in order to study the quantum fluctuations of coherent light
in nonlinear media. So we have two alternatives, either we can use an approximate
Hamiltonian which is exactly solvable or we can use an approximate operator
solution. Operator solutions (solutions in the Heisenberg approach) of anharmonic
oscillator problems were not available since the existing methods tend to introduce
inordinate mathematical complications in a detailed study. Due to the unavailability
of the operator solutions people were bound to use rotating wave approximated
Hamiltonians to study the quantum fluctuations of coherent light in nonlinear
media. The situation has improved in the recent past and many solutions of anharmonic
oscillators have appeared. Some of these solutions are obtained as a part of
the present work. For example, we have constructed a second order operator solution
for quartic oscillator and have generalized all first order operator solutions
available for the quartic oscillator to the \( m \)-th anharmonic oscillator.
From the generalized solutions we observe that there exists an apparent discrepancy
between the solutions obtained by different techniques. Then the question arises:
Which solution should be used for physical applications. Therefore, we compare
different solutions and conclude that all correct solutions are equivalent and
the apparent discrepancy is due to the use of different ordering of the operators. 

We use these solutions to investigate the possibilities of observing different
optical phenomena in a nonlinear dielectric medium. To be precise, we have studied
quantum phase fluctuations of coherent light in third order inversion symmetric
nonlinear medium. Fluctuations in phase space quadrature for the same system
are studied and the possibility of generating squeezed state is reported. Fluctuations
in photon number are studied and the nonclassical phenomenon of antibunching
is predicted. We have generalized the results obtained for third order nonlinear
media and have studied the interaction of an intense laser beam with a general
\( (m-1) \)-th order nonlinear medium. Aharonov Anandan nonadiabatic geometric
phase is also discussed in the context of \( (m-1) \)-th order nonlinear medium. {\large }{\large \par}

\newpage
{\par\centering \textbf{LIST OF PUBLICATIONS ON SOME OF WHICH MAIN RESULTS OF
THIS THESIS WERE PUBLISHED}\par}

\begin{enumerate}
\item Enhanced and reduced phase fluctuations of coherent light coupled to a quantum
quartic anharmonic oscillator, Pathak A and Mandal S 1999 \emph{Optics and Optoelectronics
: Theory Devices and Application Vol I} Eds: NIjhawan O P, Gupta A K, Musala
A K and Singh K (New Delhi: Narosa) 188.
\item Generalized quantum anharmonic oscillator using an operator ordering approach,
Pathak A 2000 \emph{J. Phys. A} \textbf{33} 5607-5613.
\item Phase fluctuations of coherent light coupled to a nonlinear medium of inversion
symmetry, Pathak A and Mandal S 2000 \emph{Phys. Lett. A.} \textbf{272} 346-352.
\item Quantum oscillator of quartic anharmonicity: second order solution, Pathak A
and Mandal S 2001 \emph{Phys. Lett. A} \textbf{286} 261-276. 
\item Aharonov Anandan phase for the quasi exactly solvable Bose Systems, Pathak A
2001 \emph{Proc. of Nat. Conf. on Laser and Its Application} 159.
\item Classical and quantum oscillators of sextic and octic anharmonicities, Pathak
A and Mandal S 2002 \emph{Phys. Lett. A} in press 
\item \footnotemark{}%
Spatio-temporal history of decay, Pathak A, submitted to \emph{Phys. Rev. A. }
\item Perturbation theory free from secular terms for quantum mechanical anharmonic
oscillators: Frequency operators Pathak A and \textsf{Fern\'{a}ndez} F M, submitted
to \emph{Annals of Physics}.
\item Possibilities of observing Aharonov Anandan geometric phase for a generalized
anharmonic oscillator, Pathak A, submitted to \emph{Journal of Optics B: Quantum
and Semiclassical Optics.} \emph{\large }{\large \par}
\end{enumerate}\footnotetext{
Not included in this thesis.
}%

\tableofcontents

\chapter{Interaction of electromagnetic field with matter}

Since the invention of laser\index{laser} people have observed different nonlinear
optical effects in dielectric media interacting with an intense electromagnetic
field. Studies into these effects are extremely important because some of them
(such as squeezing\index{squeezing} and antibunching\index{antibunching} of
photons) do not have any classical analogue. These purely quantum mechanical
effects are also important from the application point of view. Keeping that
in mind, in the present work we have discussed a class of physical systems in
which different nonlinear and quantum optical effects may be observed. To be
more specific, we consider a physical system in which an intense electromagnetic
field is interacting with an \( (m-1) \)-th order nonlinear medium. The physical
system is modeled by a generalized anharmonic oscillator (AHO). The operator
solution of the equation of motion corresponding to our model Hamiltonian is
developed using different techniques and is exploited to study the possibilities
of observing higher harmonic generation, bunching\index{bunching} and anti-bunching
of photons, self induced transparency and squeezing\index{squeezing}. In addition
to this, fluctuations in quantum phase\index{quantum phase} of the output field
is also studied with the knowledge of field operators. Geometric phase\index{geometric phase}
of the output field is also taken care of.

\section{The model}

The present thesis is devoted to the study of the interaction of a single-mode
intense electromagnetic field with a nonlinear medium. In such a nonlinear interaction
nonclassical effects like squeezing\index{squeezing} and antibunching\index{antibunching}
of photons may be produced. In fact, here we study the possibilities of observing
these nonclassical phenomena in the physical system of our interest. We can
continue our study by using either a semiclassical theory or a quantum theory.
Actually, a semiclassical theory, where the radiation is represented by a classical
wave and the atom is quantized, can treat many problems in light-matter interaction.
However, there are cases where a semiclassical theory turns out to be inadequate.
These include spontaneous emission, Lamb shift\index{Lamb shift}, resonance
fluorescence, the anomalous gyromagnetic moment of the electron and ``nonclassical''
states of light (e.g. squeezed states\index{squeezed state}). So if we need
to study the possibilities of observing nonclassical states then we have to
use a quantum theory instead of a semiclassical one and as an essential requirement
of the quantum theory we have to quantize the electromagnetic field. In the
next subsection the electromagnetic field is quantized and the model Hamiltonian
of our physical system is constructed in the subsequent subsection.

\subsection{Quantization of the electromagnetic field\index{quantization of the electromagnetic field}}

To quantize electromagnetic field let us recall some basic results of electrodynamics.
The Maxwell equations describing the propagation of electromagnetic fields are
\begin{equation}
\label{field}
\begin{array}{lcl}
\nabla \times \mathbf{E} & = & -\frac{\partial \mathbf{B}}{\partial \mathbf{t}}\\
\nabla .\mathbf{D} & = & \rho \\
\nabla \times \mathbf{H} & = & \mathbf{J}+\frac{\partial \mathbf{D}}{\partial t}\\
\nabla .\mathbf{B} & = & 0
\end{array}
\end{equation}
 where \( \mathbf{J} \) is the current density; the other symbols have their
conventional definitions. In a homogeneous isotropic medium \( \mathbf{B} \)
and \( \mathbf{D} \) are related to \( \mathbf{H} \) and \( \mathbf{E} \)
by 
\begin{equation}
\label{field2}
\begin{array}{lcl}
\mathbf{B} & = & \mu \mathbf{H}\\
\mathbf{D} & = & \epsilon \mathbf{E}
\end{array}
\end{equation}
 where \( \mu  \) and \( \epsilon  \) are, respectively, the magnetic permeability
and the dielectric constant of the medium. If we restrict ourselves to a charge
free medium, then we have 
\begin{equation}
\label{field3}
\begin{array}{lcl}
\rho  & = & 0\\
\mathbf{J} & = & 0\, .
\end{array}
\end{equation}

Consider the electric field \( \mathbf{E} \) and magnetic field \( \mathbf{H} \)
inside a volume \( V \) bounded by a surface \( S \) of perfect conductivity.
The tangential component of \( \mathbf{E} \) , \( -\mathbf{n}\times \mathbf{n}\times \mathbf{E} \),
and the normal component of \( \mathbf{H} \), \textbf{\( \mathbf{n}.\mathbf{H} \)}
will both be zero on \( S \) (\textbf{\( \mathbf{n} \)} is the unit vector
normal to \( S \)). We can expand \( \mathbf{E} \) and \( \mathbf{H} \) in
terms of two orthogonal sets of vector fields \( \mathbf{E}_{a} \) and \( \mathbf{H}_{a} \),
respectively. These sets which were originally introduced by Slater {[}\ref{slater}{]}
and nicely explained by Yariv {[}\ref{Yariv}{]} obey the relations 
\begin{equation}
\label{field4}
\begin{array}{lcl}
k_{a}\mathbf{E}_{a} & = & \nabla \times \mathbf{H}_{a}\\
 & 
\end{array}
\end{equation}
 and 
\begin{equation}
\label{field5}
k_{a}\mathbf{H}_{a}=\nabla \times \mathbf{E}_{a}
\end{equation}
where \( k_{a} \) is to be considered, for the moment, a constant. The tangential
component of  \( \mathbf{E}_{a} \) on \( S \) is zero, i.e 
\begin{equation}
\label{field6}
\mathbf{n}\times \mathbf{E}_{a}=0\, on\, S.
\end{equation}
 If we take the curl of both sides of the equations (\ref{field4}-\ref{field5})
and use the identity 
\begin{equation}
\label{field7}
\nabla \times \nabla \times \mathbf{A}=\nabla \left( \nabla .\mathbf{A}\right) -\nabla ^{2}\mathbf{A}
\end{equation}
 then we obtain the familiar wave equations 
\begin{equation}
\label{field8}
\begin{array}{lcl}
\nabla ^{2}\mathbf{E}_{a} & = & k_{a}^{2}\mathbf{E}_{a}\\
\nabla ^{2}\mathbf{H}_{a} & = & k_{a}^{2}\mathbf{H}_{a}\, .
\end{array}
\end{equation}
 Now we can write the total resonator fields of \( \mathbf{E} \) and \( \mathbf{H} \)
as 
\begin{equation}
\label{field9}
\begin{array}{lcl}
\mathbf{E}(\mathbf{r},t) & = & -\sum _{a}\frac{\omega _{a}}{\sqrt{\epsilon }}q_{a}(t)\mathbf{E}_{\mathbf{a}}(r)\\
\mathbf{H}(\mathbf{r},t) & = & \sum _{a}\frac{1}{\sqrt{\mu }}p_{a}(t)\mathbf{H}_{\mathbf{a}}(r)
\end{array}
\end{equation}
 where \( \omega _{a}=\frac{k_{a}}{\sqrt{\mu \epsilon }} \). \( q_{a}(t) \)
and \( p_{a}(t) \) are measures of the field amplitude in \( a \)-th mode.
Substituting (\ref{field9}) in the first two Maxwell equations (\ref{field})
and using (\ref{field4}-\ref{field5}) we obtain 
\begin{equation}
\label{field10}
\begin{array}{lcl}
\dot{p}_{a} & = & -\omega ^{2}_{a}q_{a}\\
\dot{q}_{a} & = & p_{a}\, .
\end{array}
\end{equation}
 From (\ref{field10}) we have
\begin{equation}
\label{field11}
\begin{array}{lcl}
\ddot{q}_{a}+\omega ^{2}_{a}q_{a} & = & 0\\
\ddot{p}_{a}+\omega ^{2}_{a}p_{a} & = & 0\, .
\end{array}
\end{equation}
 This identifies \( \omega _{a}=\frac{k_{a}}{\sqrt{\mu \epsilon }} \) as the
radian oscillation frequency of the \( a \)-th mode.

From (\ref{field9}) we can easily show that 
\begin{equation}
\label{field12}
q_{a}(t)=-\frac{\sqrt{\epsilon }}{\omega _{a}}\int _{V}\mathbf{E}(\mathbf{r},t).\mathbf{E}_{\mathbf{a}}(\mathbf{r})dv
\end{equation}
 and 

\begin{equation}
\label{field13}
p_{a}(t)=\sqrt{\mu }\int _{V}\mathbf{H}(\mathbf{r},t).\mathbf{H}_{a}(\mathbf{r})dv.
\end{equation}
 Now we conclude that the electromagnetic field can be specified either by \( \mathbf{E}(\mathbf{r},t) \)
and \( \mathbf{H}(\mathbf{r},t) \) or, alternatively, by the dynamical variables
\( q_{a}(t) \) and \( p_{a}(t) \). The total energy Hamiltonian is 
\begin{equation}
\label{field14}
H_{0}=\frac{1}{2}\int _{V}(\mu \mathbf{H}.\mathbf{H}+\epsilon \mathbf{E}.\mathbf{E})dv
\end{equation}
 where the subscript zero is used to distinguish the free field Hamiltonian
(\( H_{0} \)) from the total Hamiltonian of the system. Substituting the expansions
of \( \mathbf{E} \) and \( \mathbf{H} \) (\ref{field9}) in (\ref{field14})
we obtain 
\begin{equation}
\label{field15}
H_{0}=\sum _{a}\frac{1}{2}\left( p^{2}_{a}+\omega _{a}^{2}q_{a}\right) .
\end{equation}
 This has the basic form of a sum of harmonic oscillator Hamiltonians. The dynamical
variables \( p_{a} \) and \( q_{a} \) constitute canonically conjugate variables.
This can be seen by considering Hamilton's equations of motion relating \( \dot{p}_{a} \)
to \( q_{a} \) and \( \dot{q}_{a} \) to \( p_{a} \). The equations are

\begin{equation}
\label{field16}
\begin{array}{lclcl}
\dot{p}_{a} & = & -\frac{\partial H}{\partial q_{a}} & = & -\omega _{a}^{2}q_{a}\\
\dot{q}_{a} & = & \frac{\partial H}{\partial p_{a}} & = & p_{a}\, .
\end{array}
\end{equation}
 These are formally identical with (\ref{field11}), obtained from Maxwell's
equations\index{Maxwell's equations}.

The quantization of the electromagnetic field is finally achieved by considering
\( p_{a} \) and \( q_{a} \) as formally equivalent to the momentum and coordinate
of a quantum mechanical harmonic oscillator, thus taking the commutator relations
connecting the dynamical variables as 
\begin{equation}
\label{field17}
\begin{array}{ccc}
\left[ p_{a},p_{b}\right]  & = & \left[ q_{a},q_{b}\right] =0\\
\left[ q_{a},p_{b}\right]  & = & i\delta _{a,b}
\end{array}
\end{equation}
 where we have chosen to work in a system of units in which \( \hbar =1 \).
At this point we can introduce bosonic creation \index{creation operator} and
annihilation operators\index{annihilation operator} as 
\begin{equation}
\label{field18}
\begin{array}{ccc}
a_{l}^{\dagger }(t) & = & \left( \frac{1}{2\omega _{l}}\right) ^{\frac{1}{2}}\left[ \omega _{l}q_{l}(t)-ip_{l}(t)\right] \\
a_{l}(t) & = & \left( \frac{1}{2\omega _{l}}\right) ^{\frac{1}{2}}\left[ \omega _{l}q_{l}(t)+ip_{l}(t)\right] \, .
\end{array}
\end{equation}
 Solving (\ref{field18}) for \( p_{l} \) and \( q_{l} \) we obtain 
\begin{equation}
\label{field19}
\begin{array}{lcl}
p_{l}(t) & = & i\left( \frac{\omega _{l}}{2}\right) ^{\frac{1}{2}}\left[ a^{\dagger }_{l}(t)-a_{l}(t)\right] \\
q_{l}(t) & = & \left( \frac{1}{2\omega _{l}}\right) ^{\frac{1}{2}}\left[ a^{\dagger }_{l}(t)+a_{l}(t)\right] \, .
\end{array}
\end{equation}
 Now we can express the free field Hamiltonian in terms of annihilation and
creation operators as 
\begin{equation}
\label{field20}
H_{0}=\sum _{l}\omega _{l}(a_{l}^{\dagger }a_{l}+\frac{1}{2}).
\end{equation}
 Thus we have quantized the electromagnetic field in vacuum. But the aim of
the present thesis is to study the interaction of a single-mode intense electromagnetic
field with a nonlinear medium. Therefore, the interaction part of the Hamiltonian
should be taken into account.

\subsubsection{Conventions used }

Before we go into further detail let us state the conventions which are used
in the present thesis.

i) We work in units in which \( \hbar =1 \). 

ii) The applied electromagnetic field is chosen to be a single mode electromagnetic
field having unit frequency (i.e \( \omega =1 \)).

iii) Mass of the harmonic oscillator equivalent to the applied field is taken
to be unity.

iv) Lower case \( x(t) \) and \( \dot{x}(t) \) are to be treated as time development
of the position and momentum variables in the classical sense and the equivalent
operator representation are to be made by using the corresponding upper case
letters \( X(t) \) and \( \dot{X}(t) \). Thus 
\begin{equation}
\label{field21}
q(t)=X(t)=\frac{1}{\sqrt{2}}\left[ a^{\dagger }+a\right] ,
\end{equation}
 
\begin{equation}
\label{field22}
p(t)=m\dot{q}(t)=m\dot{X}(t)=\dot{X}(t)=\frac{i}{\sqrt{2}}\left[ a^{\dagger }-a\right] ,
\end{equation}
 (since \( m=1 \)) and 
\begin{equation}
\label{1e}
[X(t),\dot{X}(t)]=i.
\end{equation}

\subsection{Construction of the model Hamiltonian}

With the above conventions the free field Hamiltonian (\ref{field20}) of our
system reduces to 
\begin{equation}
\label{field23}
\begin{array}{lcl}
H_{0} & = & \frac{X^{2}}{2}+\frac{\dot{X}^{2}}{2}\\
 & = & a^{\dagger }a+\frac{1}{2}.
\end{array}
\end{equation}
 Again from (\ref{field9}) we observe that 
\begin{equation}
\label{field24}
\mathbf{E}\, \propto \, q\, \propto \, \left( a^{\dagger }+a\right) 
\end{equation}
 and 
\begin{equation}
\label{field25}
\mathbf{B}\, \propto \, p\, \propto \, \left( a^{\dagger }-a\right) .
\end{equation}

Now an intense electromagnetic field interacting with a dielectric medium induces
a macroscopic polarization\index{polarization} (\textbf{\( \mathbf{P} \)})
having a general form 
\begin{equation}
\label{new1}
\mathbf{P}=\chi _{1}\mathbf{E}+\chi _{2}\mathbf{EE}+\chi _{3}\mathbf{EEE}+....
\end{equation}
 where \( \mathbf{E} \) is the electric field and \( \chi _{1} \) is the linear
susceptibility. The parameters \( \chi _{2} \) and \( \chi _{3} \) are second
and third order nonlinear susceptibilities respectively. Presence of the dielectric
medium contributes to the electromagnetic energy density. This contribution
is proportional to \( \mathbf{P}.\mathbf{E} \). Therefore, the interaction
Hamiltonian \( H_{emi} \) is 
\begin{equation}
\label{new2}
H_{emi}=\lambda _{c}\mathbf{P}.\mathbf{E}
\end{equation}
 where \( \lambda _{c} \) is the proportionality constant. Now, if the symmetry
of the medium is chosen in such a way that the only nonlinear interaction in
the medium appears due to the presence of the \( (m-1) \)-th order susceptibility
\( (\chi _{m-1}) \) and the macroscopic magnetization (if any) is neglected,
then the leading contribution to the interaction part of the electromagnetic
energy comes through the coupling of the \( (m-1) \)-th order nonlinear susceptibility.
Therefore, the interaction energy is proportional to the \( m \)-th power of
the electric field. Now using (\ref{field23}-\ref{new2}) the total Hamiltonian
of the system which is a sum of free field and interaction Hamiltonians, can
be written as 
\begin{equation}
\label{ten.1}
\begin{array}{lcl}
H & = & a^{\dagger }a+\frac{1}{2}+\frac{\lambda }{m(2)^{\frac{m}{2}}}(a^{\dagger }+a)^{m}\\
 & = & \frac{X^{2}}{2}+\frac{\dot{X}^{2}}{2}+\frac{\lambda }{m}X^{m}\, .
\end{array}
\end{equation}
 The parameter \( \lambda  \) is the coupling constant \emph{}and is a function
of the \( (m-1) \)-th order nonlinear susceptibility (\( \chi _{m-1} \)) of
the medium. The above Hamiltonian (\ref{ten.1}) represents a generalized anharmonic
oscillator of unit mass and unit frequency. The equation of motion corresponding
to (\ref{ten.1}) is 
\begin{equation}
\label{eqm}
\ddot{X}+X+\lambda X^{m-1}=0
\end{equation}
 which can not be solved exactly for \( m>2 \). Present work provides a first
order operator solution for this generalized anharmonic oscillator and the solution
is used here to study various nonclassical properties of radiation field interacting
with a nonlinear medium.

\subsubsection{Special cases of the model Hamiltonian}

The Hamiltonian (\ref{ten.1}) represents a class of physical systems. Among
those systems \( m=4 \) is a case of particular interest because even order
susceptibilities (\( \chi _{2} \)\( ,\chi _{4} \) etc.) would vanish for an
inversion symmetric medium. Hence the leading contribution to the nonlinear
polarization\index{polarization} in an inversion symmetric medium comes through
the third order susceptibility (\( \chi _{3} \)) and the corresponding Hamiltonian
is 
\begin{equation}
\label{ekpf}
\begin{array}{lcl}
H & = & \frac{X^{2}}{2}+\frac{\dot{X}^{2}}{2}+\frac{\lambda }{4}X^{4}\\
 & = & a^{\dagger }a+\frac{\lambda }{16}(a^{\dagger }+a)^{4}\, .
\end{array}
\end{equation}
This is the Hamiltonian of a quartic\index{quartic} anharmonic oscillator whose
equation of motion is 
\begin{equation}
\label{dui}
\ddot{X}+X+\lambda X^{3}=0.
\end{equation}
Equation (\ref{dui}) is very important in the context of quantum optics, nonlinear
optics, molecular physics, \( \phi ^{4} \) field theory and many other branches
of physics. But equation (\ref{dui}) is not exactly solvable due to the presence
of the off-diagonal terms in the Hamiltonian (\ref{ekpf}). These off-diagonal
interaction terms are called photon number non conserving terms\footnote{
According to Loudon {[}\ref{Loudon}{]} the off-diagonal terms are called energy
nonconserving but Gerry {[}\ref{gerry3}{]} mentioned them as photon number
nonconserving terms. The phrase 'photon number nonconserving' appears to be
more appropriate in the context of the physical process involved. So we mention
the offdiagonal terms as photon number nonconserving terms.
}. The interaction involving the operator \( a^{2} \) (\( a^{\dagger 2} \))
may be viewed as the loss (gain) of two photons. On the other hand, interaction
involving \( a^{\dagger }a \) keeps the photon number conserved (i.e one photon
is created and one photon is destroyed)\emph{.} So the terms proportional to
\( \, a^{2},\, a^{\dagger 2},\, a^{\dagger }a^{3},\, a^{\dagger 3}a,\, a^{4} \)
and \( a^{\dagger 4} \) are photon number nonconserving in nature. It is usually
assumed that the photon number nonconserving terms have no significant contribution
in the time development of the field operators. This assumption may be realized
in the following way. In the absence of interaction, the time development of
the annihilation operator \index{annihilation operator} \( a(t)=a(0)e^{-it} \).
Thus the terms proportional to \( \, a^{2},\, a^{\dagger 2},\, a^{\dagger }a^{3},\, a^{\dagger 3}a,\, a^{4} \)
and \( a^{\dagger 4} \) are rapidly oscillating compared to the terms \( a^{\dagger }a \)
and \( a^{\dagger 2}a^{2} \). These rapidly oscillating terms contribute little
to the interaction Hamiltonian. \emph{}The assumption is widely used and is
called the Rotating Wave Approximation\index{rotating wave approximation} (RWA\index{RWA}).
Under RWA (\ref{ekpf}) assumes an extremely simple and exactly solvable form 

\begin{equation}
\label{charpf}
H=a^{\dagger }a+\lambda _{g}a^{\dagger ^{2}}a^{2}.
\end{equation}

Gerry {[}\ref{Gerry}{]} and Lynch {[}\ref{Lynch}{]} used this rotating wave
approximated Hamiltonian (\ref{charpf}) to study the quantum phase fluctuations\index{quantum phase fluctuations}
of coherent light interacting with a third order nonlinear medium. Joshi \emph{et
al} {[}\ref{Joshi-Pati}{]} used the Hamiltonian (\ref{charpf}) to study the
sensitivity of nonadiabatic geometric phase\index{nonadiabatic geometric phase }
on the initial photon statistics in a dispersive fiber. Tanas {[}\ref{tanas}{]}
used the same Hamiltonian (\ref{charpf}) to study the possibilities of generating
squeezed state\index{squeezed state} in the interaction of single mode electromagnetic
field with a nonabsorbing nonlinear medium.

The RWA\index{RWA} is hardly valid if the noncatalytic nonlinearities are present
in the system {[}\ref{Hoff}{]}. For example, the term proportional to \( a^{\dagger 4} \)
is responsible for higher harmonic generation and occurs naturally corresponding
to the field interacting with a nonlinear medium. In fact, the terms proportional
to \( a^{4} \) and \( a^{\dagger 4} \) were taken into consideration by Tombesi
and Mecozzi {[}\ref{Tombesi}{]} to study the possibilities of squeezing\index{squeezing}
(of coherent light passing through a nonlinear medium). Again the nonconserving
energy terms are responsible for the well known Bloch-Siegert shift\index{Bloch-Siegert shift}
if the field frequency is low (e.g RF and MW) {[}\ref{Bloch-Siegert}{]}. 

Keeping these facts in mind, in the present work, we include the offdiagonal
terms in the model Hamiltonian and observe that important physical information
is lost under RWA\index{RWA} calculations. In spite of working with \( \chi _{3} \)
medium only, here we study \( (m-1) \)-th order nonlinear medium in general
with particular emphasis to the third order nonlinear medium, i.e. to the case
of a quartic\index{quartic} anharmonic oscillator. 

Depending upon the order of the anharmonicity the anharmonic oscillators are
identified by different names such as quartic\index{quartic} and sextic\index{sextic}
anharmonic oscillators. Hamiltonian of these oscillators, their equations of
motion and classical counter parts of the equations of motion are required by
us at different stages of the development of the present work. So here we give
a brief table, containing this information.

\vspace{0.3cm}
{\centering \begin{tabular}{|c|c|c|c|c|c|}
\hline 
{\scriptsize Anharmonic}&
{\scriptsize Classical }&
{\scriptsize Classical equation}&
{\scriptsize Quantum }&
{\scriptsize Quantum equation}&
{\scriptsize Nonvanishing }\\
{\scriptsize oscillator}&
{\scriptsize Hamiltonian}&
{\scriptsize of motion}&
{\scriptsize Hamiltonian}&
{\scriptsize of motion}&
{\scriptsize susceptibility}\\
\hline 
{\scriptsize Generalized}&
{\scriptsize \( \frac{x^{2}}{2}+\frac{\dot{x}^{2}}{2}+\frac{\lambda }{m}x^{m} \)}&
{\scriptsize \( \ddot{x}+x+\lambda x^{m-1}=0 \)}&
{\scriptsize \( \frac{X^{2}}{2}+\frac{\dot{X}^{2}}{2}+\frac{\lambda }{m}X^{m} \)}&
{\scriptsize \( \ddot{X}+X+\lambda X^{m-1}=0 \)}&
{\scriptsize \( \chi _{m-1} \)}\\
\hline 
{\scriptsize Quartic\index{quartic}}&
{\scriptsize \( \frac{x^{2}}{2}+\frac{\dot{x}^{2}}{2}+\frac{\lambda }{4}x^{4} \)}&
{\scriptsize \( \ddot{x}+x+\lambda x^{3}=0 \)}&
{\scriptsize \( \frac{X^{2}}{2}+\frac{\dot{X}^{2}}{2}+\frac{\lambda }{4}X^{4} \)}&
{\scriptsize \( \ddot{X}+X+\lambda X^{3}=0 \)}&
{\scriptsize \( \chi _{3} \)}\\
\hline 
{\scriptsize Sextic\index{sextic}}&
{\scriptsize \( \frac{x^{2}}{2}+\frac{\dot{x}^{2}}{2}+\frac{\lambda }{6}x^{6} \)}&
{\scriptsize \( \ddot{x}+x+\lambda x^{5}=0 \)}&
{\scriptsize \( \frac{X^{2}}{2}+\frac{\dot{X}^{2}}{2}+\frac{\lambda }{6}X^{6} \)}&
{\scriptsize \( \ddot{X}+X+\lambda X^{5}=0 \)}&
{\scriptsize \( \chi _{5} \)}\\
\hline 
{\scriptsize Octic}{\small \index{octic}}{\scriptsize }&
{\scriptsize \( \frac{x^{2}}{2}+\frac{\dot{x}^{2}}{2}+\frac{\lambda }{8}x^{8} \)}&
{\scriptsize \( \ddot{x}+x+\lambda x^{7}=0 \)}&
{\scriptsize \( \frac{X^{2}}{2}+\frac{\dot{X}^{2}}{2}+\frac{\lambda }{8}X^{8} \)}&
{\scriptsize \( \ddot{X}+X+\lambda X^{7}=0 \)}&
{\scriptsize \( \chi _{7} \)}\\
\hline 
\end{tabular}\scriptsize \par}
\vspace{0.3cm}

{\par\centering Table 1.1\par}

The model Hamiltonian (\ref{ten.1}) is not exactly solvable in a closed analytical
form. So we have two alternatives, either we have to approximate the Hamiltonian
itself by neglecting some terms (as it is done in RWA\index{RWA}) or we have
to find approximate solution of the equation of motion (\ref{eqm}) corresponding
to the Hamiltonian (\ref{ten.1}). The first option is already tested for \( \chi _{3} \)
medium by various authors {[}\ref{Gerry}-\ref{tanas}{]}. Therefore, here we
approach the problem by using the second option. The approximate solution can
be obtained by different approaches. In the following we discuss the different
possible approaches to the AHO model.

\section{Different approaches to the model}

In the last section interaction of an intense electromagnetic field with an
\( (m-1) \)-th order nonlinear medium is modeled by a generalized anharmonic
oscillator. The equation of motion corresponding to this model Hamiltonian is
not exactly solvable and, in-fact, this provides one of the simplest examples
of quantum mechanical systems which can not be solved without making use of
approximation techniques. Approximate solution of the AHO problem can be obtained
in three different pictures in quantum mechanics. These approaches are called
Schr\"{o}dinger picture\index{Schrodinger picture}, Heisenberg picture\index{Heisenberg picture}
and interaction picture\index{Intercation picture}. At first we discuss these
approaches briefly and after that we choose the appropriate one for the further
development of the present study.

\subsection{Schr\"{o}dinger picture\index{Schrodinger picture}}

The Schr\"{o}dinger picture deals with the time development of the wave function.
Basically, the Schr\"{o}dinger picture is used to solve the quantum oscillators
as eigenvalue\index{eigenvalue} problems. In these problems, energy eigenvalues
are expressed as the sum of different orders of anharmonic constant \( \lambda  \).
However, the energy eigenvalues are found to diverge for large anharmonic constant.
In case of small \( \lambda  \), the eigenvalues for different orders can be
summed up and the convergence of these sums are ensured by the Borel summability
{[}\ref{graffi}, \ref{borel-pade}{]} and/or Stieljet conditions {[}\ref{Loeffel}{]}.
Schr\"{o}dinger approach\index{Schrodinger picture} is found very successful
to treat the quantum anharmonic oscillator problem and there exists an extensive
amount of literature on this approach {[}\ref{graffi}-\ref{C-num}{]}. Schr\"{o}dinger
approach \index{Schrodinger picture} is also successfully used to enrich the
subject of large order perturbation theory {[}\ref{bendern1}-\ref{lowdin}{]},
divergent expansion of quantum mechanics {[}\ref{graffi}, \ref{simonn2} and
\ref{Simon}{]}, Laplace transformation representation of energy eigenvalues
{[}\ref{ivanov}{]} and computational physics {[}\ref{weniger}-\ref{fernandez}{]}.

\subsection{Heisenberg picture\index{Heisenberg picture}}

In this approach, one solves Heisenberg equation\index{Heisenberg equation}
of motion to obtain the time evolution of the operators. But the noncommuting
nature of the operators poses serious difficulties to obtain approximate operator
solutions of the quantum anharmonic oscillators. Only few methods are available
till date. In 1967 Aks {[}\ref{aks1}{]} obtained an operator solution of a
quantum quartic\index{quartic} anharmonic oscillator (QQAHO\index{QQAHO})
by using the renormalization technique. In 1970 Aks and Caharat {[}\ref{aks2}{]}
reported an extension of the earlier work by using the method of Bogoliubov
and Krylov\index{Bogoliubov-Krylov method}. But the recent interest in the
problem started with the work of Bender and Bettencourt {[}\ref{Bender1}-\ref{Bender2}{]}
who obtained the solution of a QQAHO\index{QQAHO} using multiscale perturbation
theory (MSPT\index{MSPT}) in 1996. Later on Mandal {[}\ref{Mandal}{]} proposed
a Taylor series\index{Taylor series} approach, Egusquiza and Basagoiti {[}\ref{Egus}{]}
developed renormalization technique, Kahn and Zarmi {[}\ref{kahn}{]} used a
near identity transform method\index{near identity transform method} to solve
the QQAHO\index{QQAHO} problem. Fern\'{a}ndez {[}\ref{F01a},\ref{F01b}{]}
used an eigenvalue approach\index{eigenvalue approach} to solve QQAHO\index{QQAHO}. 

Some of the recently derived solutions are obtained as a part of the present
work {[}\ref{pathak2}-\ref{APFM01}{]} and are used here to study quantum optical
and nonlinear optical effects. Therefore, these solutions are much important
in the context of the present thesis. Keeping that in mind, here we give a list
of the solutions of AHO problem available in Heisenberg picture\index{Heisenberg picture}.

\vspace{0.3cm}
{\centering \begin{tabular}{|l|c|l|l|}
\hline 
Authors&
Year&
Method&
Oscillator\\
\hline 
\hline 
Aks {[}\ref{aks1}{]}&
1967&
Renormalization&
Quartic\index{quartic}\\
\hline 
Aks and Caharat {[}\ref{aks2}{]}&
1969&
Bogoliubov and Krylov&
Quartic\\
\hline 
Bender and Bettencourt {[}\ref{Bender1}{]}&
1996&
MSPT\index{MSPT}&
Quartic\\
\hline 
Mandal {[}\ref{Mandal}{]}&
1998&
Taylor series\index{Taylor series}&
Quartic\\
\hline 
Egusquiza and Basagoiti {[}\ref{Egus}{]}&
1998&
Renormalization Group\index{renormalization group}&
Quartic\\
\hline 
Kahn and Zarmi {[}\ref{kahn}{]}&
1999&
Near identity transform\index{near identity transform method}&
Quartic\\
\hline 
Speliotopoulos {[}\ref{S00}{]}&
2000&
nonperturbative &
Quartic\\
\hline 
Pathak {[}\ref{pathak2}{]}&
2000&
MSPT&
Generalized\\
\hline 
Pathak and Mandal {[}\ref{PM01}{]}&
2001&
Taylor series\index{Taylor series}&
Quartic\\
\hline 
Pathak and Mandal {[}\ref{Pathak}{]}&
2001&
Taylor Series&
Sextic{\footnotesize \index{sextic}} and Octic\index{octic}\\
\hline 
 Fern\'{a}ndez {[}\ref{F01a}{]}&
2001&
eigenvalue approach\index{eigenvalue approach}&
Quartic\\
\hline 
 Fern\'{a}ndez {[}\ref{F01b}{]}&
2001&
eigenvalue approach&
Particular values of \( m \)\\
\hline 
Pathak and Fern\'{a}ndez {[}\ref{APFM01}{]}&
2001&
All the existing methods&
Generalized\\
\hline 
\end{tabular}\par}
\vspace{0.3cm}

{\par\centering Table 1.2\par}

\subsection{Interaction picture\index{Intercation picture}}

In this approach the time dependence of the physically measurable quantity is
described by letting both observable and state vary with time. The observables
are rotated in one direction with a transformation generated by part of the
Hamiltonian, and the states are rotated in the opposite direction by a transformation
generated by the other part (the interaction part) of the Hamiltonian. In case
of anharmonic oscillator there is a simple relation between the annihilation
operator \index{annihilation operator} obtained in Heisenberg approach and
that obtained in interaction picture. Only one solution {[}\ref{APFM01}{]}
of general anharmonic oscillator in this picture is reported till date.

\section{Quantum fluctuations and appropriate approach for our study}

Second order variance \( (\triangle A)^{2}=<A^{2}>-<A>^{2} \) is a measure
of quantum fluctuations associated with an arbitrary quantum mechanical observable
\( A \). In the present work quantum fluctuations (variance) are calculated
with respect to initial the coherent state\index{coherent state} \( |\alpha > \).
There are substantial quantum fluctuations even in a pure coherent state. For
example, the quadrature fluctuation in \( |\alpha > \) is 
\begin{equation}
\label{qf1}
(\triangle X)^{2}=\frac{1}{2}\, and\, (\triangle \dot{X})^{2}=\frac{1}{2}
\end{equation}
 while the fluctuation in photon number is 
\begin{equation}
\label{qf2}
(\triangle N)^{2}=<N>=|\alpha |^{2}
\end{equation}
where \( N=a^{\dagger }a \), is the number operator. Further randomization
in the case of general classical states which are random superposition of coherent
states\index{coherent state} can only increase theses fluctuations. But there
are some phenomena in which quantum fluctuations reduce below the coherent state\index{coherent state}
level. Let us give some examples, \\
\textbf{Example 1:} An electromagnetic field is said to be electrically squeezed
field if uncertainties in the quadrature phase observable \( X \) is less than
the coherent state\index{coherent state} level (i.e. \( (\triangle X)^{2}<\frac{1}{2} \)).
Correspondingly a magnetically squeezed field is one for which \( (\triangle \dot{X})^{2}<\frac{1}{2}. \)
\\
\textbf{Example 2:} In the phenomenon of antibunching\index{antibunching} (single
mode) the fluctuation in photon number is reduced below the Poisson level (\ref{qf2}),
\begin{equation}
\label{2two}
(\triangle N)^{2}<\, <N>.
\end{equation}
 \textbf{Example 3:} The usual parameters used for the calculation of the quantum
phase fluctuations \index{quantum phase fluctuations} are defined as {[}\ref{Gerry}-\ref{Lynch}{]}
\begin{equation}
\label{chkuripf}
U\left( \theta ,t,|\alpha |^{2}\right) =(\Delta N)^{2}\left[ (\Delta S)^{2}+(\Delta C)^{2}\right] \left/ \left[ <S>^{2}+<C>^{2}\right] \right. 
\end{equation}

\begin{equation}
\label{chekushpf}
S\left( \theta ,t,|\alpha |^{2}\right) =(\Delta N)^{2}(\Delta S)^{2}
\end{equation}
 and

\begin{equation}
\label{chbishpf}
Q\left( \theta ,t,|\alpha |^{2}\right) =S\left( \theta ,t,|\alpha |^{2}\right) \left/ <C>^{2}\right. 
\end{equation}
 where \( S \) and \( C \) are sine and cosine operators respectively (explicit
definitions are given in chapter 3). Reduction of these quantum phase fluctuation
parameters from their initial values is possible for some particular values
of \( \theta  \) and \( t \). When \( S(\theta ,t,|\alpha |^{2}) \) gets
reduced to below its coherent state\index{coherent state} value then at least
one nonclassical effect (either magnetically squeezed electromagnetic field
or antibunching\index{antibunching} of photons or both) is observed.

Reduced fluctuation states can not be represented by \( P \)-representation.
Therefore, they are in the paradigm of nonclassical states. In the present thesis
we study the possibilities of observing these nonclassical states in the interaction
of a coherent electromagnetic field with a nonlinear medium. From the above
examples it is clear that in order to study the possibilities of observing nonclassical
states we have to calculate quantum fluctuations in different observables (e.g.
\( N,S,C,X,\dot{X}\,  \)etc.). Again if we have to calculate quantum fluctuations
in any of these observables then we have to know the time evolution of creation
\index{creation operator} and annihilation operator\index{annihilation operator}s.
Therefore, we have to obtain operator solution of the equation of motion of
the anharmonic oscillator. This is why, we have to work either in Heisenberg
picture \index{Heisenberg picture} or interaction picture. In the next chapter
we obtain the operator solutions and use them to study different quantum optical
and nonlinear optical phenomena in the subsequent chapters.

Chapter 2 deals with the operator solution of the equations of motion (\ref{eqm})
corresponding to our model Hamiltonian (\ref{ten.1}). The works reported in
this chapter are arranged according to the time of their appearance to give
a clear exposure of the time development of the subject as well as that of our
work. For example, at first we report classical and quantum solutions of quartic,
sextic{\footnotesize \index{sextic}} and octic\index{octic} anharmonic oscillators
by using Taylor series \index{Taylor series} technique. While working with
Taylor series\index{Taylor series}, we observe that the approach is lengthy
and time consuming. But many intrinsic symmetries of the problem is made transparent
to us. For example, we observe that first order correction to the frequency
operator\index{frequency operator} is always a function of the unperturbed
Hamiltonian (\( H_{0} \))\footnote{
An extensive proof of this observation is given by Speliotopoulos {[}\ref{S00}{]}.
}. Using these observations and imposing a physical condition that a correct
solution should give correct frequency shift we have succeeded to generalize
the results of Bender and Bettencourt {[}\ref{Bender1}{]} and have obtained
the solution of the generalized AHO {[}\ref{pathak2}{]}. Later on we have generalized
the quartic oscillator solutions obtained by various other techniques and observed
that an apparent discrepancy is present between the solutions obtained by different
techniques. Then the question arises: Which solution should be used for our
physical calculations? Therefore, at the end of chapter 2 we compare different
solutions and conclude that all the correct solutions are equivalent and the
apparent discrepancy is due to the use of different ordering of the operators.
Showing this equivalence we become mathematically equipped to investigate the
possibilities of observing different nonlinear optical phenomena in a nonlinear
dielectric medium.

In chapter 3 we study quantum phase fluctuations\index{quantum phase fluctuations}
of coherent light. We begin by asking the question: How can one write down a
quantum mechanical operator corresponding to the phase of a harmonic oscillator
or equivalently a single mode of electromagnetic field? This question is the
statement of the quantum phase\index{quantum phase} problem. Since Dirac had
introduced this problem in 1927 {[}\ref{Dirac}{]} many people tried to provide
a satisfactory Hermitian phase operator\index{phase operator}. As a result
of these attempts there exist different definitions of the phase operator\index{phase operator}.
In chapter 3 we shortly review a few of them and choose Pegg Barnett approach
{[}\ref{PB phys rev a}, \ref{PB europhys letters}{]} to calculate the quantum
phase fluctuation parameters \( S(\theta ,t,|\alpha |^{2}),Q(\theta ,t,|\alpha |^{2}) \)
and \( U(\theta ,t,|\alpha |^{2}) \) by using the operator solutions of quartic\index{quartic}
anharmonic oscillator derived in chapter2. These fluctuation parameters were
already studied by Gerry {[}\ref{Gerry}{]} and Lynch {[}\ref{Lynch}{]} for
RW approximated Hamiltonian (\ref{charpf}) of the same system. Here we observe
that enhancement as well as reduction of \( S(\theta ,t,|\alpha |^{2}),\, Q(\theta ,t,|\alpha |^{2}) \)
and \( U(\theta ,t,|\alpha |^{2}) \) with \( |\alpha |^{2} \) (number of photons
present before the interaction) is possible for different values of the free
evolution time \( t \) and phase \( \theta  \) of the input coherent light.
This observation is in sharp contrast with the earlier results {[}\ref{Gerry},
\ref{Lynch}{]}. Thus the importance of inclusion of photon number nonconserving
terms in the model Hamiltonian is established. This interesting observation
of chapter 3 has inspired us to study the other nonlinear optical effects with
the same model.

In chapter 4 we study the possibilities of generating electromagnetically squeezed
light in a third order nonlinear medium. At first an analytic expression for
the quantum fluctuation in phase quadrature \( X \) is obtained (up to the
second power in \( \lambda  \)). We observe that an electrically squeezed electromagnetic
field can be produced due to the interaction of an intense beam of coherent
light with a third order nonlinear medium. A few special cases are discussed. 

In chapter 5 quantum statistical properties of the radiation field is discussed.
To be more precise, bunching\index{bunching} and antibunching\index{antibunching}
of photons and the photon number distribution\index{photon number distribution}
(PND\index{PND}) is studied. We observe that we can obtain nonclassical phenomenon
of antibunching\index{antibunching} due to the interaction of an electromagnetic
field with third order nonlinear medium of inversion symmetry for particular
values of interaction time and phase of the input coherent state\index{coherent state}.

In chapter 6 our studies on third order nonlinear medium is extended to the
\( (m-1) \)-th order nonlinear medium in general. We study the nonadiabatic
geometric phase\index{nonadiabatic geometric phase} or Aharonov Anandan phase\index{Aharonov Anandan phase}
for our physical system. Our works of chapter 4 and 5 are extended to the case
of \( (m-1) \)-th order nonlinear medium. Possibilities of generating nonclassical
states in an \( (m-1) \)-th order nonlinear medium are also discussed.

Finally in chapter 7 an outlook on the present work is given. In particular
we talk about the limitations of our endeavor and future scope of investigations
along the line of thought indicated by us.

\chapter{Mathematical preliminaries}

Physical scientists have always tried to understand the Nature in terms of simple
models. The simple harmonic oscillator (SHO) is perhaps the most useful one
among them. A particle subject to a restoring force proportional to its displacement
gives rise to the model of a SHO. The Hamiltonian corresponding to a classical
SHO of unit mass and unit frequency is given by 
\begin{equation}
\label{1a}
H=\frac{p^{2}}{2}+\frac{x^{2}}{2}\, .
\end{equation}
 The Hamilton's equations for the SHO are 
\begin{equation}
\label{1b}
\begin{array}{lcl}
\dot{p} & = & -\frac{\partial H}{\partial x}=-x\\
\dot{x} & = & \frac{\partial H}{\partial p}=p\, .
\end{array}
\end{equation}
 Therefore, the equation of motion of the SHO is 
\begin{equation}
\label{1ee}
\ddot{x}+x=0
\end{equation}
 where equation (\ref{1b}) is used. The solution of the equation (\ref{1ee})
is found to be 
\begin{equation}
\label{1d}
x_{0}(t)=x(0)\cos t+\dot{x}(0)\sin t.
\end{equation}
 The parameters \( x(0) \) and \( \dot{x}(0) \) are the initial position and
momentum of the oscillator. The subscript '\( 0 \)' denotes the zeroth order
(i.e \( \lambda =0 \)) solution. Thus the position of the oscillator at a later
time \( t \) is completely known in terms of the initial position and momentum.
Nevertheless, the momentum is also known from the Hamilton's equations (\ref{1b}).
Hence, the oscillator problem is completely solved.

The equation of motion of a quantum SHO may simply be obtained by imposing \( x \)
and \( \dot{x} \) as noncommuting operators and the solution of a quantum SHO
in Heisenberg picture \index{Heisenberg picture} is the operator equivalent
of the equation (\ref{1d}). 

Now for the real physical problems, anharmonicity and/or damping are to be incorporated
in the model Hamiltonian and hence in the equation of motion. However, in the
previous chapter we have seen that only anharmonic term appears in the model
Hamiltonian of our physical system. The general form of the Hamiltonian of a
classical anharmonic oscillator having unit mass and unit frequency is given
by 
\begin{equation}
\label{1g}
H\, =\, \frac{p^{2}}{2}+\frac{x^{2}}{2}+\frac{\lambda }{m}x^{m}
\end{equation}
 where \( m\, (\geq 3) \) is an integer and \( \lambda  \) is the anharmonic
constant. Depending upon the problem of physical interest, different types of
anharmonic oscillators appear. Of course, the anharmonic constant \( \lambda  \)
is different for different types of anharmonic oscillators. The equation of
motion corresponding to the Hamiltonian (\ref{1g}) is 
\begin{equation}
\label{1h}
\ddot{x}+x+\lambda x^{m-1}=0.
\end{equation}
 where the equation (\ref{1b}) is used. We need an operator solution of (\ref{eqm})
to study the quantum fluctuations of coherent light in nonlinear media. But
equation of motion (\ref{eqm}) and its classical counterpart (\ref{1h}) are
not exactly solvable in a closed analytical form\footnote{
Exact solution of (\ref{1h}) can be obtained in terms of elliptic functions.
However, that solution can not be used to predict the trajectory of the particle
executing anharmonic motion. So the solution obtained in terms of elliptic functions
are not exact in a strict sense.
}. Keeping this in view, present chapter is devoted to the study of classical
and quantum anharmonic oscillator problem. This problem is already investigated
by several authors. To begin with, we recall the problem of a classical quartic\index{quartic}
oscillator (Duffing oscillator).

\section{Classical and quantum solutions of the quartic\index{quartic} oscillator}

For \( m=4 \), equation (\ref{1h}) has an exact solution in terms of the elliptic
function {[}\ref{Nayfeh}{]}. This solution is available only in the phase plane.
For positive anharmonic constant \( (\lambda >0), \) the solution is periodic
with a fixed center. The period is also obtained in terms of the elliptic functions.
However, this approach is not useful in many occasions. For example, it is not
possible to predict the trajectory of the particle executing quartic\index{quartic}
anharmonic motion. For this reason, the solution in the phase plane is hardly
useful except in the nonlinear dynamical studies. Keeping the above facts in
view, a large number of approximate methods have been devised for the purpose
of getting an analytical solution of the classical Duffing oscillator problem.
These include perturbation technique {[}\ref{Nayfeh}{]}, variation of parameters
{[}\ref{Ross}{]} and Taylor series \index{Taylor series} approaches {[}\ref{Marganeau}{]}.
The ordinary perturbation technique, a pedestrian approach, leads to the unwanted
secular terms {[}\ref{Nayfeh}{]}. The removal of secular terms from the solution
is a serious problem. There are some methods which are successfully used to
sum up the secular terms for all orders. These include ``tucking in technique\index{tucking in technique}''
{[}\ref{Bellmann}{]}, multiscale perturbation theory (MSPT\index{MSPT}) and
renormalization technique {[}\ref{Nayfeh}{]}.

The noncommuting nature of the operators poses serious difficulties for the
purpose of getting approximate operator solutions to the quantum anharmonic
oscillators. However, few methods are available for the purpose. We have already
mentioned these methods {[}\ref{aks1}-\ref{S00}{]} in 1.2.2. In the next two
subsections of the present thesis we derive solutions of classical and quantum
quartic\index{quartic} anharmonic oscillators respectively. Actually, at first
we develop a second order solution for the QQAHO\index{QQAHO} using the Taylor
series \index{Taylor series} approach and after that, we construct a first
order solution for \( m-th \) oscillator in general using various other techniques.
Solutions obtained by different techniques are compared in the subsequent subsections.
The solutions are used to study the interaction of the strong electromagnetic
field with a dielectric medium in the later part of the thesis.

\subsection{{\small Classical and quantum solution of quartic\index{quartic} oscillator }\small }

Let us attempt a second order solution of the QQAHO\index{QQAHO} by using the
Taylor series \index{Taylor series} approach devised by Mandal {[}\ref{Mandal}{]}.
At first we develop the solution of the QAHO in classical picture and after
that a straight forward generalization is made for getting the corresponding
solution in quantum picture.  

The differential equation of the classical Duffing oscillator of unit mass and
unit frequency is 
\begin{equation}
\label{1}
\ddot{x}\, +x\, +\lambda x^{3}\, =\, 0
\end{equation}
 where the parameter \( \lambda  \) is the anharmonic constant of the system.
The solution of the equation (\ref{1}) may be written as the sum of different
orders of anharmonicities. Thus the corresponding solution is 
\begin{equation}
\label{2}
x(t)\, =x_{0}(t)\, +x_{1}(t)\, +x_{2}(t)+\ldots 
\end{equation}
 where \( x_{0}(t) \), \( x_{1}(t) \) and \( x_{2}(t) \) are zeroth, first
and second order solutions respectively. The first and second order solutions
are proportional to \( \lambda  \) and \( \lambda ^{2} \) respectively. The
zeroth and first order solutions are available from the works of Mandal {[}\ref{Mandal}{]}
and are given by 
\begin{equation}
\label{2a}
x_{0}(t)=x(0)\cos t+\dot{x}(0)\sin t
\end{equation}
 and 
\begin{equation}
\label{2b}
\begin{array}{lcl}
x_{1}(t) & = & -\frac{\lambda x^{3}(0)}{32}(\cos t-\cos 3t+12t\sin t)\\
 & + & \frac{3\lambda x^{2}(0)\dot{x}(0)}{32}(\sin 3t-7\sin t+4t\cos t)\\
 & - & \frac{3\lambda x(0)\dot{x}^{2}(0)}{32}(\cos 3t-\cos t+4t\sin t)\\
 & - & \frac{\dot{x}^{3}(0)}{32}(\sin 3t+9\sin t-12t\cos t)
\end{array}
\end{equation}
 respectively. The purpose of the present subsection is to find the second order
solution \( x_{2}(t) \) of the classical quartic\index{quartic} oscillator
and to introduce the Taylor series \index{Taylor series} method which is also
used here to deal with the higher anharmonic oscillators. The solution of the
classical Duffing oscillator (without forcing term) is given by the following
Taylor series \index{Taylor series} {[}\ref{Marganeau}{]}
\begin{equation}
\label{3}
x(t)=x(0)\, +t\dot{x}(0)\, +\frac{t^{2}}{2!}\ddot{x}(0)+\ldots \, \, .
\end{equation}
The assumed Taylor series \index{Taylor series} is expanded about the origin.
Furthermore, we assume that \( t \) is sufficiently small and is such that
the above series (\ref{3}) expansion is possible. Now  we express higher order
time derivatives of \( x(t) \) at \( t=0 \) 
\begin{equation}
\label{4}
\begin{array}{lcl}
\ddot{x}^{\cdot }(0) & = & -\dot{x}(0)-3\lambda x^{2}(0)\dot{x}(0)\\
\dot{x}^{\cdots }(0) & = & x(0)+4\lambda x^{3}(0)-6\lambda x(0)\dot{x}^{2}(0)+3\lambda ^{2}x^{5}(0)\\
\ddot{x}^{\cdots }(0) & = & \dot{x}(0)+24\lambda x^{2}(0)\dot{x}(0)-6\lambda \dot{x}^{3}(0)+27\lambda ^{2}x^{4}(0)\dot{x}(0)\\
\ldots  & \ldots  & \ldots \\
\ldots  & \ldots  & \ldots 
\end{array}
\end{equation}
where we neglect the terms beyond \( \lambda ^{2} \) and always substitute
\( \ddot{x} \) by \( -x-\lambda x^{3} \). The relations (\ref{4}) are substituted
back in equation (\ref{3}) and the coefficients of \( x^{5}(0),\, x^{4}(0)\dot{x}(0),\, x^{3}(0)\dot{x}^{2}(0),\, x^{2}(0)\dot{x}^{3}(0),\, x(0)\dot{x}^{4}(0)\,  \)
and \( \dot{x}^{5}(0) \) are collected. The corresponding coefficients are
\begin{equation}
\label{5}
\begin{array}{lcl}
C_{1} & = & \lambda ^{2}(3-51+846-15078+300705-6633081+\ldots )\\
C_{2} & = & \lambda ^{2}(27-639+13230-284094+\ldots )\\
C_{3} & = & \lambda ^{2}(126-3888+98820-2440224+\ldots )\\
C_{4} & = & \lambda ^{2}(378-14688+429084-11416464+\ldots )\\
C_{5} & = & \lambda ^{2}(756-33156+1023948-27952668+\ldots )\\
and &  & \\
C_{6} & = & \lambda ^{2}(756-33156+1023948-27952668+\ldots )
\end{array}
\end{equation}
where the contributions from Taylor expansion part are not taken into account.
These contributions are taken care of in the later part of this work. Constructing
difference equations for the above series and solving them we can obtain the
\( r-th \) term of the coefficient \( C_{n} \). For example, the \( r-th \)
term of \( C_{1} \) is given by 
\begin{equation}
\label{6}
\begin{array}{lcl}
t^{\prime }_{r} & = & \left( -1\right) ^{r+1}\times \frac{\lambda ^{2}}{1024}\left( 25^{r+1}+216r\times 9^{r}+288r^{2}+240r-25\right) .\\
 &  & 
\end{array}
\end{equation}
Now we can write the \( r-th \) term of the coefficient of \( x^{5}(0) \)
as 

\begin{equation}
\label{7}
t_{r}=t^{\prime }_{r}\times \frac{t^{2r+2}}{(2r+2)!},
\end{equation}
 where the factor \( \frac{t^{2r+2}}{(2r+2)!} \) comes from the Taylor expansion
part. The net coefficient of \( x^{5}(0) \) is obtained as 
\begin{equation}
\label{8}
\begin{array}{lcl}
K_{1}=\sum ^{\infty }_{r=0}t_{r} & = & \frac{\lambda ^{2}}{1024}(\cos 5t-36\, t\sin 3t-24\, \cos 3t\\
 & - & 72\, t^{2}\cos t+96\, t\sin t+23\, \cos t).
\end{array}
\end{equation}
The parameter \( K_{1} \) exhibits fifth and third harmonic generations. The
remaining coefficients are also obtained by using the similar procedure as adopted
for the evaluation of \( K_{1}. \) The corresponding coefficients are

\begin{equation}
\label{ch9}
\begin{array}{lcl}
K_{2} & = & \frac{\lambda ^{2}}{1024}(5\sin 5t+108t\cos 3t-132\sin 3t-72t^{2}\sin t+599\sin t-336t\cos t)\\
K_{3} & = & \frac{2\lambda ^{2}}{1024}(-5\cos 5t+90\cos 3t+36t\sin 3t-72t^{2}\cos t-85\cos t+264t\sin t)\\
K_{4} & = & \frac{2\lambda ^{2}}{1024}(-5\sin 5t+36t\cos 3t+6\sin 3t-72t^{2}\sin t+427\sin t-456t\cos t)\\
K_{5} & = & \frac{\lambda ^{2}}{1024}(5\cos 5t+108t\sin 3t+108\cos 3t-72t^{2}\cos t-113\cos t+240t\sin t)\\
and &  & \\
K_{6} & = & \frac{\lambda ^{2}}{1024}(\sin 5t-36t\cos 3t+48\sin 3t-72t^{2}\sin t+271\sin t-384t\cos t)\, .
\end{array}
\end{equation}
Hence the desired second order solution is given by
\begin{equation}
\label{10}
\begin{array}{lcl}
x_{2}(t) & = & K_{1}x^{5}(0)+K_{2}x^{4}(0)\dot{x}(0)+K_{3}x^{3}(0)\dot{x}^{2}(0)\\
 & + & K_{4}x^{2}(0)\dot{x}^{3}(0)+K_{5}x(0)\dot{x}^{4}(0)+K_{6}\dot{x}^{5}(0).
\end{array}
\end{equation}
The total solution (\ref{2}) is simply the sum of (\ref{2a}), (\ref{2b})
and (\ref{10}). For \( \dot{x}(0)=0 \) and \( x(0)=A \), the total solution
reduces to an extraordinarily simple form 
\begin{equation}
\label{11class}
\begin{array}{lcl}
x(t) & = & A\cos t-\frac{\lambda A^{3}}{32}(\cos t-\cos 3t+12t\sin t)+\frac{\lambda ^{2}A^{5}}{1024}(\cos 5t\\
 & - & 36\, t\sin 3t-24\, \cos 3t-72\, t^{2}\cos t+96\, t\sin t+23\, \cos t).
\end{array}
\end{equation}
It is clear that the presence of the secular terms proportional to \( t\sin t \),
\( t\sin 3t \) and \( t^{2}\cos t \) pose a serious difficulty as \( t \)
grows. For weak coupling case \( (i.e\, \lambda \, \ll 1), \) these terms may
be summed up with the help of various summation techniques. However, for strong
coupling the summation technique would fail and would give divergent solutions.
We confine ourselves in the weak coupling regime and hence the summation for
different orders of \( \lambda  \) is possible. For weak coupling, the secular
term in the first order solution is known to produce the frequency shift of
the oscillator {[}\ref{Mandal}{]}. In the present case similar frequency shift
may be observed in following way. Equation (\ref{11class}) can be rearranged
as 
\begin{equation}
\label{12class}
\begin{array}{lcl}
x(t) & = & A\left[ 1-\frac{(\frac{3}{8}\lambda A^{2}t)^{2}}{2!}\right] \cos t-A(\frac{3}{8}\lambda A^{2}t)\sin t-\frac{\lambda A^{3}}{32}\left[ (\cos t-\frac{3}{8}\lambda A^{2}t\sin t)\right. \\
 & - & \left. (\cos 3t-\frac{9}{8}\lambda A^{2}t\sin 3t)\right] +\frac{\lambda ^{2}A^{5}}{1024}[\cos 5t-24\, \cos 3t+23\, \cos t]+\frac{21\lambda ^{2}A^{5}}{256}t\sin t.
\end{array}
\end{equation}
 Now we use the tucking in technique\index{tucking in technique} to remove
the secular terms. As we are dealing with the second order solution we can write
\begin{equation}
\label{13a}
\begin{array}{ccc}
\sin (b\lambda t) & = & b\lambda t\\
\cos (b\lambda t) & = & 1-\frac{(b\lambda t)^{2}}{2!}\\
(b'\lambda )\sin (b\lambda t) & = & b'b\lambda ^{2}t\\
(b'\lambda )\cos (b\lambda t) & = & b'\lambda 
\end{array}
\end{equation}
where \( b \) and \( b' \) are constants and the terms beyond \( \lambda ^{2} \)
are neglected. Thus the secular terms can be removed from the solution and finally
the solution (\ref{12class}) reduces to 
\begin{equation}
\label{13}
x(t)=A\cos \omega ^{/}t-\frac{\lambda A^{3}}{32}(\cos \omega ^{/}t-\cos 3\omega ^{/}t)+\frac{\lambda ^{2}A^{5}}{1024}(\cos 5\omega ^{/}t-24\cos 3\omega 't+23\cos \omega ^{/}t).
\end{equation}
The shifted frequency is given by 
\begin{equation}
\label{13b}
\omega ^{/}=1+\frac{3}{8}\lambda A^{2}-\frac{21}{256}\lambda ^{2}A^{4}
\end{equation}
which coincides exactly with the frequency obtained by using Laplace transform
{[}\ref{Pipes}{]}. Thus, the existing second order solution of a QAHO {[}\ref{Pipes}{]}
is reproduced as a special case \( (i.e\, \dot{x}(0)=0,\, x(0)=A) \) of our
solution (\ref{2}). For \( \dot{x}(0)=0,x(0)=A \) the frequency shift of the
oscillator due to quartic\index{quartic} anharmonicity is \( (\frac{A^{2}}{8}-\frac{7\lambda A^{4}}{256})3\lambda . \)
The frequency shift due to the first term \( (3\lambda A^{2}/8) \) dominates
over the frequency shift due to the second term \( (21\lambda ^{2}A^{4}/256) \)
as long as \( \lambda  \) is small. Hence, the shift in the frequency increases
with the increase of \( \lambda  \) till the contribution of the second term
compensates the contribution of the first one. Further increase of \( \lambda  \)
causes a frequency shift in the opposite direction. Thus the second order correction
to the frequency shift is quite consistent with the physical system. Now to
have an idea of how good the solutions (\ref{11class}) and (\ref{13}) are
we compare them with the exact numerical solution obtained by Mathematica (figure
1). From figure 1, we observe that the first order frequency renormalized solution
(\ref{13}) coincides exactly with the exact numerical solution obtained by
using Mathematica (for \( a=2 \) and \( \lambda =0.05 \)). However, the solution
(\ref{11class}) diverges with the increase of time (figure 1). The divergent
nature of the solution is a manifestation of the presence of secular term.

{\par\centering \resizebox*{7in}{5in}{\rotatebox{270}{\includegraphics{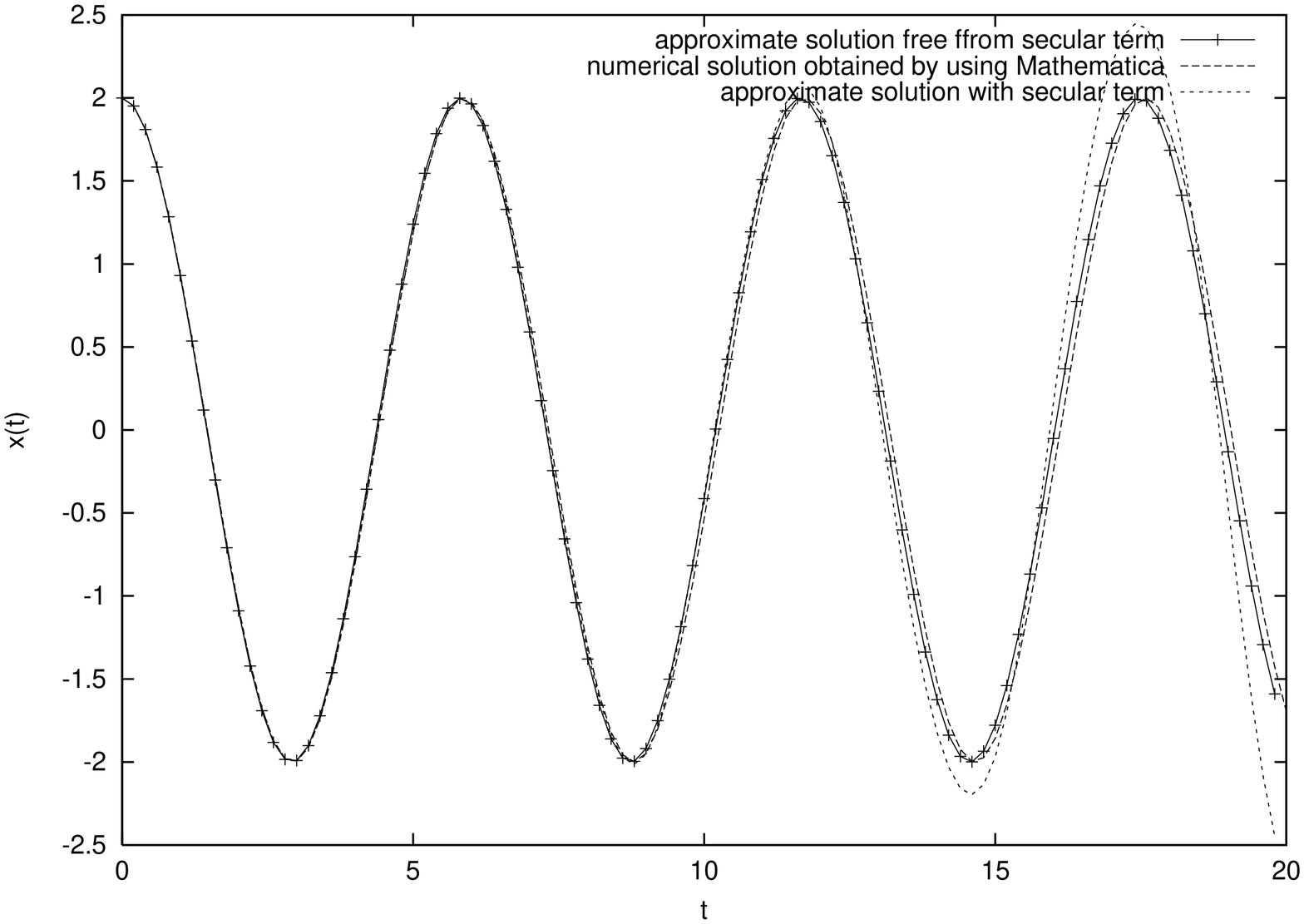}}} \par}

{\par\centering figure 1\par}

\subsection{{\small Solution of quantum quartic\index{quartic} oscillator}\small }

It is straightforward to extend the classical solution (\ref{2}) into a quantum
solution for the anharmonic oscillator by using the condition of quantization
{[}\ref{Schiff}{]}. The corresponding solution is given by 
\begin{equation}
\label{14}
\begin{array}{lcl}
X(t) & = & X(0)\, \cos t+\dot{X}(0)\, \sin t-\frac{\lambda X^{3}(0)}{32}(\cos t-\cos 3t+12t\sin t)\\
 & + & \frac{\lambda }{32}[X^{2}(0)\dot{X}(0)+X(0)\dot{X}(0)X(0)+\dot{X}(0)X^{2}(0)]\times (\sin 3t-7\sin t+4t\cos t)\\
 & - & \frac{\lambda }{32}[X(0)\dot{X}^{2}(0)+\dot{X}(0)X(0)\dot{X}(0)+\dot{X}^{2}(0)X(0)]\times (\cos 3t-\cos t+4t\sin t)\\
 & - & \frac{\lambda \dot{X}^{3}(0)}{32}(\sin 3t+9\sin t-12t\cos t)\\
 & + & \frac{\lambda ^{2}X^{5}(0)}{1024}(\cos 5t-36t\sin 3t-24\cos 3t-72t^{2}\cos t+96t\sin t+23\cos t)\\
 & + & \frac{\lambda ^{2}}{5120}[X^{4}(0)\dot{X}(0)+X^{3}(0)\dot{X}(0)X(0)+X^{2}(0)\dot{X}(0)X^{2}(0)\\
 & + & X(0)\dot{X}(0)X^{3}(0)+\dot{X}(0)X^{4}(0)]\\
 & \times  & (5\sin 5t+108t\cos 3t-132\sin 3t-72t^{2}\sin t+599\sin t-336t\cos t)\\
 & + & \frac{\lambda ^{2}}{5120}[X^{3}(0)\dot{X}^{2}(0)+X^{2}(0)\dot{X}^{2}(0)X(0)+X^{2}(0)\dot{X}(0)X(0)\dot{X}(0)+X(0)\dot{X}^{2}(0)X^{2}(0)\\
 & + & X(0)\dot{X}(0)X(0)\dot{X}(0)X(0)+X(0)\dot{X}(0)X^{2}(0)\dot{X}(0)+\dot{X}^{2}(0)X^{3}(0)\\
 & + & \dot{X}(0)X(0)\dot{X}(0)X^{2}(0)+\dot{X}(0)X^{2}(0)\dot{X}(0)X(0)+\dot{X}(0)X^{3}(0)\dot{X}(0)]\\
 & \times  & (-5\cos 5t+90\cos 3t+36t\sin 3t-72t^{2}\cos t-85\cos t+264t\sin t)\\
 & + & \frac{\lambda ^{2}}{5120}[\dot{X}^{3}(0)X^{2}(0)+\dot{X}^{2}(0)X^{2}(0)\dot{X}(0)+\dot{X}^{2}(0)X(0)\dot{X}(0)X(0)+\dot{X}(0)X^{2}(0)\dot{X}^{2}(0)\\
 & + & \dot{X}(0)X(0)\dot{X}(0)X(0)\dot{X}(0)+\dot{X}(0)X(0)\dot{X}^{2}(0)X(0)+X^{2}(0)\dot{X}^{3}(0)\\
 & + & X(0)\dot{X}(0)X(0)\dot{X}^{2}(0)+X(0)\dot{X}^{2}(0)X(0)\dot{X}(0)+X(0)\dot{X}^{3}(0)X(0)]\\
 & \times  & (-5\sin 5t+36t\cos 3t+6\sin 3t-72t^{2}\sin t+427\sin t-456t\cos t)\\
 & + & \frac{\lambda ^{2}}{5120}[X(0)\dot{X}^{4}(0)+\dot{X}(0)X(0)\dot{X}^{3}(0)+\dot{X}^{2}(0)X(0)\dot{X}^{2}(0)+\dot{X}^{3}(0)X(0)\dot{X}(0)+\dot{X}^{4}(0)X(0)]\\
 & \times  & (5\cos 5t+108t\sin 3t+108\cos 3t-72t^{2}\cos t-113\cos t+240t\sin t)\\
 & + & \frac{\lambda ^{2}\dot{X}^{5}(0)}{1024}(\sin 5t-36t\cos 3t+48\sin 3t-72t^{2}\sin t+271\sin t-384t\cos t)
\end{array}
\end{equation}
 where \( X(0) \) and \( \dot{X}(0) \) are the initial position and momentum
operators. Of course, the direct use of the Taylor series \index{Taylor series}
for the equivalent operator differential equation (\ref{eqm}) leads to the
same solution (\ref{14}). The quantization condition, \( [X(t),\dot{X}(t)]=i, \)
takes care the passage of classical solution (\ref{2}) to quantum solution
(\ref{14}) which may be written in the symmetrical form as 
\begin{equation}
\label{15}
\begin{array}{lcl}
X(t) & = & X(0)\, \cos t+\dot{X}(0)\, \sin t-\frac{\lambda X^{3}(0)}{32}(\cos t-\cos 3t+12t\sin t)\\
 & + & \frac{3\lambda }{64}[X^{2}(0)\dot{X}(0)+\dot{X}(0)X^{2}(0)]\times (\sin 3t-7\sin t+4t\cos t)\\
 & - & \frac{3\lambda }{64}[X(0)\dot{X}^{2}(0)+\dot{X}^{2}(0)X(0)]\times (\cos 3t-\cos t+4t\sin t)\\
 & - & \frac{\lambda \dot{X}^{3}(0)}{32}(\sin 3t+9\sin t-12t\cos t)\\
 & + & \frac{\lambda ^{2}X^{5}(0)}{1024}(\cos 5t-36t\sin 3t-24\cos 3t-72t^{2}\cos t+96t\sin t+23\cos t)\\
 & + & \frac{\lambda ^{2}}{2048}[X^{4}(0)\dot{X}(0)+\dot{X}(0)X^{4}(0)]\\
 & \times  & (5\sin 5t+108t\cos 3t-132\sin 3t-72t^{2}\sin t+599\sin t-336t\cos t)\\
 & + & \frac{2\lambda ^{2}}{2048}[X^{3}(0)\dot{X}^{2}(0)+\dot{X}^{2}(0)X^{3}(0)+3X(0)]\\
 & \times  & (-5\cos 5t+90\cos 3t+36t\sin 3t-72t^{2}\cos t-85\cos t+264t\sin t)\\
 & + & \frac{2\lambda ^{2}}{2048}[\dot{X}^{3}(0)X^{2}(0)+X^{2}(0)\dot{X}^{3}(0)+3\dot{X}(0)]\\
 & \times  & (-5\sin 5t+36t\cos 3t+6\sin 3t-72t^{2}\sin t+427\sin t-456t\cos t)\\
 & + & \frac{\lambda ^{2}}{2048}[X(0)\dot{X}^{4}(0)+\dot{X}^{4}(0)X(0)]\\
 & \times  & (5\cos 5t+108t\sin 3t+108\cos 3t-72t^{2}\cos t-113\cos t+240t\sin t)\\
 & + & \frac{\lambda ^{2}\dot{X}^{5}(0)}{1024}(\sin 5t-36t\cos 3t+48\sin 3t-72t^{2}\sin t+271\sin t-384t\cos t).
\end{array}
\end{equation}
The equation (\ref{15}) is our desired solution for a quantum quartic\index{quartic}
anharmonic oscillator. The solution contains terms proportional to \( X^{3} \)
and \( X^{5} \) etcetera. In the classical limit these terms produce the cubic
and fifth powers of amplitude respectively. Hence, as a limiting situation,
the solution (\ref{15}) gives rise to the solution corresponding to the classical
QAHO. It is not at all surprising since the solution (\ref{15}) is obtained
from its classical counterpart (\ref{2}) by the imposition of the quantization
condition.

The unpleasant secular terms can be removed (summed) from the solution (\ref{15})
by using the tucking in technique\index{tucking in technique}. All the secular
terms of equation (\ref{15}) are tucked in to have the following solution 
\begin{equation}
\label{16}
\begin{array}{lcl}
X(t) & = & \frac{1}{2\cos (\frac{3\lambda t}{8}-\frac{51\lambda ^{2}nt}{64})}\left\{ X(0)\cos \Psi t+\cos \Psi t\, X(0)+\dot{X}(0)\sin \Psi t+\sin \Psi t\, \dot{X}(0)\right. \\
 & - & \frac{\lambda X^{3}(0)}{32}(\cos \Psi t-\cos 3\Psi t)-\frac{\lambda }{32}(\cos \Psi t-\cos 3\Psi t)X^{3}(0)\\
 & + & \frac{3\lambda }{64}[X^{2}(0)\dot{X}(0)+\dot{X}(0)X^{2}(0)]\times (\sin 3\Psi t-7\sin \Psi t)+\frac{3\lambda }{64}(\sin 3\Psi t-7\sin \Psi t)\\
 & \times  & [X^{2}(0)\dot{X}(0)+\dot{X}(0)X^{2}(0)]-\frac{3\lambda }{64}[X(0)\dot{X}^{2}(0)+\dot{X}^{2}(0)X(0)]\times (\cos 3\Psi t-\cos \Psi t)\\
 & - & \frac{3\lambda }{64}(\cos 3\Psi t-\cos \Psi t)\times [X(0)\dot{X}^{2}(0)+\dot{X}^{2}(0)X(0)]\\
 & - & \frac{\lambda \dot{X}^{3}(0)}{32}(\sin 3\Psi t+9\sin \Psi t)-\frac{\lambda }{32}(\sin 3\Psi t+9\sin \Psi t)\dot{X}^{3}(0)\\
 & + & \frac{\lambda ^{2}X^{5}(0)}{1024}(\cos 5\Psi t-24\cos 3\Psi t+23\cos \Psi t)\\
 & + & \frac{\lambda ^{2}}{1024}(\cos 5\Psi t-24\cos 3\Psi t+23\cos \Psi t)X^{5}(0)\\
 & + & \frac{\lambda ^{2}}{2048}[X^{4}(0)\dot{X}(0)+\dot{X}(0)X^{4}(0)]\times (5\sin 5\Psi t-132\sin 3\Psi t+599\sin \Psi t)\\
 & + & \frac{\lambda ^{2}}{2048}(5\sin 5\Psi t-132\sin 3\Psi t+599\sin \Psi t)\times [X^{4}(0)\dot{X}(0)+\dot{X}(0)X^{4}(0)]\\
 & + & \frac{2\lambda ^{2}}{2048}[X^{3}(0)\dot{X}^{2}(0)+\dot{X}^{2}(0)X^{3}(0)+3X(0)]\times (-5\cos 5\Psi t+90\cos 3\Psi t-85\cos \Psi t)\\
 & + & \frac{2\lambda ^{2}}{2048}(-5\cos 5\Psi t+90\cos 3\Psi t-85\cos \Psi t)\times [X^{3}(0)\dot{X}^{2}(0)+\dot{X}^{2}(0)X^{3}(0)+3X(0)]\\
 & + & \frac{2\lambda ^{2}}{2048}[\dot{X}^{3}(0)X^{2}(0)+X^{2}(0)\dot{X}^{3}(0)]\times (-5\sin 5\Psi t+6\sin 3\Psi t+427\sin \Psi t)\\
 & + & \frac{2\lambda ^{2}}{2048}(-5\sin 5\Psi t+6\sin 3\Psi t+427\sin \Psi t)\times [\dot{X}^{3}(0)X^{2}(0)+X^{2}(0)\dot{X}^{3}(0)]\\
 & + & \frac{2\lambda ^{2}}{2048}[3\dot{X}(0)]\times (-5\sin 5\Psi t+6\sin 3\Psi t+403\sin \Psi t)\\
 & + & \frac{2\lambda ^{2}}{2048}(-5\sin 5\Psi t+6\sin 3\Psi t+403\sin \Psi t)\times [3\dot{X}(0)]\\
 & + & \frac{\lambda ^{2}}{2048}[X(0)\dot{X}^{4}(0)+\dot{X}^{4}(0)X(0)]\times (5\cos 5\Psi t+108\cos 3\Psi t-113\cos \Psi t)\\
 & + & \frac{\lambda ^{2}}{2048}(5\cos 5\Psi t+108\cos 3\Psi t-113\cos \Psi t)\times [X(0)\dot{X}^{4}(0)+\dot{X}^{4}(0)X(0)]\\
 & + & \frac{\lambda ^{2}\dot{X}^{5}(0)}{1024}(\sin 5\Psi t+48\sin 3\Psi t+271\sin \Psi t)\\
 & + & \left. \frac{\lambda ^{2}}{1024}(\sin 5\Psi t+48\sin 3\Psi t+271\sin \Psi t)\dot{X}^{5}(0)\right\} 
\end{array}
\end{equation}
 where \( n \) is the eigenvalue\index{eigenvalue} of the number operator
(\( a^{\dagger }a \)). The parameter \( n \) may be regarded as the number
of photons present in the radiation field corresponding to the electric field
operator \( X(t). \) The prefactor of the above equation (\ref{16}) takes
care of the proper normalization of \( X(t). \) The operator \( \Psi  \) is
second order frequency operator\index{frequency operator} and is given by 
\begin{equation}
\label{17}
\Psi =1+\frac{3\lambda H_{0}}{4}-\frac{\lambda ^{2}}{64}\left( 69H^{2}_{0}-12X^{4}(0)+\frac{51}{4}\right) 
\end{equation}
where \( H_{0}=\frac{\dot{X}^{2}(0)}{2}+\frac{X^{2}(0)}{2} \) is the unperturbed
Hamiltonian. Now, the diagonal element of \( \Psi  \) in the number state basis
is 
\begin{equation}
\label{18}
\psi _{n}=<n|\Psi |n>=1+\frac{3\lambda }{4}(n+\frac{1}{2})-\frac{\lambda ^{2}}{64}(51n^{2}+51n+21)
\end{equation}
where the following relations are used
\begin{equation}
\label{19}
\begin{array}{lcl}
H_{0}|n> & = & (n+\frac{1}{2})\, |n>\\
X(0)|n> & = & \frac{1}{\sqrt{2}}\left[ \sqrt{n}|n-1>+\sqrt{n+1}|n+1>\right] \, for\, n\neq 0
\end{array}.
\end{equation}
 The consequences of our solution (\ref{16}) may be compared critically with
the existing results. For example, the energy and hence the frequency shift
of a QQAHO\index{QQAHO} is already known from the second order perturbation
theory. The energy of a quartic\index{quartic} oscillator in \( n-th \) state
is given by {[}\ref{Powell}{]} 
\begin{equation}
\label{20}
E_{n}=(n+\frac{1}{2})+\frac{3\lambda }{8}(n^{2}+n+\frac{1}{2})-\frac{\lambda ^{2}}{128}(34n^{3}+51n^{2}+59n+21).
\end{equation}
Therefore, the energy difference between two consecutive levels is 
\begin{equation}
\label{15b}
\Delta E=E_{n}-E_{n-1}=1+\frac{3\lambda n}{4}-\frac{\lambda ^{2}}{64}(51n^{2}+21).
\end{equation}
 In order to calculate the dipole moment matrix element \( <n-1|X(t)|n> \)
and hence the shifted frequency of the quantum oscillator, we use equations
(\ref{16}) and (\ref{18}). The corresponding calculated frequency shift coincides
exactly with that of the frequency shift obtained by the perturbation method
(\ref{15b}). The frequency shift for the vacuum field\index{vacuum field}
\( (i.e\, n=0) \) is evidently clear. It actually arises due to the vacuum
field interaction. Here we can note that the Lamb shift\index{Lamb shift} is
also explained in terms of the self interacting field (for a detailed discussion
see reference {[}\ref{Milloni}{]}). Actually, our model Hamiltonian permits
such an interaction through the quartic\index{quartic} anharmonicity. Here
we can note that a first order calculation is unable to predict the  frequency
shift for a vacuum field\index{vacuum field}.

\subsection{{\small Remarks on the solutions }\small }

We obtain an analytical second order solution for a classical quartic\index{quartic}
anharmonic oscillator by using the Taylor series \index{Taylor series}method.
The solution is found to agree with the existing solutions obtained by various
methods. The frequency shift of a QAHO increases linearly with the increase
of anharmonic constant as long as the first order solution is considered {[}\ref{Mandal},
\ref{Nayfeh}{]}. These type of monotonic increase of the frequency shifts seem
to be inconsistent. We obtain a correction to such monotonic increase of frequency
shift. The solutions presented here are perturbative. We prefer perturbative
solution instead of an exact numerical one. Because they can provide more physical
insight into the problem. 

The classical solution is quantized in order to obtain the solution corresponding
to a quantum quartic\index{quartic} oscillator. Interestingly, the frequency
shift of the quantum oscillator coincides with that of the frequency shift obtained
by using a second order perturbation theory. The second order solution exhibits
Lamb shift \index{Lamb shift} which is totally absent in a first order solution.
Hence, we observe that our solutions reproduce the existing results for classical
and quantum oscillators of quartic\index{quartic} anharmonicity. 

From the mathematical point of view, the solutions presented in this section
are of great academic interests. In addition to these, the classical solution
finds several applications in nonlinear mechanics. The solution corresponding
to the quantum oscillator is useful in quantum optics and in the field theory.
For example, the solutions are used to study the quantum statistical properties
(e.g. squeezing\index{squeezing}, higher order squeezing\index{squeezing},
bunching\index{bunching} and antibunching\index{antibunching} of photons)
of radiation field interacting with a cubic nonlinear media in the subsequent
chapters of the present thesis. Moreover, the quantum phase\index{quantum phase}
of the output field is also studied with the knowledge of field operators \( X(t),\, \dot{X}(t),\, a(t)\,  \)
and \( a^{\dagger }(t) \).

\section{Higher anharmonic oscillators : sextic\index{sextic} and octic\index{octic}
oscillators}

In addition to the quartic\index{quartic} oscillator, people have studied the
higher anharmonic oscillators also {[}\ref{weniger}, \ref{Laksman},\ref{R. Dutt}{]}.
These studies are made to obtain the eigenvalues as function of anharmonicity.
However, the operator solution of higher anharmonic oscillators were not reported
since recent past. Recently we have reported first order operator solutions
{[}\ref{Pathak}{]} of sextic and octic\index{octic} anharmonic oscillators
by using the Taylor series \index{Taylor series}method. We have also developed
a first order solution for the generalized anharmonic oscillator {[}\ref{pathak2}{]}.
Fern\'{a}ndez obtained first order correction to the frequency operator\index{frequency operator}
for a few particular values of \( m \) {[}\ref{F01a}{]}. In the next two subsection
of the present thesis, we present analytical solutions of classical and quantum
oscillators of sextic\index{sextic} and octic\index{octic} anharmonicities
by using the Taylor series \index{Taylor series}approach. In the later part
of this thesis we discuss the generalized anharmonic oscillator and compare
the results with that obtained in the Taylor series \index{Taylor series} approach.
The sextic\index{sextic} and octic anharmonic oscillators have potential applications
in the studies of nonlinear mechanics, molecular physics, quantum optics and
in field theory. The solutions are used to obtain the frequency shifts of sextic\index{sextic}
and octic\index{octic} oscillators. The computed shifts are compared and found
to have exact coincidence with the frequency shifts calculated by using a first
order perturbation theory.

\subsection{Classical and quantum solutions of the sextic\index{sextic} oscillator }

For \( m=6, \) the Hamiltonian (\ref{1g}) and the equation of motion (\ref{1h})
correspond to the case of a sextic\index{sextic} anharmonic oscillator. The
solution for such an oscillator is obtained as the sum of different orders of
anharmonicities. The corresponding solution is 
\begin{equation}
\label{2higher}
x(t)\, =x_{0}(t)\, +x_{1}(t)\, +\, \ldots 
\end{equation}
 where \( x_{0}(t) \) and \( x_{1}(t) \) are zeroth and first order solutions
respectively. The zeroth order solution (\ref{1d}) is simply obtained for \( \lambda =0 \)
in equation (\ref{1h}). The purpose of the present section is to find the first
order solution \( x_{1}(t). \) By using the Taylor series \index{Taylor series}
approach introduced in connection with the quartic\index{quartic} oscillator
we can obtain first order solution of the classical sextic\index{sextic} anharmonic
oscillator as
\begin{equation}
\label{se10}
x_{1}(t)=\sum ^{5}_{i=0}L_{i}x^{5-i}(0)\dot{x}^{i}
\end{equation}
 where
\begin{equation}
\label{se9}
\begin{array}{lcl}
L_{0} & = & \frac{\lambda }{384}(\cos 5t+15\cos 3t-16\cos t-120\, t\sin t)\\
L_{1} & = & \frac{\lambda }{384}(5\sin 5t+45\sin 3t-280\sin t+120t\cos t)\\
L_{2} & = & \frac{2\lambda }{384}(-5\cos 5t-15\cos 3t+20\cos t-120t\sin t)\\
L_{3} & = & \frac{2\lambda }{384}(-5\sin 5t+15\sin 3t-140\sin t+120t\cos t)\\
L_{4} & = & \frac{\lambda }{384}(5\cos 5t-45\cos 3t+40\cos t-120t\sin t)\\
and &  & \\
L_{5} & = & \frac{\lambda }{384}(\sin 5t-15\sin 3t-80\sin t+120t\cos t).
\end{array}
\end{equation}
 The total solution (\ref{2}) is simply the sum of (\ref{1d}) and (\ref{se10}).

Now to have a check of the solution (\ref{2higher}), we consider a special
case \( \dot{x}(0)=0 \) and \( x(0)=a \). Hence the total solution reduces
to an extraordinarily simple form as 
\begin{equation}
\label{se11}
x(t)=a\cos t+\frac{\lambda a^{5}}{384}(\cos 5t+15\cos 3t-16\cos t-120t\sin t).
\end{equation}
 It is clear that the presence of a secular term proportional to \( t\sin t \)
makes life difficult as \( t \) grows. We confine ourselves to the weak coupling
regime and assume \( C \) is a constant such that 
\begin{equation}
\label{13a.1}
\begin{array}{lcl}
\sin \left( C\lambda t\right)  & = & C\lambda t\\
\cos \left( C\lambda t\right)  & = & 1
\end{array}.
\end{equation}
 Equation (\ref{13a.1}) is actually a first order form of (\ref{13a}). Now,
the equation (\ref{se11}) can be rearranged as 
\begin{equation}
\label{se12}
x(t)=a\cos [t(1+\frac{5}{16}\lambda a^{4})]+\lambda a^{5}(-\frac{1}{24}\cos t+\frac{5}{128}\cos 3t+\frac{1}{384}\cos 5t)
\end{equation}
 where the equation (\ref{13a.1}) is used. Thus the secular term present in
the first order solution (\ref{se11}) is summed up for all orders and the frequency
shift of the oscillator is obtained. The corresponding frequency shift of the
sextic\index{sextic} anharmonic oscillator is \( \frac{5}{16}\lambda a^{4} \).
The shifted frequency of the oscillator may be viewed as the renormalization
of frequency by the anharmonic interaction. Interestingly, the frequency shift
obtained in the solution (\ref{se12}) exactly coincides with that obtained
by Dutt and Lakshmanan {[}\ref{R. Dutt}{]}.

Now, a quantum solution of the sextic\index{sextic} anharmonic oscillator may
be obtained by replacing \( x(t) \) and \( \dot{x}(t) \) by noncommuting operators
\( X(t) \) and \( \dot{X}(t) \) respectively and imposing the commutation
relation (\ref{1e}). Corresponding operator solution of the sextic\index{sextic}
oscillator is

\begin{equation}
\label{se14}
\begin{array}{lcl}
X(t) & = & X(0)\cos t+\dot{X}(0)\sin t+\frac{\lambda }{384}X^{5}(0)\times (\cos 5t+15\cos 3t-16\cos t-120t\sin t)\\
 & + & \frac{\lambda }{1920}\left[ X^{4}(0)\dot{X}(0)+X^{3}(0)\dot{X}(0)X(0)\right. \\
 & + & \left. X^{2}(0)\dot{X}(0)X^{2}(0)+X(0)\dot{X}(0)X^{3}(0)+\dot{X}(0)X^{4}(0)\right] \\
 & \times  & (5\sin 5t+45\sin 3t-280\sin t+120t\cos t)\\
 & + & \frac{\lambda }{1920}\left[ X^{3}(0)\dot{X}^{2}(0)+X^{2}(0)\dot{X}^{2}(0)X(0)+X^{2}(0)\dot{X}(0)X(0)\dot{X}(0)\right. \\
 & + & X(0)\dot{X}^{2}(0)X^{2}(0)+X(0)\dot{X}(0)X(0)\dot{X}(0)X(0)+X(0)\dot{X}(0)X^{2}(0)\dot{X}(0)\\
 & + & \left. \dot{X}^{2}(0)X^{3}(0)+\dot{X}(0)X(0)\dot{X}(0)X^{2}(0)+\dot{X}(0)X^{2}(0)\dot{X}(0)X(0)+\dot{X}(0)X^{3}(0)\dot{X}(0)\right] \\
 & \times  & \left( -5\cos 5t-15\cos 3t+20\cos t-120t\sin t\right) \\
 & + & \frac{\lambda }{1920}\left[ \dot{X}^{3}(0)X^{2}(0)+\dot{X}^{2}(0)X^{2}(0)\dot{X}(0)+\dot{X}^{2}(0)X(0)\dot{X}(0)X(0)+\dot{X}(0)X^{2}(0)\dot{X}^{2}(0)\right. \\
 & + & \left. \dot{X}(0)X(0)\dot{X}(0)X(0)\dot{X}(0)+\dot{X}(0)X(0)\dot{X}^{2}(0)X(0)+X^{2}(0)\dot{X}^{3}(0)\right. \\
 & + & \left. X(0)\dot{X}(0)X(0)\dot{X}^{2}(0)+X(0)\dot{X}^{2}(0)X(0)\dot{X}(0)+X(0)\dot{X}^{3}(0)X(0)\right] \\
 & \times  & \left( -5\sin 5t+15\sin 3t-140\sin t+120t\cos t\right) \\
 & + & \frac{\lambda }{1920}\left[ X(0)\dot{X}^{4}(0)+\dot{X}(0)X(0)\dot{X}^{3}(0)+\dot{X}^{2}(0)X(0)\dot{X}^{2}(0)\right. \\
 & + & \left. \dot{X}^{3}(0)X(0)\dot{X}(0)+\dot{X}^{4}(0)X(0)\right] \\
 & \times  & \left( 5\cos 5t-45\cos 3t+40\cos t-120t\sin t\right) \\
 & + & \frac{\lambda \dot{X}^{5}(0)}{384}\left( \sin 5t-15\sin 3t-80\sin t+120t\cos t\right) 
\end{array}
\end{equation}
 where \( X(0) \) and \( \dot{X}(0) \) are the initial position and momentum
operators. The solution (\ref{se14}) may be written in symmetrical form as
\begin{equation}
\label{se15}
\begin{array}{lcl}
X(t) & = & X(0)\cos t+\dot{X}(0)\sin t\\
 & + & \frac{2\lambda }{768}X^{5}(0)(\cos t+15\cos 3t-16\cos t-120t\sin t)\\
 & + & \frac{\lambda }{768}[X^{4}(0)\dot{X}(0)+\dot{X}(0)X^{4}(0)]\\
 & \times  & (5\sin 5t+45\sin 3t-280\sin t+120t\cos t)\\
 & + & \frac{2\lambda }{768}[X^{3}(0)\dot{X}^{2}(0)+\dot{X}^{2}(0)X^{3}(0)+3X(0)]\\
 & \times  & (-5\cos 5t-15\cos 3t+20\cos t-120t\sin t)\\
 & + & \frac{2\lambda }{768}[\dot{X}^{3}(0)X^{2}(0)+X^{2}(0)\dot{X}^{3}(0)+3\dot{X}(0)]\\
 & \times  & (-5\sin 5t+15\sin 3t-140\sin t+120t\cos t)\\
 & + & \frac{\lambda }{768}[X(0)\dot{X}^{4}(0)+\dot{X}^{4}(0)X(0)]\\
 & \times  & (5\cos 5t-45\cos 3t+40\cos t-120t\sin t)\\
 & + & \frac{2\lambda \dot{X}^{5}(0)}{768}(\sin 5t-15\sin 3t-80\sin t+120t\cos t).\\
 &  & 
\end{array}
\end{equation}
 Equation (\ref{se15}) is our desired solution for a quantum sextic\index{sextic}
anharmonic oscillator. Now we rearrange our derived solution (\ref{se15}) with
the help of (\ref{13a.1}) as 
\begin{equation}
\label{se16}
\begin{array}{lcl}
X(t) & = & \frac{1}{2}\left\{ X(0)\cos \left( t+\frac{5\lambda t}{4}\left\{ H^{2}_{0}+\frac{1}{4}\right\} \right) +\cos \left( t+\frac{5\lambda t}{4}\left\{ H^{2}_{0}+\frac{1}{4}\right\} \right) X(0)\right. \\
 & + & \dot{X}(0)\sin \left( t+\frac{5\lambda t}{4}\left\{ H^{2}_{0}+\frac{1}{4}\right\} \right) +\sin \left( t+\frac{5\lambda t}{4}\left\{ H^{2}_{0}+\frac{1}{4}\right\} \right) \dot{X}(0)\\
 & + & \frac{2\lambda X^{5}(0)}{384}(\cos 5t+15\cos 3t-16\cos t)\\
 & + & \frac{\lambda }{384}[X^{4}(0)\dot{X}(0)+\dot{X}(0)X^{4}(0)](5\sin 5t+45\sin 3t-280\sin t)\\
 & + & \frac{2\lambda }{384}[X^{3}(0)\dot{X}^{2}(0)+\dot{X}^{2}(0)X^{3}(0)+3X(0)](-5\cos 5t-15\cos 3t+20\cos t)\\
 & + & \frac{2\lambda }{384}[\dot{X}^{3}(0)X^{2}(0)+X^{2}(0)\dot{X}^{3}(0)+3\dot{X}(0)](-5\sin 5t+15\sin 3t-140\sin t)\\
 & + & \frac{\lambda }{384}[X(0)\dot{X}^{4}(0)+\dot{X}^{4}(0)X(0)](5\cos 5t-45\cos 3t+40\cos t)\\
 & + & \left. \frac{2\lambda \dot{X}^{5}(0)}{384}(\sin 5t-15\sin 3t-80\sin t)\right\} \\
 &  & 
\end{array}
\end{equation}
 where \( H_{0}=\frac{\dot{X}^{2}(0)}{2}+\frac{X^{2}(0)}{2} \) is the unperturbed
Hamiltonian. Now the matrix element \( <n-1|X(t)|n> \) and hence the frequency
shift of the oscillator may be calculated with the help of (\ref{19}). Now,
the dipole matrix element in terms of the number eigenket \( |n> \) is 
\begin{equation}
\label{se18}
\begin{array}{lcl}
<n-1|X(t)|n> & = & \frac{\cos (\frac{5\lambda t}{4}n)}{2}\left\{ <n-1|X(0)|n>\cos \left( t+\frac{5\lambda t}{4}\left\{ n^{2}+\frac{1}{2}\right\} \right) \right. \\
 & + & \cos \left( t+\frac{5\lambda t}{4}\left\{ n^{2}+\frac{1}{2}\right\} \right) <n-1|X(0)|n>\\
 & + & <n-1|\dot{X}(0)|n>\sin \left( t+\frac{5\lambda t}{4}\left\{ n^{2}+\frac{1}{2}\right\} \right) \\
 & + & \sin \left( t+\frac{5\lambda t}{4}\left\{ n^{2}+\frac{1}{2}\right\} \right) <n-1|\dot{X}(0)|n>\\
 & + & <n-1|\left[ \frac{2\lambda X^{5}(0)}{384}(\cos 5t+15\cos 3t-16\cos t)\right. \\
 & + & \frac{\lambda }{384}[X^{4}(0)\dot{X}(0)+\dot{X}(0)X^{4}(0)](5\sin 5t+45\sin 3t-280\sin t)\\
 & + & \frac{2\lambda }{384}[X^{3}(0)\dot{X}^{2}(0)+\dot{X}^{2}(0)X^{3}(0)+3X(0)](-5\cos 5t-15\cos 3t+20\cos t)\\
 & + & \frac{2\lambda }{384}[\dot{X}^{3}(0)X^{2}(0)+X^{2}(0)\dot{X}^{3}(0)+3\dot{X}(0)](-5\sin 5t+15\sin 3t-140\sin t)\\
 & + & \frac{\lambda }{384}[X(0)\dot{X}^{4}(0)+\dot{X}^{4}(0)X(0)](5\cos 5t-45\cos 3t+40\cos t)\\
 & + & \left. \left. \frac{2\lambda \dot{X}^{5}(0)}{384}(\sin 5t-15\sin 3t-80\sin t)\right] |n>\right\} .\\
 &  & 
\end{array}
\end{equation}
 Hence, we obtain the frequency shift of the oscillator as \( \frac{5\lambda }{4}(n^{2}+\frac{1}{2}). \)
Note that the frequency shift of the oscillator is observed even when there
are no photons (i.e vacuum field\index{vacuum field}) present in the radiation
field. The vacuum field interacts with the medium and causes the frequency shift.
This phenomena is a direct outcome of pure quantum electrodynamic effect and
has no classical analogue. The frequency shift of the oscillator for a vacuum
field\index{vacuum field} is actually a second order effect and is discussed
in the context of the second order operator solution of the quartic\index{quartic}
oscillator. The solution (\ref{se16}) is consistent with the generalized solutions
obtained in the later part of this thesis. At present, the frequency shift in
(\ref{se18}) may be compared with the frequency shift calculated by using the
first order perturbation technique {[}\ref{Powell}{]}. As a check, we calculate
the energy of a sextic\index{sextic} oscillator in the \( n-th \) eigenstate
which is given by 
\begin{equation}
\label{se15a}
E_{n}=(n+\frac{1}{2})+\frac{5\lambda }{48}(4n^{3}+6n^{2}+8n+3).
\end{equation}
 The energy difference between two consecutive levels is 
\begin{equation}
\label{se15b}
\Delta E=E_{n}-E_{n-1}=1+\frac{5\lambda }{4}(n^{2}+\frac{1}{2}).
\end{equation}
 Thus our calculated frequency shift and hence the solution agrees nicely. Finally
we normalize the equation (\ref{se16}) with the help of the equation (\ref{se18})
as 
\begin{equation}
\label{se19}
\begin{array}{lcl}
X(t) & = & \frac{1}{2cos(\frac{5\lambda t}{4}n)}\left\{ X(0)cos\left( t+\frac{5\lambda t}{4}\left\{ H^{2}_{0}+\frac{1}{4}\right\} \right) +cos\left( t+\frac{5\lambda t}{4}\left\{ H^{2}_{0}+\frac{1}{4}\right\} \right) X(0)\right. \\
 & + & \dot{X}(0)sin\left( t+\frac{5\lambda t}{4}\left\{ H^{2}_{0}+\frac{1}{4}\right\} \right) +sin\left( t+\frac{5\lambda t}{4}\left\{ H^{2}_{0}+\frac{1}{4}\right\} \right) \dot{X}(0)\\
 & + & \frac{2\lambda X^{5}(0)}{384}(cos5t+15\, cos3t-16\, cost)\\
 & + & \frac{\lambda }{384}[X^{4}(0)\dot{X}(0)+\dot{X}(0)X^{4}(0)](5sin5t+45sin3t-280sint)\\
 & + & \frac{2\lambda }{384}[X^{3}(0)\dot{X}^{2}(0)+\dot{X}^{2}(0)X^{3}(0)+3X(0)](-5cos5t-15cos3t+20cost)\\
 & + & \frac{2\lambda }{384}[\dot{X}^{3}(0)X^{2}(0)+X^{2}(0)\dot{X}^{3}(0)+3\dot{X}(0)](-5sin5t+15sin3t-140sint)\\
 & + & \frac{\lambda }{384}[X(0)\dot{X}^{4}(0)+\dot{X}^{4}(0)X(0)](5cos5t-45cos3t+40cost)\\
 & + & \left. \frac{2\lambda \dot{X}^{5}(0)}{384}(sin5t-15sin3t-80sint)\right\} .\\
 &  & 
\end{array}
\end{equation}
 The equation (\ref{se19}) is the desired solution for quantum sextic\index{sextic}
anharmonic oscillator where the secular terms for all orders are summed up.

\subsection{Classical and quantum solutions of the octic\index{octic} oscillator }

Depending upon the nature of nonlinearity, the higher anharmonic oscillators
come into the picture. For example, the Hamiltonian and the equation of motion
of classical octic anharmonic oscillator may be obtained if we put \( m=8 \)
in the equations (\ref{1g}) and (\ref{1h}) respectively. Using the similar
procedure (as adopted for sextic\index{sextic} oscillator), the solution (up
to the linear power of \( \lambda  \)) for classical octic\index{octic} oscillator
follows immediately as 
\begin{equation}
\label{oc2}
\begin{array}{lcl}
x(t) & = & x(0)cost+\dot{x}(0)sint\\
 & + & \frac{\lambda x^{7}(0)}{3072}(cos7t+14cos5t+126cos3t-141cost-840tsint)\\
 & + & \frac{\lambda \dot{x}(0)x^{6}(0)}{3072}(7sin7t+70sin5t+378sint-2373sint+840tcost)\\
 & + & \frac{\lambda \dot{x}^{2}(0)x^{5}(0)}{3072}(-21cos7t-126cos5t-126cos3t+273cost-2520tsint)\\
 & + & \frac{\lambda \dot{x}^{3}(0)x^{4}(0)}{3072}(-35sin7t-70sin5t+630sin3t-3815sint+2520tcost)\\
 & + & \frac{\lambda \dot{x}^{4}(0)x^{3}(0)}{3072}(35cos7t-70cos5t-630cos3t+665cost-2520tsint)\\
 & + & \frac{\lambda \dot{x}^{5}(0)x^{2}(0)}{3072}(21sin7t-126sin5t+126sin3t-2415sint+2520tcost)\\
 & + & \frac{\lambda \dot{x}^{6}(0)x(0)}{3072}(-7cos7t+70cos5t-378cos3t+315cost-840tsint)\\
 & + & \frac{\lambda \dot{x}^{7}(0)}{3072}(-sin7t+14sin5t-126sin3t-525sint+840tcost)\\
 & 
\end{array}
\end{equation}
 To check the validity of our solution, we consider a special case \( x(0)=b \)
and \( \dot{x}(0)=0. \) After a little algebra we have 
\begin{equation}
\label{oc3}
x(t)=bcos(1+\frac{35\lambda }{128}b^{6})t+\frac{\lambda b^{7}}{3072}(-141cost+126cos3t+14cos5t+cos7t).
\end{equation}
 Thus the frequency shift of the classical octic\index{octic} oscillator is
proportional to the sixth power of amplitude as long as the first order solution
is concerned. The solution (\ref{oc3}) as well as the frequency shift have
exact coincidence with the solution obtained by the procedure introduced by
Bradbury and Brintzenhoff {[}\ref{Bradbury}{]}. The solution of the classical
octic\index{octic} oscillator (\ref{oc2}) is now used to obtain the corresponding
solution for its quantum counter part. It is found that the solution contains
the secular terms similar to those appeared in (\ref{14}). The problem of secular
terms are taken care by the same procedure as it is done in the case of a sextic\index{sextic}
anharmonic oscillator and the final solution appears as 
\begin{equation}
\label{oc5}
\begin{array}{lcl}
X(t) & = & \frac{1}{2cos[\frac{35\lambda }{64}(6n^{2}+3)t]}\left\{ X(0)cos\Omega t+cos\Omega tX(0)+\dot{X}(0)sin\Omega t+sin\Omega t\dot{X}(0)\right. \\
 & + & \frac{2\lambda X^{7}(0)}{3072}(cos7t+14cos5t+126cos3t-141cost)\\
 & + & \frac{\lambda }{3072}[\dot{X}(0)X^{6}(0)+X^{6}(0)\dot{X}(0)](7sin7t+70sin5t+378sint-2373sint)\\
 & + & \frac{\lambda }{3072}[\dot{X}^{5}(0)X^{2}(0)+X^{2}(0)\dot{X}^{5}(0)+10X^{3}(0)]\\
 & \times  & (-21cos7t-126cos5t-126cos3t+273cost)\\
 & + & \frac{\lambda }{3072}[\dot{X}^{4}(0)X^{3}(0)+X^{3}(0)\dot{X}^{4}(0)+9(X^{2}(0)\dot{X}(0)+\dot{X}(0)X^{2}(0))]\\
 & \times  & (-35sin7t-70sin5t+630sin3t-3815sint)\\
 & + & \frac{\lambda }{3072}[\dot{X}^{3}(0)X^{4}(0)+X^{4}(0)\dot{X}^{3}(0)+9(\dot{X}^{2}(0)X(0)+X(0)\dot{X}^{2}(0))]\\
 & \times  & (35cos7t-70cos5t-630cos3t+665cost)\\
 & + & \frac{\lambda }{3072}[\dot{X}^{5}(0)X^{2}(0)+X^{2}(0)\dot{X}^{5}(0)+10\dot{X}^{3}(0)]\\
 & \times  & (21sin7t-126sin5t+126sin3t-2415sint)\\
 & + & \frac{\lambda }{3072}[\dot{X}^{6}(0)X(0)+X(0)\dot{X}^{6}(0)](-7cos7t+70cos5t-378cos3t+315cost)\\
 & + & \left. \frac{2\lambda \dot{X}^{7}(0)}{3072}(-sin7t+14sin5t-126sin3t-525sint)\right\} \\
 & 
\end{array}
\end{equation}
 where the first order frequency operator\index{frequency operator} is \( \Omega =1+\frac{35\lambda }{64}(4H_{0}^{3}+5H_{0}) \).
Now, the matrix element \( <n-1|X(t)|n> \) is calculated to obtain the shifted
frequency as \( \omega ^{\prime }=1+\frac{35\lambda }{16}(n^{3}+2n) \). Interestingly,
the frequency shift \( \frac{35\lambda }{16}(n^{3}+2n) \) depends on \( n^{3}. \)
It is clear that the vacuum field \index{vacuum field} has no role to play
in the frequency shift of the quantum octic\index{octic} oscillator. To have
a possible verification of the solution (\ref{oc5}), we calculate the frequency
shift by using Rayleigh- Schr\"{o}dinger perturbation theory. A first order
calculation under this theory shows that the energy of the octic\index{octic}
oscillator in the \( n-th \) state is given by 
\begin{equation}
\label{oc6}
E_{n}=(n+\frac{1}{2})+\frac{35\lambda }{64}(\frac{3}{2}+4n+5n^{2}+2n^{3}+n^{4}).
\end{equation}
 The energy difference between two consecutive levels is \( \Delta E\, =\, E_{n}-E_{n-1}=1+\frac{35\lambda }{16}(n^{3}+2n). \)
Thus, the frequency shift coincides exactly with the frequency shift obtained
by us.

\subsection{Remarks on the solutions: }

We obtain analytical solutions (first order) for sextic\index{sextic} (\ref{se10})
and octic\index{octic} (\ref{oc2}) oscillators in classical pictures by using
the Taylor series \index{Taylor series}method. The solutions exhibit the presence
of the usual secular terms. The secular terms for all orders are tucked in to
obtain the frequency shifts of those oscillators. It is found that the calculated
shifts agree satisfactorily with the shifts obtained by other methods.

The position and momentum of classical oscillators are replaced by their corresponding
operators and the commutation relation is imposed. Hence, the solutions for
quantum sextic\index{sextic} and octic\index{octic} oscillators in the realm
of the Heisenberg picture \index{Heisenberg picture} is obtained. Interestingly,
the vacuum field \index{vacuum field} has no role to play in the first order
frequency shift of the quantum oscillators of quartic\index{quartic} and octic\index{octic}
anharmonicities. However, the vacuum field \index{vacuum field} causes the
frequency shift of a quantum sextic\index{sextic} oscillator. 

Till now we have discussed the Taylor series \index{Taylor series}method for
obtaining first order operator solution of the quartic\index{quartic}, sextic\index{sextic}
and octic\index{octic} anharmonic oscillators. A close look into the difference
equations and their solutions developed in the earlier sections shows some symmetries
among them. Those symmetries encouraged us to search methods of obtaining solutions
of \( m-th \) anharmonic oscillator in general. In the next section of this
chapter we develop a zeroth order MSPT\index{MSPT} solution for the \( m-th \)
anharmonic oscillator and generalize the first order operator solutions of quartic\index{quartic}
oscillator obtained by various other techniques to obtain the first order operator
solution of generalized anharmonic oscillator in the subsequent sections. One
more thing we would like to note here, the Taylor series \index{Taylor series}
and MSPT\index{MSPT} methods for obtaining solutions of (\ref{eqm}) for particular
values of \( m \) are difficult and tedious. So in the next few sections of
this chapter we shall try to develop simpler techniques for solving (\ref{eqm})
in general.

\section{Generalized quantum anharmonic oscillator using an operator ordering approach}

In the present section we study the generalized quantum anharmonic oscillator
problem in the Heisenberg representation by making use of two simple normal
ordering\index{normal order} theorems for the expansion of \( (a+a^{\dagger })^{m} \).
We observe that the merit of the present work is its simplicity. For example,
the present method provides, on the one hand, a straight forward mathematical
frame work to construct expressions for the energy eigenvalues as well as frequency
shifts and, on the other hand, generalizes some of the recent results on the
topic by Bender and Bettencourt {[}\ref{Bender1}-\ref{Bender2}{]}. In the
subsequent sections of these chapter we have shown that the above mentioned
theorems may be used to generalize the results obtained by other methods.

We have already mentioned that the equation of motion corresponding to the Hamiltonian
(\ref{ten.1}) of a generalized anharmonic oscillator having unit mass and unit
frequency is given by (\ref{eqm}) which can not be solved exactly for \( m>2 \).
However, a large number of approximation methods are available for solving (\ref{eqm})
for particular values of \( m. \) We have discussed the Taylor series method
in detail and have mentioned some other methods. MSPT\index{MSPT} is one of
those methods. Bender and Bettencourt {[}\ref{Bender1},\ref{Bender2}{]} generalized
the existing theory of MSPT\index{MSPT} into an operator approach to get the
zeroth order solution involving a quantum operator analogue of the classical
first order frequency shift. Following the interpretation of Aks {[}\ref{aks1},\ref{aks2}{]}
they interpreted it as an operator mass renormalization that expresses the first
order shift of all energy levels. They obtained MSPT\index{MSPT} operator solution
of a quantum quartic\index{quartic} oscillator. In the present work we suggest
a novel method in which MSPT\index{MSPT} results for anharmonic oscillators
in general can be found. To that end we prove in the next subsection two theorems
to construct a normal ordered\index{normal order} expansion of \( (a+a^{\dagger })^{m} \)
. In the subsequent subsections we obtain a generalized expression for the energy
eigenvalues and a generalized solution for the equation of motion (\ref{eqm}).
We also specialize our result to reproduce some of the existing results as useful
checks on the generalized solution obtained by us.

\subsection{Operator ordering theorems}

On a very general ground one knows that in quantum mechanics and quantum field
theory proper ordering of the operators plays a crucial role. Let \( f(a,a^{\dagger }) \)
be an arbitrary operator function of the usual bosonic annihilation and creation
\index{creation operator} operators \( a \) and \( a^{\dagger } \) which
satisfy the commutation relation 
\begin{equation}
\label{one}
[a,a^{\dagger }]=1.
\end{equation}
One can write \( f(a,a^{\dagger }) \) in such a way that all powers of \( a^{\dagger } \)
always appear to the left of all powers of \( a \). Then \( f(a,a^{\dagger }) \)
is said to be normal ordered\index{normal order}. In this work we want to write
\( (a+a^{\dagger })^{m} \) in the normal ordered\index{normal order} form
for integral values of \( m \). Traditionally, for a given value of \( m \)
this is achieved by using a very lengthy procedure which involves repeated application
of (\ref{one}). One of our objectives in this work is to construct a normal
ordered\index{normal order} expansion of \( (a+a^{\dagger })^{m} \) without
taking recourse to such repeated applications. 

We denote normal ordered\index{normal order} form of \( f \) by \( f_{N} \).
On the other hand \( :\, f\, : \) denote an operator obtained from \( f \)
by arranging all powers of \( a^{\dagger } \) to the left of all powers of
\( a \) without making use of the commutation relation (\ref{one}). Now if
\( f=aa^{\dagger } \) then \( f_{N}=a^{\dagger }a+1 \) and \( :\, f\, :=a^{\dagger }a \).
Therefore, we can write
\begin{equation}
\label{two}
:\, (a+a^{\dagger })^{m}:=:\, (a^{\dagger }+a)^{m}:=a^{m}+^{m}C_{1}a^{\dagger }a^{m-1}+...+^{m}C_{r}a^{\dagger ^{r}}a^{m-r}+..+a^{\dagger ^{m}}.
\end{equation}
 Thus in this notation \( :\, (a+a^{\dagger })^{m}: \) is simply a binomial
expansion in which powers of the \( a^{\dagger } \) are always kept to the
left of the powers of the \( a \). To write \( (a^{\dagger }+a)_{N}^{m} \)
we can proceed by using the following theorems.

\subsubsection{Theorem 1. }

\begin{equation}
\label{th1.1}
:\, (a^{\dagger }+a)^{m}:\, (a^{\dagger }+a)=:\, (a^{\dagger }+a)^{m+1}:+m\, :\, (a^{\dagger }+a)^{m-1}:\, \, .
\end{equation}

Proof: From (\ref{two}) \( :\, (a^{\dagger }+a)^{m}:\, (a^{\dagger }+a) \)
can be written in the form 
\begin{equation}
\label{chfour}
\begin{array}{cl}
 & :\, (a^{\dagger }+a)^{m}:\, (a^{\dagger }+a)\\
= & \left[ a^{\dagger ^{m+1}}+(\, ^{m}C_{1}+^{m}C_{0})a^{\dagger ^{m}}a+(\, ^{m}C_{2}+^{m}C_{1})a^{\dagger ^{m-1}}a^{2}+....+(\, ^{m}C_{r}+^{m}C_{r-1})a^{\dagger ^{m-r+1}}a^{r}+...\right] \\
+ & (\, ^{m}C_{1}a^{\dagger ^{m-1}}+2\, ^{m}C_{2}a^{\dagger ^{m-2}}a+...+r\, ^{m}C_{r}a^{\dagger ^{m-r}}a^{r-1}+...)\\
= & :\, (a^{\dagger }+a)^{m+1}:+m\, :\, (a^{\dagger }+a)^{m-1}:\, \, .
\end{array}
\end{equation}
 In the above we have made use of the identities 
\begin{equation}
\label{pocha}
\begin{array}{lcl}
\, \, \, \, \, \, (a^{r}a^{\dagger })_{N} & = & a^{\dagger }a^{r}+ra^{r-1},\\
\, \, \, \, \, \, r\, ^{n}C_{r} & = & n\, ^{n-1}C_{r-1},\\
and\, \, \, \, ^{m+1}C_{r+1} & = & (\, ^{m}C_{r}+^{m}C_{r+1})\, .
\end{array}
\end{equation}
 Note that first identity in (\ref{pocha}) can be proved with the help of the
general operator ordering theorems {[}\ref{Louisell}{]} while the other two
are trivial.

\subsubsection{Theorem 2 : }

For any integral values of \( m \) 
\begin{equation}
\label{th2.1}
(a^{\dagger }+a)_{N}^{m}=\sum ^{m}_{r=0,2,4..}t_{r}\, ^{m}C_{r}:\, (a^{\dagger }+a)^{m-r}:
\end{equation}
with
\begin{equation}
\label{th2.2}
t_{r}=\frac{r!}{2^{\frac{r}{2}}(\frac{r}{2})!}
\end{equation}

Proof\footnote{
F M Fernandez {[}\ref{APFM01}{]} has pointed out that this theorem can be proved
by expanding both sides of the Campbell-Baker-Hausdroff formula {[}\ref{Louisell}{]}
\[
e^{\xi (a+a^{\dagger })}=e^{\xi a^{\dagger }}e^{\xi a}e^{\xi ^{2}}\]
 about \( \xi =0 \). 
}:

Using theorem 1, we can write
\begin{equation}
\label{six}
\begin{array}{lcl}
(a^{\dagger }+a)_{N} & = & :\, (a^{\dagger }+a):\\
(a^{\dagger }+a)_{N}^{2} & = & :\, (a^{\dagger }+a)^{2}:+1\\
(a^{\dagger }+a)_{N}^{3} & = & :\, (a^{\dagger }+a)^{3}:+3:\, (a^{\dagger }+a):\\
(a^{\dagger }+a)_{N}^{4} & = & :\, (a^{\dagger }+a)^{4}:+6\, :\, (a^{\dagger }+a)^{2}:+3\\
(a^{\dagger }+a)_{N}^{5} & = & :\, (a^{\dagger }+a)^{5}:+10\, :\, (a^{\dagger }+a)^{3}:+15\, :\, (a^{\dagger }+a):\\
(a^{\dagger }+a)_{N}^{6} & = & :\, (a^{\dagger }+a)^{6}:+15:\, (a^{\dagger }+a)^{4}:+45:\, (a^{\dagger }+a)^{2}:+15\\
(a^{\dagger }+a)_{N}^{7} & = & :\, (a^{\dagger }+a)^{7}:+21:\, (a^{\dagger }+a)^{5}:+105:\, (a^{\dagger }+a)^{3}:\\
 & + & 105:\, (a^{\dagger }+a):\\
(a^{\dagger }+a)_{N}^{8} & = & :\, (a^{\dagger }+a)^{8}:+28:\, (a^{\dagger }+a)^{6}:+210:\, (a^{\dagger }+a)^{4}:\\
 & + & 420:\, (a^{\dagger }+a)^{2}:+105\\
(a^{\dagger }+a)_{N}^{9} & = & :\, (a^{\dagger }+a)^{9}:+36:\, (a^{\dagger }+a)^{7}:+378:\, (a^{\dagger }+a)^{5}:\\
 & + & 1260:\, (a^{\dagger }+a)^{3}:+945:\, (a^{\dagger }+a):\, \, .
\end{array}
\end{equation}
 From (\ref{six}) we venture to identify the general form of the above expansion
for a given value of \( m \) as
\begin{equation}
\label{seven}
(a^{\dagger }+a)_{N}^{m}=\sum ^{m}_{r=0,2,4..}t_{r}\, ^{m}C_{r}:\, (a^{\dagger }+a)^{m-r}:\, \, .
\end{equation}
It is easy to check that (\ref{seven}) gives all the expansions of (\ref{six}).
So (\ref{seven}) is true for \( m=1,2,..,9 \). To ensure the general validity
of (\ref{seven}) we can use the method of induction. First we assume that (\ref{seven})
is true for a particular value of \( m \). Now, 
\begin{equation}
\label{nine}
\begin{array}{lcl}
(a^{\dagger }+a)_{N}^{m+1} & = & (a^{\dagger }+a)_{N}^{m}\, (a^{\dagger }+a)\\
 & = & \left[ :\, (a^{\dagger }+a)^{m}:+^{m}C_{2}:\, (a^{\dagger }+a)^{m-2}:+...\right. \\
 & + & \left. t_{r-2}\, ^{m}C_{r-2}:(a^{\dagger }+a)^{m-r+2}:+t_{r}\, ^{m}C_{r}:\, (a^{\dagger }+a)^{m-r}:+...\right] (a^{\dagger }+a)\\
 & = & \sum _{r}\left[ t_{r-2}\, ^{m}C_{r-2}(m-r+2)+t_{r}\, ^{m}C_{r}\right] \, :\, (a^{\dagger }+a)^{m-r+1}:\\
 & = & \sum _{r}t_{r}\, ^{m+1}C_{r}:\, (a^{\dagger }+a)^{m+1-r}:\, \, .\\
 &  & \\
 &  & 
\end{array}
\end{equation}
 Equation (\ref{nine}) exhibits that if (\ref{seven}) holds for any arbitrary
\( m \) it must hold for \( m+1 \). This establishes that (\ref{seven}) gives
the normal order \index{normal order} expansion of \( (a^{\dagger }+a)^{m} \)
for any arbitrary integer \( m \). We use this normal ordered\index{normal order}
expansion to study the anharmonic oscillator problem.

\subsection{Energy eigenvalues }

The first order energy eigenvalue\index{eigenvalue} \( E_{1} \) is \( <n|H|n> \),
where \( |n> \) is the number state. In the literature there are two lengthy
procedures {[}\ref{Powell}{]} to obtain \( E_{1} \). The first one is the
usual normal ordering\index{normal order} method. This method involves iterative
use of (\ref{one}) and the number of iterations increases very fast as \( m \)
increases. For a given value of \( m \) we need   \( \left[ 2^{m}-(m+1)\right]  \)
iterations. Just to get a feeling of the number of iterations required for large
values of \( m \) we can look at table 1 below. 

\vspace{0.375cm}
{\centering \begin{tabular}{|c|c|}
\hline 
\( m \)&
no. of iterations required\\
\hline 
\hline 
2&
1\\
\hline 
6&
57\\
\hline 
10&
1013\\
\hline 
20&
1048555\\
\hline 
100&
1267650600228229401496703205275\\
\hline 
\end{tabular}\par}
\vspace{0.375cm}

{\par\centering Table 2.1\par}

The table shows that for large \( m \) the construction of the expansion  \( (a^{\dagger }+a)_{N}^{m} \)
becomes formidable. In the second procedure one proceeds by making repeated
application of \( x \) operator on the number state. As in the normal ordering\index{normal order}
method this procedure is also equally lengthy. In the following we implement
theorem 2 to derive an uncomplicated method to construct an expansion for \( E_{1} \)
for the Hamiltonian (\ref{ten.1}).

Using the result in (\ref{th2.1}) we can write the expression for \( E_{1} \)
for any integer value of \( m \) in the form
\begin{equation}
\label{12}
\begin{array}{lcl}
E_{1} & = & (n+\frac{1}{2})+\frac{\lambda }{2^{\frac{m}{2}}m}<n|\sum ^{m}_{r=0,2,4..}t_{r}\, ^{m}C_{r}:\, (a^{\dagger }+a)^{m-r}:\, |n>\\
 & = & (n+\frac{1}{2})+\frac{\lambda }{2^{\frac{m}{2}}m}<n|\sum ^{m}_{r=0,2,4..}t_{r}\, ^{m}C_{r}\, ^{m-r}C_{\frac{m-r}{2}}\, a^{\dagger ^{\frac{m-r}{2}}}a^{\frac{m-r}{2}}|n>\\
 & = & (n+\frac{1}{2})+\frac{\lambda }{2^{\frac{m}{2}}m}\sum ^{m}_{r=0,2,4..}t_{r}\, ^{m}C_{r}\, ^{m-r}C_{\frac{m-r}{2}}\, ^{n}C_{\frac{m-r}{2}}\, (\frac{m-r}{2})!
\end{array}
\end{equation}
 The expression (\ref{12}) involves summations which are easy to evaluate and
thereby avoids the difficulties associated with earlier iterative procedures.
Although somewhat forced, it may be tempting to compare the simplicity sought
in our approach with the use of logarithm in a numerical calculation or use
of integral transforms in solving a partial differential equation. It is of
interest to note that expression similar to (\ref{12}) can also be constructed
for the first order energy eigenvalue\index{eigenvalue} of a Hamiltonian in
which the anharmonic term is a polynomial in \( X \).

\subsection{MSPT\index{MSPT} solution of the generalized quantum anharmonic oscillator}

We want to obtain MSPT operator solution of equation (\ref{eqm}) for arbitrary
integral values of \( m \). The essential idea behind our approach is that
the quantum operator analogue of the classical first order frequency shift is
an operator function ( \( \Omega (H_{0}) \) ) of the unperturbed Hamiltonian
\( H_{0} \) and secondly a correct solution should reproduce the first order
energy spectrum. Now the general form of the zeroth order solution of (\ref{eqm})
is given by 
\begin{equation}
\label{sol1}
\begin{array}{lcl}
X_{0}(t) & = & \frac{1}{G(n)}\left[ X(0)cos(t+\lambda \Omega (H_{0})t)+cos(t+\lambda \Omega (H_{0})t)X(0)\right. \\
 & + & \left. \dot{X}(0)sin(t+\lambda \Omega (H_{0})t)+sin(t+\lambda \Omega (H_{0})t)\dot{X}(0)\right] 
\end{array}
\end{equation}
 where \( G(n) \) is a normalization factor. So our task is to find out \( \Omega (H_{0}) \)
and \( G(n) \) in general. From equation (\ref{12}) we obtain the energy difference
for two consecutive energy levels as 
\begin{equation}
\label{sol2}
\begin{array}{lcl}
\omega _{n,n-1} & = & (E_{1})_{m,n}-(E_{1})_{m,n-1}\\
 & = & 1+\lambda \omega (m,n)\\
 & = & 1+\frac{\lambda }{2^{\frac{m}{2}}m}\sum ^{(m-2)}_{r=0,2,4..}t_{r}\, ^{m}C_{r}\, ^{m-r}C_{\frac{m-r}{2}}\, ^{n-1}C_{\frac{m-r-2}{2}}\, (\frac{m-r}{2})!\, .
\end{array}
\end{equation}

Since the correct quantum operator solution has to give the first order energy
spectrum, we should have
\begin{equation}
\label{sol3}
\begin{array}{lcl}
<n-1|X_{0}(t)|n> & = & <n-1|X_{0}(0)|n>cos[t+\lambda \omega (m,n)t]\\
 & + & <n-1|\dot{X}_{0}(0)|n>sin[t+\lambda \omega (m,n)t]\, .
\end{array}
\end{equation}
 Equation (\ref{sol1}) and (\ref{sol3}) impose restrictions on our unknown
functions \( \Omega (H_{0}) \) and \( G(n) \). The condition imposed on \( \Omega (H_{0}) \)
is 

\begin{equation}
\label{sol4}
<n|\Omega (H_{0})|n>+<n-1|\Omega (H_{0})|n-1>=2\omega (m,n)
\end{equation}
or, 
\begin{equation}
\label{sol5}
\Omega (n+\frac{1}{2})+\Omega (n-\frac{1}{2})=2\omega (m,n)\, \, .
\end{equation}
For a particular \( m \) the right hand side is a known polynomial in \( n \)
and our job is simply to find out \( \Omega (n+\frac{1}{2}) \). We obtain this
as 
\begin{equation}
\label{sol6}
\Omega (n+\frac{1}{2})=2\left[ \sum ^{n}_{k=0}(-1)^{n-k}\omega (m,k)\right] +(-1)^{n+\frac{m}{2}}\frac{t_{m}}{2^{\frac{m-2}{2}}m}.
\end{equation}
 Substituting the functional form of \( \Omega (n+\frac{1}{2}) \) or \( \Omega (H_{0}) \)
in (\ref{sol1}) if we impose condition (\ref{sol3}) we get 
\begin{equation}
\label{sol6.1}
G(n)=2cos\left[ \frac{\lambda t}{2}\left( \Omega (n+\frac{1}{2})-\Omega (n-\frac{1}{2})\right) \right] .
\end{equation}
 The results in (\ref{sol6}) and (\ref{sol6.1}) when substituted in (\ref{sol1})
solve the generalized quantum anharmonic oscillator problem.

\subsection{Specific results and their comparison with the existing spectra:}

Although we have solved the quantum anharmonic oscillator in general, just now
it is not possible to compare our solution directly with any other result since
the present study happens to be the first operator solution of the generalized
anharmonic oscillator. Later on we develop other techniques to solve the anharmonic
oscillator in general and compare those result with the present one. Here we
calculate some specific results from our general expressions and compare them
with the existing results. 

For \( m=4 \) we have, 
\begin{equation}
\label{quartic1}
\Omega (n+\frac{1}{2})=\frac{3n}{4}+\frac{3}{8}=\frac{3}{4}(n+\frac{1}{2})\, .
\end{equation}
 Therefore, 
\begin{equation}
\label{sol8}
\begin{array}{lcl}
\, \, \, \, \, \, \Omega (H_{0}) & = & \frac{3}{4}H_{0}\\
and\, \, \, \, \, G(n) & = & 2cos(\frac{3\lambda t}{8})\, .
\end{array}
\end{equation}
 In terms of (\ref{sol8}) the total solution for the quantum quartic\index{quartic}
anharmonic oscillator is 
\begin{equation}
\label{quartic2}
\begin{array}{lcl}
X(t)|_{m=4} & = & \frac{1}{2cos(\frac{3\lambda t}{8})}\left[ X(0)cos[t+\frac{3\lambda t}{4}H_{0}]+cos[t+\frac{3\lambda t}{4}H_{0}]X(0)\right. \\
 & + & \left. \dot{X}(0)sin[t+\frac{3\lambda t}{4}H_{0}]+sin[t+\frac{3\lambda t}{4}H_{0}]\dot{X}(0)\right] 
\end{array}
\end{equation}
 This coincides exactly with the solution given by Bender and Bettencourt {[}\ref{Bender1},\ref{Bender2}{]}
and also coincides with the Taylor series \index{Taylor series} solution (\ref{16}).
This also gives the correct classical frequency shift {[}\ref{bdr}{]} in the
limit \( x(0)=a,\, \dot{x}(0)=0 \). Similarly we have 
\begin{equation}
\label{sol9}
\begin{array}{lcl}
X(t)|_{m=6} & = & \frac{1}{2cos(\frac{5\lambda t}{4}n)}\left[ X(0)cos[t+\frac{5\lambda }{4}(H_{0}^{2}+\frac{1}{4})t]+cos[t+\frac{5\lambda }{4}(H_{0}^{2}+\frac{1}{4})t]X(0)\right. \\
 & + & \left. \dot{X}(0)sin[t+\frac{5\lambda }{4}(H_{0}^{2}+\frac{1}{4})t]+sin[t+\frac{5\lambda }{4}(H_{0}^{2}+\frac{1}{4})t]\dot{X}(0)\right] 
\end{array},
\end{equation}
\[
\begin{array}{lcl}
x(t)|_{m=8} & = & \frac{1}{2cos[\frac{35\lambda }{64}(6n^{2}+3)t]}\\
 & \times  & \left[ X(0)cos[t+\frac{35\lambda }{64}(4H_{0}^{3}+5H_{0})t]+cos[t+\frac{35\lambda }{64}(4H_{0}^{3}+5H_{0})t]X(0)\right. \\
 & + & \left. \dot{X}(0)sin[t+\frac{35\lambda }{64}(4H_{0}^{3}+5H_{0})t]+sin[t+\frac{35\lambda }{64}(4H_{0}^{3}+5H_{0})t]\dot{X}(0)\right] ,
\end{array}\]
 and 
\begin{equation}
\label{sol11}
\begin{array}{lcl}
x(t)|_{m=10} & = & \frac{1}{2cos[\frac{63\lambda t}{8}(n^{3}+2n)]}\left[ X(0)cos[t+\frac{63\lambda }{16}(H_{0}^{4}+\frac{7}{2}H^{2}_{0}+\frac{9}{16})t]\right. \\
 & + & cos[t+\frac{63\lambda }{16}(H_{0}^{4}+\frac{7}{2}H^{2}_{0}+\frac{9}{16})t]X(0)+\dot{X}(0)sin[t+\frac{63\lambda }{16}(H_{0}^{4}+\frac{7}{2}H^{2}_{0}+\frac{9}{16})t]\\
 & + & \left. sin[t+\frac{35\lambda }{64}\frac{63\lambda }{16}(H_{0}^{4}+\frac{7}{2}H^{2}_{0}+\frac{9}{16})t]\dot{X}(0)\right] \, .
\end{array}
\end{equation}
 The solutions for sextic\index{sextic} and octic\index{octic} oscillator
exactly coincides with the solution obtained by us using Taylor series \index{Taylor series}
approach {[}\ref{Pathak}{]} and all solutions give the correct classical frequency
shifts in the appropriate limit {[}\ref{R. Dutt}, \ref{Bradbury}{]}.

\subsection{Remarks on the solutions}

We conclude by noting that depending on the nature of nonlinearity in a physical
problem the treatment of higher anharmonic oscillators assumes significance.
But studies in such oscillators (for \( m>4 \)) are not undertaken in the Heisenberg
approach presumably because the existing methods tends to introduce inordinate
mathematical complications in a detailed study. In the present work we contemplate
to circumvent them by proving a theorem for the expansion of \( (a^{\dagger }+a)_{N}^{m} \).
Thus the results of the present section are expected to serve a useful purpose
for physicists working in nonlinear mechanics, molecular physics, quantum optics
and quantum field theory. In fact, we use these solutions to study nonlinear
and quantum optical effects in detail in the later part of the present thesis.
\vspace{1.5cm}

\section{Generalized quantum anharmonic oscillator using the time evolution operator
approach}

Generalized quantum anharmonic oscillator can be studied through some alternative
approaches. We have started our discussion on the possible alternative approaches
from the MSPT\index{MSPT} approach. In the present section we study the time
evolution operator approach. In this approach we find out the perturbed time
evolution operator in the interaction picture \( U_{I}(t) \) as a function
of \( a,a^{\dagger } \) and \( t \) and then we use that to find out the time
evolution of the creation \index{creation operator} and annihilation operators\index{annihilation operator}
which in turn solve the equation of motion (\ref{eqm}) of the generalized anharmonic
oscillator. The secular terms are removed from the solution by using the tucking
in technique\index{tucking in technique}, discussed in the context of Taylor
series \index{Taylor series} solution.

\subsection{Time evolution operator in the interaction picture:}

In the interaction picture potential \( V_{I} \) and the time evolution operator
\( U_{I}(t) \) are respectively 
\begin{equation}
\label{ev1}
V_{I}(t)=\exp (iH_{0}t)V\exp (-iH_{0}t)
\end{equation}
 and 
\begin{equation}
\label{ev2}
U_{I}(t)=1-i\int ^{t}_{0}dt_{1}V_{I}(t_{1})+(-i)^{2}\int ^{t}_{0}dt_{1}V_{I}(t_{1})\int ^{t_{1}}_{0}dt_{2}V_{I}(t_{2})+...\, \, .
\end{equation}
 Where the suffix \( I \) stands for the interaction picture. Now to obtain
the compact expressions for \( V_{I}(t) \) and \( U_{I}(t) \) (up to first
order) for the model Hamiltonian (\ref{ten}) of our interest we have to use
following operator ordering theorems \\
\textbf{Theorem 3:} When \( f\left( a,a^{\dagger }\right)  \) can be expanded
as a power series in \( a \) and \( a^{\dagger } \) then 
\begin{equation}
\label{ev3}
\exp (\varsigma a^{\dagger }a)f\left( a,a^{\dagger }\right) \exp (-\varsigma a^{\dagger }a)=f\left( a\exp (-\varsigma ),a^{\dagger }\exp (\varsigma )\right) 
\end{equation}
 where \( \varsigma  \) is a c-number {[}\ref{Louisell}{]}. 

With the help of theorem 2 (see section 2.3.1.2) and theorem 3 we can write
\begin{equation}
\label{ev4}
\begin{array}{lcl}
V_{I}(t) & = & \exp (ia^{\dagger }at)\frac{\lambda }{m(2)^{\frac{m}{2}}}(a^{\dagger }+a)^{m}\exp (-ia^{\dagger }at)\\
 & = & \frac{\lambda }{m(2)^{\frac{m}{2}}}\sum ^{\frac{m}{2}}_{r}t_{2r}\, ^{m}C_{2r}\sum ^{(m-2r)}_{p}\, ^{(m-2r)}C_{p}a^{\dagger ^{p}}a^{m-2r-p}\exp \left( it(2p-m+2r)\right) 
\end{array}.
\end{equation}
 Now the first order time evolution operator in the interaction picture is 
\begin{equation}
\label{ev5}
\begin{array}{lcl}
U_{I}(t) & = & 1-i\int ^{t}_{0}dt_{1}V_{I}(t_{1})\\
 & = & 1-i\frac{\lambda }{m(2)^{\frac{m}{2}}}\int ^{t}_{0}\left( \sum ^{\frac{m}{2}}_{r}t_{2r}\, ^{m}C_{2r}\sum ^{(m-2r)}_{p}\, ^{(m-2r)}C_{p}a^{\dagger ^{p}}a^{m-2r-p}\exp \left( it_{1}(2p-m+2r)\right) \right) dt_{1}\\
 & = & 1-\frac{\lambda }{m(2)^{\frac{m}{2}}}\left( \sum ^{\frac{m}{2}}_{r}t_{2r}\, ^{m}C_{2r}\sum ^{(m-2r)}_{p\neq \frac{m-2r}{2}}\, ^{(m-2r)}C_{p}a^{\dagger ^{p}}a^{m-2r-p}\frac{\left[ \exp \left( it(2p-m+2r)\right) -1\right] }{2p-m+2r}\right) \\
 & - & i\frac{\lambda t}{m(2)^{\frac{m}{2}}}\left( \sum ^{\frac{m}{2}}_{r}t_{2r}\, ^{m}C_{2r}\, ^{(m-2r)}C_{\frac{m-2r}{2}}a^{\dagger ^{\frac{m-2r}{2}}}a^{\frac{m-2r}{2}}\right) \, .\\
 &  & 
\end{array}
\end{equation}
 The time evolution of the annihilation operator\index{annihilation operator}
\( a \) in the Heisenberg picture \index{Heisenberg picture} \( a_{H} \)
and in the interaction picture \( a_{I} \) are related by 
\begin{equation}
\label{ev5.1}
a_{H}(t)=U^{\dagger }_{I}(t)\exp (ia^{\dagger }at)a(0)\exp (-ia^{\dagger }at)U_{I}(t)=\exp (-it)a_{I}(t).
\end{equation}
 Now the time evolution of the annihilation operator\index{annihilation operator}
in the interaction picture is 
\begin{equation}
\label{ev6}
\begin{array}{lcl}
a_{I}(t) & = & U^{\dagger }_{I}(t)a(0)U_{I}(t)\\
 & = & a-\frac{\lambda }{m(2)^{\frac{m}{2}}}\left( \sum ^{\frac{m}{2}}_{r}t_{2r}\, ^{m}C_{2r}\sum ^{(m-2r)}_{p\neq \frac{m-2r}{2}}\, ^{(m-2r)}C_{p}a^{\dagger ^{p-1}}a^{m-2r-p}\frac{p\left[ \exp \left( it(2p-m+2r)\right) -1\right] }{2p-m+2r}\right) \\
 & - & i\frac{\lambda t}{m(2)^{\frac{m}{2}}}\left( \sum ^{\frac{m}{2}}_{r}\frac{m-2r}{2}t_{2r}\, ^{m}C_{2r}\, ^{(m-2r)}C_{\frac{m-2r}{2}}a^{\dagger ^{\frac{m-2r}{2}-1}}a^{\frac{m-2r}{2}}\right) .
\end{array}
\end{equation}
 So we have obtained the time evolution of the annihilation operator\index{annihilation operator}.

\subsection{First order frequency operator\index{frequency operator}: }

In the solution of (\ref{eqm}) frequency becomes a operator function. In the
first order frequency operator\index{frequency operator} is a function of the
unperturbed Hamiltonian \( H_{0} \). The frequency operator is known for various
values of \( m \). Here we give a compact expression for the first order frequency
operator\index{frequency operator} \( \Omega (H_{0}) \) and write the expression
for some particular \( m \) to compare our result with the existing results.
To do so we have to use the following identity 
\begin{equation}
\label{ev8}
\begin{array}{lcl}
a^{\dagger ^{n}}a^{n} & = & a^{\dagger }a(a^{\dagger }a-1)(a^{\dagger }a-2)...(a^{\dagger }a-n+1)\\
 & = & (H_{0}-\frac{1}{2})(H_{0}-\frac{3}{2})....(H_{0}-n+\frac{1}{2})\\
 & = & \frac{\Gamma (H_{0}+\frac{1}{2})}{\Gamma (H_{0}-n+\frac{1}{2})}\, .\\
 &  & 
\end{array}
\end{equation}
 The secular terms present in (\ref{ev6}) can be tucked in into the expression
of \( a \) by using the tucking in technique \index{tucking in technique}introduced
in (\ref{13a}). Now using (\ref{ev6} and \ref{13a.1}) we can write
\begin{equation}
\label{ev11}
\begin{array}{lcl}
a(t) & = & \exp (-i\lambda \Omega t)a\\
 & - & \frac{\lambda }{m(2)^{\frac{m}{2}}}\left( \sum ^{\frac{m}{2}}_{r}t_{2r}\, ^{m}C_{2r}\sum ^{(m-2r)}_{p\neq \frac{m-2r}{2}}\, ^{(m-2r)}C_{p}a^{\dagger ^{p-1}\exp \left( i\Omega t(2p-m+2r-1)\right) }\right. \\
 & \times  & \left. a^{m-2r-p}\frac{p\left[ \exp \left( it(2p-m+2r)\right) -1\right] }{2p-m+2r}\right) 
\end{array}
\end{equation}
 where 
\begin{equation}
\label{ev12}
\begin{array}{lcl}
\Omega  & = & \frac{1}{m(2)^{\frac{m}{2}}}\left( \sum ^{\frac{m}{2}}_{r}\frac{m-2r}{2}t_{2r}\, ^{m}C_{2r}\, ^{(m-2r)}C_{\frac{m-2r}{2}}a^{\dagger ^{\frac{m-2r}{2}-1}}a^{\frac{m-2r}{2}-1}\right) \\
 & = & \frac{1}{m(2)^{\frac{m}{2}}}\left( \sum ^{\frac{m}{2}}_{r}\frac{m-2r}{2}t_{2r}\, ^{m}C_{2r}\, ^{(m-2r)}C_{\frac{m-2r}{2}}\frac{\Gamma (H_{0}+\frac{1}{2})}{\Gamma (H_{0}-\frac{m-2r}{2}+\frac{3}{2})}\right) 
\end{array}.
\end{equation}
 Thus we have a compact expression for the first order correction to the frequency
operator\index{frequency operator} (\ref{ev12}) and an expression for the
time evolution of annihilation operator \index{annihilation operator} which
is free from secular term (\ref{ev11}). Expression (\ref{ev11}) essentially
solves the equation of motion (\ref{eqm}) of the generalized anharmonic oscillator
since 
\begin{equation}
\label{ev7}
X(t)=\frac{1}{\sqrt{2}}(a^{\dagger }(t)+a(t)).
\end{equation}
 Here we must note that the frequency operator\index{frequency operator} of
the annihilation operator \index{annihilation operator} is not exactly the
same with the operator usually used to write the first order frequency correction
of the position operator. Relationship and equivalence criterion for the frequency
operators\index{frequency operator} obtained by using different techniques
are discussed at the end of this chapter.

\section{Generalized quantum anharmonic oscillator using renormalization group \index{renormalization group}
technique}

In this method a time parameter, additional to the initial value point is introduced,
in such a way that the perturbation expansion is valid in the vicinity of the
introduced time parameter. The coupling constants, constants of motion and/or
initial conditions are suitably changed by the introduced time parameter. But
the solution should not depend on the introduced time parameter so the first
derivative of the perturbed solution with respect to introduced time parameter
should be zero and this condition is known as RG condition. Solution of the
differential equation imposed by RG condition can give provide us an operator
solution of the anharmonic oscillator free from secular terms.

From the perturbation expansion (\ref{ev6}) the secular part of the annihilation
operator \index{annihilation operator} up to the first order is 
\begin{equation}
\label{rg1}
a_{sec}=-i\frac{\lambda (t-\tau )}{m(2)^{\frac{m}{2}}}\left( \sum ^{\frac{m}{2}}_{r}\frac{m-2r}{2}t_{2r}\, ^{m}C_{2r}\, ^{(m-2r)}C_{\frac{m-2r}{2}}a^{\dagger ^{\frac{m-2r}{2}-1}}a^{\frac{m-2r}{2}}\right) .
\end{equation}
 Now imposing the RG condition \( \frac{da}{d\tau }=0 \), we obtain 
\begin{equation}
\label{rg2}
\frac{da}{d\tau }=-i\frac{\lambda }{m(2)^{\frac{m}{2}}}\left( \sum ^{\frac{m}{2}}_{r}\frac{m-2r}{2}t_{2r}\, ^{m}C_{2r}\, ^{(m-2r)}C_{\frac{m-2r}{2}}a^{\dagger ^{\frac{m-2r}{2}-1}}a^{\frac{m-2r}{2}}\right) .
\end{equation}
 Since \( a^{\dagger }a \) and \( [a,a^{\dagger }] \) are constants under
the flow of \( \tau  \), we are allowed to solve equation (\ref{rg2}) as 
\begin{equation}
\label{rg3}
a(\tau )=\exp \left( -i\frac{\lambda \tau }{m(2)^{\frac{m}{2}}}\left( \sum ^{\frac{m}{2}}_{r}\frac{m-2r}{2}t_{2r}\, ^{m}C_{2r}\, ^{(m-2r)}C_{\frac{m-2r}{2}}a^{\dagger ^{\frac{m-2r}{2}-1}}a^{\frac{m-2r}{2}-1}\right) \right) a(0).
\end{equation}
 Therefore the first order correction to frequency operator\index{frequency operator}
is 
\begin{equation}
\label{rg4}
\begin{array}{lcl}
\Omega _{1} & = & \frac{1}{m(2)^{\frac{m}{2}}}\left( \sum ^{\frac{m}{2}}_{r}\frac{m-2r}{2}t_{2r}\, ^{m}C_{2r}\, ^{(m-2r)}C_{\frac{m-2r}{2}}a^{\dagger ^{\frac{m-2r}{2}-1}}a^{\frac{m-2r}{2}-1}\right) \\
 & = & \frac{1}{m(2)^{\frac{m}{2}}}\left( \sum ^{\frac{m}{2}}_{r}\frac{m-2r}{2}t_{2r}\, ^{m}C_{2r}\, ^{(m-2r)}C_{\frac{m-2r}{2}}\frac{\Gamma (H_{0}+\frac{1}{2})}{\Gamma (H_{0}-\frac{m-2r}{2}+\frac{3}{2})}\right) 
\end{array}.
\end{equation}
 This is in exact accordance with the first order frequency operator \index{frequency operator}
derived by the time evolution operator method (\ref{ev12}).

\section{Generalized quantum anharmonic oscillator using near-identity transformation
technique\index{near identity transform method}}

The near-identity transformation relates the full solution for the annihilation
operator \index{annihilation operator} \( a \) to the zeroth order term \( b \)
as 
\begin{equation}
\label{nit1}
a(t)=b(t)+\lambda T_{1}(b^{\dagger }(t),b(t))+\lambda ^{2}T_{2}(b^{\dagger }(t),b(t))+.......
\end{equation}
 and the equation for the time dependence of the zeroth order term in the normal
form is 
\begin{equation}
\label{nit2}
\frac{db(t)}{dt}=U_{0}(b^{\dagger }(t),b(t))+\lambda U_{1}(b^{\dagger }(t),b(t))+\lambda ^{2}U_{2}(b^{\dagger }(t),b(t))+....
\end{equation}
 Now using the Heisenberg equation\index{Heisenberg equation} of motion for
the Hamiltonian (\ref{ten.1} ) we obtain 
\begin{equation}
\label{nit3}
\begin{array}{lcl}
\frac{da}{dt} & = & ia+i\frac{\lambda }{2^{\frac{m}{2}}}(a+a^{\dagger })^{m-1}\\
 & = & ia+i\frac{\lambda }{2^{\frac{m}{2}}}\sum ^{\frac{m-1}{2}}_{r=0}t_{2r}\, ^{m-1}C_{2r}\, ^{(m-2r-1)}C_{p}a^{\dagger ^{p}}a^{m-2r-1-p}
\end{array}.
\end{equation}
 Inserting equation (\ref{nit1} and \ref{nit2}) into (\ref{nit3}) we obtain
a relation between \( U_{n} \) and \( T_{n} \) in each order \( n \). These
relations can not determine \( U_{n} \) and \( T_{n} \) uniquely. To reduce
this nonuniqueness problem we follow the work of Kahn and Zarmi {[}\ref{kahn}{]}
and choose \( T_{n} \) not to depend on time explicitly. In the first order
only the resonant terms of the form \( (b^{\dagger }b)^{k}b \) acts as the
source terms for the explicitly time dependent part of the annihilation operator
\index{annihilation operator} \( a \). Thus from the expansion (\ref{nit3})
we have 
\begin{equation}
\label{nit4}
U_{1}=\frac{1}{2^{\frac{m}{2}}}\sum ^{(\frac{m}{2}-1)}_{r=0}t_{2r}\, ^{m-1}C_{2r}\, ^{(m-2r-1)}C_{\frac{m-2r}{2}-1}b^{\dagger ^{\frac{m-2r}{2}-1}}(t)b^{\frac{m-2r}{2}}(t).
\end{equation}
 Substituting (\ref{nit4}) in (\ref{nit2}) we obtain the dynamical equation
of \( b(t) \) as 
\[
\begin{array}{lcl}
\frac{db(t)}{dt} & = & -ib(t)+\frac{\lambda }{2^{\frac{m}{2}}}\sum ^{(\frac{m}{2}-1)}_{r=0}t_{2r}\, ^{m-1}C_{2r}\, ^{(m-2r-1)}C_{\frac{m-2r}{2}-1}b^{\dagger ^{\frac{m-2r}{2}-1}}(t)b^{\frac{m-2r}{2}}(t)\\
 &  & 
\end{array}.\]
 Therefore we have 
\begin{equation}
\label{nit6}
b=\exp \left( -it\left( 1+\frac{\lambda }{2^{\frac{m}{2}}}\sum ^{(\frac{m}{2}-1)}_{r=0}t_{2r}\, ^{m-1}C_{2r}\, ^{(m-2r-1)}C_{\frac{m-2r}{2}-1}b_{0}^{\dagger ^{\frac{m-2r}{2}-1}}b_{0}^{\frac{m-2r}{2}-1}\right) \right) b_{0}.
\end{equation}
 Now since the perturbation starts at \( t=0 \) so \( b(0)=a(0)+\cal O(\lambda ) \)
and the zeroth order term \( b \) of the annihilation operator \index{annihilation operator}
is 
\[
b(t)=\exp \left( -it\left( 1+\frac{\lambda }{2^{\frac{m}{2}}}\sum ^{(\frac{m}{2}-1)}_{r=0}t_{2r}\, ^{m-1}C_{2r}\, ^{(m-2r-1)}C_{\frac{m-2r}{2}-1}a^{\dagger ^{\frac{m-2r}{2}-1}}a^{\frac{m-2r}{2}-1}\right) \right) a(0).\]
 Therefore the first order correction to the frequency operator\index{frequency operator}
is 
\begin{equation}
\label{nit7}
\begin{array}{lcl}
\Omega _{1} & = & \frac{1}{2^{\frac{m}{2}}}\sum ^{(\frac{m}{2}-1)}_{r=0}t_{2r}\, ^{m-1}C_{2r}\, ^{(m-2r-1)}C_{\frac{m-2r}{2}-1}a^{\dagger ^{\frac{m-2r}{2}-1}}a^{\frac{m-2r}{2}-1}\\
 & = & \frac{1}{2^{\frac{m}{2}}}\sum ^{(\frac{m}{2}-1)}_{r=0}t_{2r}\, ^{m-1}C_{2r}\, ^{(m-2r-1)}C_{\frac{m-2r}{2}-1}\frac{\Gamma (H_{0}+\frac{1}{2})}{\Gamma (H_{0}-\frac{m-2r}{2}+\frac{3}{2})}\\
 & = & \frac{1}{m(2)^{\frac{m}{2}}}\left( \sum ^{\frac{m}{2}}_{r}\frac{m-2r}{2}t_{2r}\, ^{m}C_{2r}\, ^{(m-2r)}C_{\frac{m-2r}{2}}\frac{\Gamma (H_{0}+\frac{1}{2})}{\Gamma (H_{0}-\frac{m-2r}{2}+\frac{3}{2})}\right) 
\end{array}.
\end{equation}
 This is exactly equivalent with the frequency operator\index{frequency operator}
obtained by using the evolution operator technique (\ref{ev12}) and the renormalization
group \index{renormalization group} technique (\ref{rg4}).

\section{Generalized quantum anharmonic oscillator using eigenvalue approach\index{eigenvalue approach}}

This straightforward operator method is based on a generalization of the harmonic
oscillator algebra for the creation \index{creation operator} and annihilation
boson operators. This method was first used by Fern\'{a}ndez for the purpose
of obtaining the first order correction to the frequency operator\index{frequency operator}
for particular values of \( m \). Later on we have generalized his idea to
obtain a generalized expression for the first order frequency operator {[}\ref{APFM01}{]}.
The generalized treatment is given below.

In this approach the Hamiltonian operator \( H \) is given and we look for
an operator \( b \) such that {[}\ref{F01a}-\ref{S00}{]} 
\begin{equation}
\label{eq:[H,b]}
\begin{array}{rcl}
\left[ H,b\right]  & = & -\Omega b\\
\left[ H,\Omega \right] = & 0,\; \Omega ^{\dagger }=\Omega 
\end{array}
\end{equation}
Under such conditions it is possible to obtain eigenfunctions common to \( H \)
and \( \Omega  \): 
\begin{eqnarray}
H\Psi _{n} & = & E_{n}\Psi _{n},\nonumber \\
\Omega \Psi _{n} & = & \Omega _{n}\Psi _{n}.\label{eq:HPsi} 
\end{eqnarray}
 It follows from the equations (\ref{eq:[H,b]} and \ref{eq:HPsi} ) that 
\begin{equation}
\label{eq:hyper}
(E_{n}-E_{m}+\Omega _{n})\langle \Psi _{n}|b|\Psi _{m}\rangle =0.
\end{equation}
 Since the operator \( b \) is nonzero, there must be a pair of states \( \Psi _{m} \)
and \( \Psi _{n} \) such that \( \langle \Psi _{m}|b|\Psi _{n}\rangle \neq 0 \);
therefore, 
\begin{equation}
\label{eq:Omega_n}
\Omega _{n}=E_{m}-E_{n.}
\end{equation}
 The time evolution of the operator \( b \) is given by 
\begin{equation}
\label{eq:b(t)}
b(t)=e^{itH}be^{-itH}=e^{-it\Omega }b;
\end{equation}
 therefore, 
\begin{equation}
\label{eq:b(t)_mn}
\langle \Psi _{n}|b(t)|\Psi _{m}\rangle =e^{-it\Omega _{n}}\langle \Psi _{n}|b|\Psi _{m}\rangle =e^{it(E_{n}-E_{m})}\langle \Psi _{n}|b|\Psi _{m}\rangle .
\end{equation}
 If we can write an operator \( O \) in terms of \( b \) and \( b^{\dagger } \)
as \( O=O(b,b^{\dagger }) \) then we have 
\begin{equation}
\label{eq:O(t)}
e^{itH}Oe^{-itH}=O(e^{-it\Omega }b,b^{\dagger }e^{it\Omega })
\end{equation}

If we rewrite equation (\ref{eq:b(t)}) as \( e^{it(H+\Omega )}b=be^{itH} \)
and expand both sides in Taylor series \index{Taylor series} about \( t=0 \)
we obtain \( (H+\Omega )^{k}b=bH^{k} \). Consequently, for any operator function
\( F(H) \) that can be expressed as an \( H \)-power series we have {[}\ref{S00}{]}
\begin{equation}
\label{eq:bF(H)}
bF(H)=F(H+\Omega )b,
\end{equation}
 which is consistent with equation (\ref{eq:Omega_n}). In order to generalize
this result we can define 
\begin{equation}
\label{eq:F(k,H)}
b^{k}F(H)=F(k,H)b^{k},\; F(0,H)=F(H).
\end{equation}
 A straightforward calculation shows that 
\begin{equation}
\label{eq:recF(k,H)}
F(k,H)=F(k-1,H+\Omega (H)).
\end{equation}

It is not difficult to prove that \( b^{\dagger }b \) is a constant of the
motion 
\begin{equation}
\label{eq:[H,b+b]}
[H,b^{\dagger }b]=b^{\dagger }\Omega b-b^{\dagger }\Omega b=0
\end{equation}
 but that \( bb^{\dagger } \) is not 
\begin{equation}
\label{eq:[H,bb+]}
[H,bb^{\dagger }]=-\Omega bb^{\dagger }+bb^{\dagger }\Omega =[\Omega (H+\Omega (H)-\Omega (H+\Omega ))-\Omega (H)]bb^{\dagger },
\end{equation}
 unless \( \Omega (H+\Omega )=\Omega (H) \).

In general, it is not possible to solve equations (\ref{eq:[H,b]}) except for
some trivial models. One can, however, obtain approximate solutions by means
of perturbation theory. If we write \( H=H_{0}+\lambda H^{\prime } \) and expand
\begin{equation}
\label{eq:op_series}
\Omega =\sum _{j=0}^{\infty }\Omega _{j}\lambda ^{j},\; b=\sum _{j=0}^{\infty }b_{j}\lambda ^{j}
\end{equation}
 then we obtain the coefficients \( \Omega _{j} \) and \( b_{j} \) from {[}\ref{F01a},\ref{F01b}{]}
\begin{equation}
\label{eq:b_j}
\begin{array}{lcl}
\left[ H_{0},\Omega _{j}\right]  & = & [\Omega _{j-1},H^{\prime }],\\
\left[ H_{0},b_{j}\right] +\Omega _{0}b_{j} & = & [b_{j-1},H^{\prime }]-\sum _{k=1}^{j}\Omega _{k}b_{j-k}
\end{array}.
\end{equation}

For concreteness and simplicity we consider dimensionless anharmonic oscillators
(\ref{ten.1}). In two recent papers Fern\'{a}ndez {[}\ref{F01a}, \ref{F01b}{]}
obtained \( \Omega _{1}(H_{0}) \) for several values of \( m \), and \( \Omega _{1}(H_{0}) \)
and \( \Omega _{2}(a,a^{\dagger }) \) for \( m=4 \). These operators enable
us to derive energy differences in the form of \( \lambda  \)-power series:
 
\begin{equation}
\label{sundar}
\Omega _{n}=\sum ^{\infty }_{j=0}\Omega _{n,j}\lambda ^{j},\, \, E_{n}=\sum ^{\infty }_{j=0}E_{n,j}\lambda ^{j}
\end{equation}
In the present case \( \Omega _{0}=1 \) so that \( \Omega _{1} \) commutes
with \( H_{0} \). For sufficiently small \( \lambda  \) we have \( b=a+\mathcal{O}(\lambda ) \)
and we may choose \( m=n+1 \) in equation (\ref{eq:Omega_n}), so that 
\begin{equation}
\label{eq:DeltaE_nj}
\Omega _{n,j}=E_{n+1,j}-E_{n,j}.
\end{equation}
 For the first order we have: 
\begin{equation}
\label{eq:Omega_n1}
\Omega _{n,1}=<n|\Omega _{1}|n>,
\end{equation}
 where \( |n\rangle  \) is an unperturbed state. The operator \( \Omega _{1} \)
is diagonal and does not contribute to the correction of second order; therefore
\begin{equation}
\label{eq:Omega_n2}
\Omega _{n,2}=<n|\Omega _{2}|n>;.
\end{equation}
 All the expressions of first and second order derived by means of the method
just outlined {[}\ref{F01a}, \ref{F01b}{]} already satisfy equations (\ref{eq:DeltaE_nj}-\ref{eq:Omega_n2}).
Fern\'{a}ndez has obtained \( \Omega _{1} \) for different \( m \) in this
approach. Here we establish a general form of \( \Omega _{1} \) valid for all
integer values of \( m \). To do so we use the fact that \( \Omega _{1} \)
is always a function of the unperturbed Hamiltonian \( H_{0} \). Now 
\begin{equation}
\label{fer1}
\begin{array}{lcl}
<n|\Omega _{1}|n> & = & E_{n+1}-E_{n}\\
 & = & \frac{1}{2^{\frac{m}{2}}m}\sum ^{(\frac{m}{2}-1)}_{r=0}t_{2r}\, ^{m}C_{2r}\, ^{m-2r}C_{\frac{m-2r}{2}}\, ^{n}C_{\frac{m-2r}{2}-1}\, (\frac{m-2r}{2})!\\
 & = & \frac{1}{2^{\frac{m}{2}}m}\sum ^{(\frac{m}{2}-1)}_{r=0}t_{2r}\, ^{m}C_{2r}\, ^{m-2r}C_{\frac{m-2r}{2}}\frac{n!}{(n-\frac{m-2r}{2}+1)!(\frac{m-2r}{2}-1)!}\, (\frac{m-2r}{2})!\\
 & = & <n|\frac{1}{2^{\frac{m}{2}}m}\sum ^{(\frac{m}{2}-1)}_{r=0}t_{2r}\, ^{m}C_{2r}\, ^{m-2r}C_{\frac{m-2r}{2}}\frac{\Gamma (a^{\dagger }a+1)}{\Gamma (a^{\dagger }a-\frac{m-2r}{2}+2)!}\, (\frac{m-2r}{2})|n>\\
 & = & <n|\frac{1}{2^{\frac{m}{2}}m}\sum ^{(\frac{m}{2}-1)}_{r=0}t_{2r}\, ^{m}C_{2r}\, ^{m-2r}C_{\frac{m-2r}{2}}\frac{\Gamma (H_{0}+\frac{1}{2})}{\Gamma (H_{0}-\frac{m-2r}{2}+\frac{3}{2})!}\, (\frac{m-2r}{2})|n>
\end{array}
\end{equation}
Therefore we have 
\begin{equation}
\label{fer2}
\Omega _{1}=\frac{1}{2^{\frac{m}{2}}m}\sum ^{(\frac{m}{2}-1)}_{r=0}t_{2r}\, ^{m}C_{2r}\, ^{m-2r}C_{\frac{m-2r}{2}}\frac{\Gamma (H_{0}+\frac{1}{2})}{\Gamma (H_{0}-\frac{m-2r}{2}+\frac{3}{2})!}\, (\frac{m-2r}{2})
\end{equation}
This coincides exactly with the first order frequency operator\index{frequency operator}
obtained by other techniques.

\section{Comparison among the different approaches}

We have seen that different approaches may be used to obtain the first order
frequency operator\index{frequency operator} and the first order position operator.
But all the results obtained in these methods are not the same. In this section
we have checked the equivalence of different methods and have established some
relationships between apparently different but equivalent frequency operators\index{frequency operator}.
To begin with, let us compare the first order cases in the next subsection and
higher order cases in the subsequent subsection.

\subsection{First order case}

In the preceding sections we have obtained the same frequency operator \index{frequency operator}
\( \Omega  \) from perturbation theory in the interaction picture, renormalization
group\index{renormalization group}, near-identity transformation, and eigen
operator techniques. In these techniques we usually write 
\begin{equation}
\label{comp1}
X(t)=A\left[ \exp (-i\Omega t)a(0)+a^{\dagger }(0)\exp (i\Omega t)\right] 
\end{equation}
where \( \Omega =\Omega (H_{0})=1+\lambda \Omega _{1}(H_{0})+\mathcal{O}(\lambda ^{2}) \)
and \( A \) is the normalization factor. However, this form of the frequency
operator\index{frequency operator} appears to be different from the one derived
from the Taylor series \index{Taylor series} method {[}\ref{Mandal}, \ref{PM01}{]}
and multiple-scale analysis {[}\ref{Bender1}, \ref{Bender2} and \ref{pathak2}{]}.
In these approaches position operator \( X(t) \) is usually expressed as
\begin{equation}
\label{eq:x(t)P}
X(t)=\cos (\omega t)X(0)+X(0)\cos (\omega t)+\sin (\omega t)\dot{X}(0)+\dot{X}(0)\sin (\omega t)+\mathcal{O}(\lambda ),
\end{equation}
 where \( \omega =\omega (H_{0})=1+\lambda \omega _{1}(H_{0})+\mathcal{O}(\lambda ^{2}) \).
Taking into account a particular form of equation (\ref{eq:bF(H)}): \( aF(H_{0})=F(H_{0}+1)a,\; F(H_{0})a^{\dagger }=a^{\dagger }F(H_{0}+1) \),
we easily rewrite equation (\ref{eq:x(t)P}) as 
\begin{eqnarray}
X(t)= &  & \sqrt{2}\cos \left( \lambda t\frac{\omega _{1}-\omega _{1}^{\prime }}{2}\right) \left[ \exp \left( -it-i\lambda t\frac{\omega _{1}+\omega _{1}^{\prime }}{2}\right) a\right. \nonumber \\
 &  & \left. +a^{\dagger }\exp \left( it+i\lambda t\frac{\omega _{1}+\omega _{1}^{\prime }}{2}\right) \right] +O(\lambda ^{2}),\label{eq:x(t)PF} 
\end{eqnarray}
 where \( \omega _{1}^{\prime }=\omega _{1}(H_{0}+1) \). Upon comparing equations
(\ref{comp1}) and (\ref{eq:x(t)PF}), and taking into account that \( \cos \left( \lambda t\frac{\omega _{1}-\omega _{1}^{\prime }}{2}\right) =1+\mathcal{O}(\lambda ^{2}) \),
we conclude that 
\begin{equation}
\label{eq:comp_Omega}
\Omega _{1}(H_{0})=\frac{\omega _{1}(H_{0})+\omega _{1}(H_{0}+1)}{2}.
\end{equation}
 The inverse transformation\index{inverse transformation} 
\begin{equation}
\label{eq:comp_Omega2}
\omega _{1}(H_{0})=\sum _{j=0}c_{j}\left. \frac{d^{j}\Omega _{1}(\xi )}{d\xi ^{j}}\right| _{\xi =H_{0}},\; c_{0}=1,\; c_{j}=\frac{2}{j!}\left. \frac{d^{j}}{d\xi ^{j}}\left( 1+e^{\xi }\right) ^{-1}\right| _{\xi =0}
\end{equation}
 also gives a suitable analytical expression because both \( \Omega _{1} \)
and \( \omega _{1} \) are polynomial functions of \( H_{0} \).

\subsection{Corrections of higher order}

The calculation of perturbation corrections of higher order and comparison of
expressions coming from different methods is considerably more difficult. We
can, however, draw some useful conclusions from available results. For example,
Aks and Carhart {[}\ref{aks2}{]} showed that the Hamiltonian operator \( H(p(t),x(t)) \),
in terms of their perturbation expressions free from secular terms for \( x(t) \)
and \( p(t) \), is diagonal and gives the correct energy through first order
perturbation theory. Unfortunately, they did not verify whether their second-order
results were consistent with this criterion.

Egusquiza and Valle Basagoiti obtained an expression of the form
\begin{equation}
\label{eq:EB_a}
\begin{array}{lcl}
a(\tau ) & = & a\exp \left\{ \frac{-3i\tau \lambda (H_{0}-\frac{1}{2})}{4}+\frac{3i\tau \lambda ^{2}\left[ 5\left( H_{0}-\frac{1}{2}\right) ^{2}-1\right] }{64}\right\} \\
 & = & \exp \left\{ \frac{-3i\tau \lambda (H_{0}+\frac{1}{2})}{4}+\frac{3i\tau \lambda ^{2}\left[ 5\left( H_{0}+\frac{1}{2}\right) ^{2}-1\right] }{64}\right\} a
\end{array}
\end{equation}
 by means of renormalization {[}\ref{Egus}{]}. The second-order part of their
frequency operator\index{frequency operator}
\begin{equation}
\label{eq:EB_Omega_2}
\Omega _{2}^{EB}=-\frac{3}{64}\left( 5H_{0}^{2}+5H_{0}+\frac{1}{4}\right) .
\end{equation}
 yields 
\begin{equation}
\label{DeltaE_n2B}
\langle n-1|\Omega _{2}^{EB}|n-1\rangle =-\frac{3}{64}(5n^{2}-1)
\end{equation}
 which does not agree with the well-known result 
\begin{equation}
\label{eq:DeltaE_n2_x4}
E_{n,2}-E_{n-1,2}=-\frac{3}{64}(17n^{2}+7)
\end{equation}
 that one easily obtains by means of Rayleigh-Schr\"{o}dinger perturbation
theory {[}\ref{F01a}{]}.

By means of the method outlined in Section 2.8 Fern\'{a}ndez obtained\cite{F01b}
\begin{equation}
\label{eq:Omega2_x4}
\begin{array}{lcl}
\Omega _{2} & = & \frac{3}{16}\left( \frac{1}{4}a^{\dagger 4}+a^{\dagger 3}a+\frac{3}{2}a^{\dagger 2}+a^{\dagger }a^{3}+\frac{1}{4}a^{4}+\frac{3}{2}a^{2}\right) \\
 & - & \frac{3}{64}\left( 17a^{\dagger 2}a^{2}+51a^{\dagger }a+24\right) \\
 & = & -\frac{1}{64}\left( 69H_{0}^{2}+51H_{0}-12X^{4}+\frac{153}{4}\right) 
\end{array}
\end{equation}
 which is consistent with equations (\ref{eq:Omega_n2}) and (\ref{eq:DeltaE_n2_x4}).
This expression shows off-diagonal terms and agrees with the perturbation expansion
of Speliotopoulos's frequency operator\index{frequency operator} equation {[}\ref{S00}{]}
when \( \lambda =4\epsilon  \). There seems to be a conceptual oversight in
Egusquiza and Valle Basagoiti's correction of second order to the frequency
operator\index{frequency operator} because it does not show offdiagonal terms
and does not give the correct energy difference {[}\ref{Egus}{]}. To be more
precise, the second order frequency operator\index{frequency operator} obtained
by them differs from all other existing results {[}see equation 2.10 of ref.
\ref{Egus}{]} and the discrepancy is not due to a difference in operator ordering.
This fact may be established from their expression for the energy difference:
\begin{equation}
\label{egu1}
E_{n}-E_{n-1}=\left( 1+\frac{3\lambda }{4}n-\frac{3\lambda ^{2}}{64}(5n^{2}-1)+O(\lambda ^{3})\right) .
\end{equation}
They obtained (\ref{egu1}) as a consequence of equation 2.10 of ref. {[}\ref{Egus}{]}.
Equation (\ref{egu1}) is equation 2.11 of ref. {[}\ref{Egus}{]} and it differs
from the energy difference obtained by simple perturbation calculation which
is 
\begin{equation}
\label{egu2}
E_{n}-E_{n-1}=\left( 1+\frac{3\lambda }{4}n-\frac{\lambda ^{2}}{64}(51n^{2}+21)+O(\lambda ^{3})\right) .
\end{equation}
 Possibly the relevant secular term chosen by them (see equation 2.8 of {[}\ref{Egus}{]})
is incomplete and this causes the discrepancy in the expression for frequency
operator\index{frequency operator} obtained by them and others.

By means of a completely different approach we obtain \cite{PM01}
\begin{equation}
\label{eq:omega2_x4}
\omega _{2}=-\frac{1}{64}\left( 69H_{0}^{2}-12x^{4}+\frac{51}{4}\right) 
\end{equation}
 which is consistent with their recipe for the second-order energy difference
{[}\ref{pathak2}{]}
\begin{equation}
\label{eq:DeltaE_n2_x4_P}
2\left( E_{n,2}-E_{n-1,2}\right) =\langle n|\omega _{2}|n\rangle +\langle n-1|\omega _{2}|n-1\rangle .
\end{equation}
 Equations (\ref{eq:DeltaE_n2_x4}) and (\ref{eq:DeltaE_n2_x4_P}) suggest that
the diagonal parts of the operators (\ref{eq:Omega2_x4}) and (\ref{eq:omega2_x4})
are related by 
\begin{equation}
\label{eq:comp_Omega2d}
\Omega _{2}^{d}(H_{0})=\frac{\omega _{2}^{d}(H_{0})+\omega _{2}^{d}(H_{0}+1)}{2},
\end{equation}
 as one easily verifies by means of the operators themselves.

Aks and Carhart {[}\ref{aks2}{]} forced the frequency operator \index{frequency operator}
to be diagonal at all perturbation orders. They proved that their results give
the Rayleigh-Schr\"{o}dinger perturbation series of first order where we know
that the frequency operator\index{frequency operator} is diagonal and correctly
given by all the approaches. We have seen that from second order onwards the
frequency operator\index{frequency operator} is a polynomial function of \( H \),
and, therefore, it should exhibit off-diagonal terms for the unperturbed states.
It is not clear to us if the results of Aks and Carhart {[}\ref{aks2}{]} are
correct at second order because they did not give the corresponding expression
of the coordinate explicitly. Moreover, it is difficult to compare the results
of Aks and Carhart with others because the way they split the frequency operator\index{frequency operator}
into the different corrections to the coordinate operator differs markedly from
the other techniques discussed here.

Here we can also check that if we put \( H=H_{0}+\frac{\lambda }{4}x^{4} \)
in the expression of \( \lambda (H) \) obtained by Speliotopoloulos {[}\ref{S00}{]}
and take terms up to the second order in \( \lambda  \) (where \( \lambda =\frac{\epsilon _{1}}{\epsilon _{0}} \))
then we obtain the second order frequency operator\index{frequency operator}
obtained by Fern\'{a}ndez {[}\ref{F01b}{]}. So the results of Fern\'{a}ndez
and Speliotopoloulos are in exact accordance with each other. Actually according
to the work of Speliotopoloulos frequency operator \index{frequency operator}
is a function of the total Hamiltonian \( H=H_{0}+\lambda V \). Now if we express
the frequency operator in power series of \( \lambda  \) as \( \Omega =\Omega _{0}+\lambda \Omega _{1}(H_{0}+\lambda V)+\lambda ^{2}\Omega _{2}(H_{0}+\lambda V)+.... \)
then we observe that the first order correction to the frequency operator (\( \Omega _{1} \))
should be a function of \( H_{0} \) only since even the lowest power of \( V \)
will give terms of the order of \( \lambda ^{2} \) . So it is expected that
the second order frequency operator \index{frequency operator} will be a function
of \( H_{0} \) and \( V \) but no cross term will be present but third and
higher order frequency operator\index{frequency operator} will contain cross
terms like \( H^{p}_{0}V \). This is really in accordance with the available
results and we can recall that the second order frequency operator\index{frequency operator}
for the quartic\index{quartic} oscillator obtained by us is 
\[
\Omega _{2}=-\frac{1}{64}(69H^{2}_{0}-12x^{4}+\frac{51}{4}).\]
 This matches with the argument given above.

Thus we conclude that apparent differences among perturbation corrections of
first and second orders to the frequency operator\index{frequency operator}
derived by different alternative approaches are simply due to different arrangements
and ordering of products of noncommuting quantum-mechanical observable. In some
cases it is not difficult to obtain the relationships between the different
forms of the frequency operator\index{frequency operator} as shown above. Therefore
we can use any of these solutions to investigate the quantum fluctuations of
coherent light in nonlinear media.

\chapter{Phase fluctuations \index{quantum phase fluctuations}of coherent light coupled
to a nonlinear medium of inversion symmetry}

The quantum phase\index{quantum phase} problem may be stated as: How can one
write down a quantum mechanical operator corresponding to the phase of a harmonic
oscillator or equivalently a single mode of electromagnetic field? The concept
of phase plays a very important role in the understanding of basic physics.
So people are trying to answer this question from different points of view and
are debating on this issue since Dirac {[}\ref{Dirac}{]} initiated the search
of a quantum mechanical operator corresponding to the phase of a harmonic oscillator
in 1927. Due to this seventy-five years long search and healthy debate an extensive
amount of literature on this topic is now available but a satisfactory Hermitian
phase operator\index{phase operator} is still unavailable. However, the observable
nature of the phase demands that the corresponding operator should be a Hermitian
one. There are many good reviews on this topic {[}\ref{Neito}-\ref{Dubin}{]}.
Some of them are written in a lucid manner {[}\ref{Neito},\ref{Barnet}{]}
and some are written from much more mathematical and technical points of view
{[}\ref{Dubin}{]}. 

A complete description of a harmonic oscillator is possible from the idea of
a phase. Again the interference and diffraction pattern of electromagnetic waves
can be explained convincingly with the help of a phase concept. Thus the phase
in classical physics is well understood for a long time. We do not have such
resemblance of phases in quantum mechanics. But it is widely believed that the
quantum phase\index{quantum phase} is one of the distinguishing features between
quantum and classical physics. Actually quantum phase\index{quantum phase}
is responsible for all quantum mechanical interference phenomena. Even the discrete
eigenvalues in quantum theory can be viewed as a quantum phase condition for
the Schr\"{o}dinger equation {[}\ref{Neito}{]}. The examples of quantum phase
are from double slit experiment {[}\ref{Ggoswami}{]} to the dark resonances
{[}\ref{Scully M O}{]} and from Aharonov-Bohm {[}\ref{Gasiorowicz}{]} effect
to Bose-Einstein condensate {[}\ref{Javanainen}{]}.

Quantum mechanical phases can be classified into two main parts depending upon
their origin. The phase having geometrical origin is quite familiar to us and
is called geometric quantum phase \index{quantum phase} (e.g Berry's phase\index{Berry's phase}
{[}\ref{Berry}{]}, Aharonov Anandan phase {[}\ref{Aharonov}{]}). The other
type of phase arises due to the dynamical nature of the system. Geometric phase
is discussed in chapter 6. In this chapter we briefly review the quantum phase
problem \index{quantum phase} and study the quantum phase fluctuations\index{quantum phase fluctuations}
of coherent light coupled to a third order nonlinear medium of inversion symmetry.

\section{The quantum phase\index{quantum phase}: Dirac approach}

Dirac started his work with the assumption that the annihilation operator \( a \)
can be factored out into a Hermitian function \( f(N) \) of the number operator
\( N \) and a unitary operator \( U \) {[}\ref{Dirac}{]}. The later one (\( i.e.U \))
defines the Hermitian phase operator\index{phase operator} \( \phi  \) in
the following way

\begin{equation}
\label{ek.tin}
U=exp(i\phi ).
\end{equation}
 Hence in the Dirac formalism 
\begin{equation}
\label{ani}
a=exp(i\phi )f(N).
\end{equation}
 Now the functional form of \( f(N) \) may easily be obtained by using

\begin{equation}
\label{ek.char}
N=a^{\dagger }a=f^{2}.
\end{equation}
 Thus the explicit expression for \( a \) is given by 
\begin{equation}
\label{five}
a=exp(i\phi )N^{\frac{1}{2}}
\end{equation}
 where equations (\ref{ani}) and (\ref{ek.char}) have been used. The expression
(\ref{five}) looks like the polar decomposition of a complex number \( (a). \)
The usual commutation relation between \( a \) and \( a^{\dagger } \) is 
\begin{equation}
\label{sixqph}
aa^{\dagger }-a^{\dagger }a=1.
\end{equation}
 From equations (\ref{five}) and (\ref{sixqph}) we have 
\begin{equation}
\label{7a}
exp(i\phi )Nexp(-i\phi )-N=1
\end{equation}
 or, 
\begin{equation}
\label{7qph}
exp(i\phi )N-Nexp(i\phi )=exp(i\phi ).
\end{equation}
 Equation (\ref{7qph}) is valid only if the following commutation relation
holds good

\begin{equation}
\label{8qph}
[N,\phi ]=i.
\end{equation}
 The justification of the above condition (\ref{8qph}) may be simply given
in the following way. If equation (\ref{8qph}) is true then from the method
of induction we have 
\begin{equation}
\label{9qph}
[N,\phi ^{n}]=in\phi ^{n-1}.
\end{equation}
 Multiplying both sides of equation (\ref{9qph}) by \( \frac{i^{n}}{n!} \)
and summing from \( 0 \) to \( \infty  \) we arrive at equation (\ref{7qph}).
Now the uncertainty relation corresponding to equation (\ref{8qph}) is given
by

\begin{equation}
\label{10qph}
\triangle N\triangle \phi \geq \frac{1}{2}.
\end{equation}
 The equations (\ref{8qph}) and (\ref{10qph}) have many problems. Let us see
them one by one.

\subsection{Problems with the Dirac approach}

\textbf{Problem-1}: The uncertainty relation (\ref{10qph}) allows the uncertainty
in \( \phi  \) (i.e \( \Delta  \)\( \phi  \)) greater than \( 2\pi  \) for
\( \triangle N<\frac{1}{4\pi } \). This is impossible and is inconsistent with
the sense of a phase. This problem is known as the periodicity problem. In this
context we can recall that the uncertainty relation for \( L_{z} \) and \( \phi  \)
also has the same problem {[}\ref{ljudge}-\ref{L.Judge}{]}. \textbf{}\\
\textbf{Problem- 2:} The Hermitian nature of \( \phi  \) (as it is expected
in quantum mechanics) demands that the operator \( U \) should be an unitary
one. However \( U \) is not an unitary operator. \textbf{}\\
\textbf{Problem 3}: In 1963 Louisell {[}\ref{louisell}{]} showed that equation
(\ref{8qph}) leads to confusing results when one attempts to evaluate the matrix
element of \( \phi  \) in a representation in which \( N \) is diagonal. Since,
\[
<l|N\phi -\phi N|n>=i<l|n>\]
 so we have 
\begin{equation}
\label{qph18}
<l|\phi |n>=-i\frac{\delta _{n,l}}{(n-l)}
\end{equation}
 where \( N|n>=n|n> \) has been used. If the eigenvalues of \( N \) are integer
then the right hand side of equation (\ref{qph18}) becomes undefined. Only
in the limit of large values of \( n \) and \( l \) (Correspondence limit)
\( \phi  \) takes on a definite meaning. Thus \( \phi  \) has a definite meaning
only in the classical limit. This is quite consistent with the classical idea
of phases.

\subsection{\textmd{Why does the Hermitian phase operator}\index{phase operator} \textmd{not
exist?}}

In the previous section we have seen that the Hermitian phase operator\index{phase operator}
does not exist. But a question remained unanswered: Why does a Hermitian phase
operator\index{phase operator} not exist? J. Sarfatt, one of the pioneer in
quantum phase\index{quantum phase} problem, addressed this question for the
first time. He wrote a nice paper in connection with the complexity of the introduction
of quantum phase {[}\ref{Saraftt}{]}. In Sarfatt words {[}\ref{Saraftt}{]}'...Recent
work by L. Susskind, J. Glogower and J. Sarafatt shows that it is impossible
to define a phase operator because of the existence of a lowest state for the
number operator of the oscillator. Thus the uncertainty relation \( \triangle N\triangle \phi \geq 1 \)
is meaningless....'

The main cause for the nonunitarity of \( U \) is that the number state matrix
representation of the diagonal number operator is bound from below. So a bare
phase operator does not exist. This observation provoked people to introduce
geometrical functions of phase operators\index{phase operator}.

\section{\textmd{Periodic function of phase: Louisell's approach }}

Louisell {[}\ref{louisell}{]} proposed a self consistent way to avoid these
difficulties. Actually we can write equation (\ref{9qph}) in an alternative
form as 
\begin{equation}
\label{9.1qph}
[N,\phi ^{n}]=in\phi ^{n-1}=i\frac{d\phi ^{n}}{d\phi }.
\end{equation}
 Therefore for any polynomial function \( P(\phi ) \) of \( \varphi  \) we
 have a commutation relation

\begin{equation}
\label{qph19}
[N,P(\phi )]=i\frac{dP(\phi )}{d\phi }.
\end{equation}
 Louisell started his work with the fundamental commutation relation (\ref{qph19})
and supposing that \( P(\phi ) \) is a periodic function of period \( 2\pi . \)
He considered \( \sin (\phi ) \) and \( \cos (\phi ) \) as the Hermitian operators
which satisfy the following commutation relations,

\begin{equation}
\label{20qph}
[N,\cos (\phi )]=-i\sin (\phi )
\end{equation}

and 
\begin{equation}
\label{21qph}
[N,\cos (\phi )]=i\sin (\phi ).
\end{equation}
 Therefore, the uncertainty relations take the following forms

\begin{equation}
\label{22qph}
\triangle N\triangle \cos (\phi )\geq \frac{1}{2}|<\sin (\phi )>|
\end{equation}

\begin{equation}
\label{23qph}
\triangle N\triangle \sin (\phi )\geq \frac{1}{2}|<\cos (\phi )>|.
\end{equation}
 It is to be noted that the above uncertainty relations (\ref{22qph} and \ref{23qph})
are free from problem 1. Instead of a bare phase operator\index{phase operator},
Louisell defined a meaningful phase operator\index{phase operator} in terms
of the periodic function of phase. But the following questions remain unanswered:

1) What are the explicit forms of sine and cosine operators?

2) Are they Hermitian?

\section{\textmd{sine and cosine operators: Susskind and Glogower approach}}

The questions asked in the previous subsection were properly addressed by Susskind
and Glogower (SG) {[}\ref{S  G}{]}. They reaffirmed that London's result {[}\ref{London1},
\ref{London2}{]} (i.e \( U \) is not unitary) are indeed correct and obtained
an explicit form of sine operator \( S \) and cosine operator \( C \) as 
\begin{equation}
\label{37qph}
S=\frac{1}{2i}(E_{-}-E_{+})=\frac{1}{2i}\left[ \frac{1}{(N+1)^{\frac{1}{2}}}a-a^{\dagger }\frac{1}{(N+1)^{\frac{1}{2}}}\right] 
\end{equation}
 and 
\begin{equation}
\label{36qph}
C=\frac{1}{2}(E_{-}+E_{+})=\frac{1}{2}\left[ \frac{1}{(N+1)^{\frac{1}{2}}}a+a^{\dagger }\frac{1}{(N+1)^{\frac{1}{2}}}\right] .
\end{equation}
 It is easy to check that the operators \( S \) (\ref{37qph}) and \( C \)
(\ref{36qph}) are Hermitian. Therefore, SG approach is fully capable of giving
the answers to the questions remained unsolved after Louisell's work. Now the
commutator of \( C \) and \( S \) may be calculated by using (\ref{36qph}
and \ref{37qph}) and is given by

\begin{equation}
\label{38qph}
[C,S]=\frac{i}{2}P^{0},
\end{equation}
 where \( P^{0} \) is the projection onto the ground state. The noncommuting
nature of \( C \) and \( S \) (\ref{38qph}) is one of the serious drawbacks
of SG approach. This is because the cosine and sine functions of a real phase
should commute. Moreover, the sum of the squares of the \( S \) and \( C \)
operators is given by 
\begin{equation}
\label{qph39}
C^{2}+S^{2}=1-\frac{1}{2}P^{0}.
\end{equation}

Therefore, in SG approach we got satisfactory answers to the previous questions.
However, we ended up with two new problems which essentially go against our
perception.

\section{\textmd{Measured phase operators}\index{phase operator}\textmd{: Pegg and
Barnett approach}}

In 1986 Pegg and Barnett {[}\ref{PB J.PHYS}{]} proposed a Hermitian phase operator
in an extended Hilbert space\index{Hilbert space} containing negative number
states. Such a proposal seems to be unphysical. So very soon they gave up this
proposal and started their new venture with a new proposal. They proposed that
the infinite dimensional harmonic oscillator Hilbert space\index{Hilbert space}
should be truncated to \( (s+1) \) dimensions and expectation values of the
quantum mechanical operators should be taken in this truncated Hilbert space\index{truncated Hilbert space}
only. After doing that one should take the limit \( s\rightarrow \infty  \).
This prescription of working in a truncated Hilbert space\index{truncated Hilbert space}
was not new in Physics. Popov-Yarunin, Sinha and many others already did some
work on the finite dimensional Hilbert space\index{finite dimensional Hilbert space}.
But the Pegg-Barnett prescription made the problem much more user friendly and
it also defined a Hermitian phase operator\index{phase operator} in the truncated
Hilbert space\index{truncated Hilbert space}. So this formalism got a lot of
attention of the people.

Pegg and Barnett started their work by defining a state having a well defined
phase {[}\ref{PB phys rev a}-\ref{PB europhys letters}{]} as  
\begin{equation}
\label{pbth}
|\theta >=lim_{s\rightarrow \infty }(s+1)^{-\frac{1}{2}}\sum _{n=0}^{s}\exp (in\theta )|n>
\end{equation}
 where \( |n> \) is the number state spanning the \( (s+1) \)-dimensional
space. In this space they selected a reference phase state

\begin{equation}
\label{44qph}
|\theta _{0}>=(s+1)^{-\frac{1}{2}}\sum ^{s}_{n=0}\exp (in\theta _{0})|n>
\end{equation}
 and found the subset of states \( |\theta _{m}> \), defined by replacing \( \theta _{0} \)
by \( \theta _{m} \) in equation (\ref{44qph}), which are orthogonal to this
state. Thus we have

\begin{equation}
\label{qph45}
<\theta _{m}|\theta _{0}>=(s+1)^{-1}\sum _{n=0}^{s}\exp (in(\theta _{0}-\theta _{m})).
\end{equation}
 One can easily prove that the phase states \( |\theta _{0}> \) and \( |\theta _{m}> \)
are orthogonal to each other if the following relation holds
\begin{equation}
\label{qph46}
\theta _{m}=\theta _{0}+\frac{2m\pi }{s+1}\, (m=0,1,....s).
\end{equation}
 The state (\ref{44qph}) is an over-complete one and \( |\theta _{m}> \) forms
a complete set of orthogonal basis vectors spanning the state space. Therefore,
the number state \( |n> \) can be expanded in terms of \( |\theta _{m}> \)
as follows 
\begin{equation}
\label{47qph}
|n>=\sum ^{s}_{m=0}|\theta _{m}><\theta _{m}|n>=(s+1)^{-\frac{1}{2}}\sum ^{s}_{m=0}\exp (-in\theta _{m})|\theta _{m}>.
\end{equation}
 In this approach phase and number basis states are so related that a system
in a number state is equally likely to be in any state \( |\theta _{m}> \)
and a system prepared in a phase state is equally likely to be found in any
number state {[}\ref{PB europhys letters}{]}. Now the problem is to find a
unitary operator \( \widehat{exp}_{\theta }(i\phi ) \) whose eigenstates are
the phase states \( |\theta _{m}> \):

\begin{equation}
\label{48qph}
\widehat{\exp _{\theta }}(i\phi )|\theta _{m}>=\exp (i\theta _{m})|\theta _{m}>
\end{equation}

\begin{equation}
\label{49qph}
\widehat{\exp _{\theta }}(-i\phi )|\theta _{m}>=\exp (-i\theta _{m})|\theta _{m}>.
\end{equation}
 Here, the caret indicates that the whole expression is an operator. From equation
(\ref{47qph} and \ref{48qph}) we have

\begin{equation}
\label{50qph}
\widehat{\exp _{\theta }}(i\phi )|n>=(s+1)^{-\frac{1}{2}}\sum ^{s}_{m=0}\exp [-i(n-1)\theta _{m}]|\theta _{m}>=|n-1>.
\end{equation}
 For the vacuum state (\( n=0 \)) the resulting state is 
\begin{equation}
\label{51qph}
\begin{array}{c}
(s+1)^{-\frac{1}{2}}\sum _{m}\exp (i\theta _{m})|\theta _{m}>=(s+1)^{-\frac{1}{2}}\exp [i(s+1)\theta _{0}]\sum _{m}\exp (-is\theta _{m})|\theta _{m}>\\
=\exp [i(s+1)\theta _{0}]|s>
\end{array}.
\end{equation}
 Now from equations (\ref{50qph} and \ref{51qph}) one can see that the number
state representation of \( \widehat{\exp _{\theta }}(i\phi ) \) is

\begin{equation}
\label{52qph}
\widehat{\exp _{\theta }}(i\phi )=|0><1|+|1><2|+.....+|s-1><s|+\exp [i(s+1)\theta _{0}]|s><0|
\end{equation}
 and \( \widehat{\exp _{\theta }}(-i\phi ) \) is just the Hermitian conjugate
to it. Thus we have unitary operator \( \widehat{\exp _{\theta }}(i\phi ) \)
and we can define cosine and sine operators in terms of that. Those operators
have properties that are much more desirable and familiar for the description
of phase. In particular in Pegg-Barnett (PB) formalism we find 
\begin{equation}
\label{54qph}
[\widehat{\cos _{\theta }}\phi ]^{2}+[\widehat{\sin _{\theta }}\phi ]^{2}=1
\end{equation}

\begin{equation}
\label{53qph}
[\widehat{\cos _{\theta }}\phi ,\widehat{\sin _{\theta }}\phi ]=0
\end{equation}

\begin{equation}
\label{55qph}
<n|[\widehat{\cos _{\theta }}\phi ]^{2}|n>=<n|[\widehat{\sin _{\theta }}\phi ]^{2}|n>=\frac{1}{2},\forall n,
\end{equation}
 where 
\begin{equation}
\label{56qph}
\widehat{\cos _{\theta }}\phi =\frac{1}{2}[\widehat{\exp _{\theta }}(i\phi )+\widehat{\exp _{\theta }}(-i\phi )]
\end{equation}
 and 
\begin{equation}
\label{57qph}
\widehat{\sin _{\theta }}\phi =\frac{1}{2i}[\widehat{\exp _{\theta }}(i\phi )-\widehat{\exp _{\theta }}(-i\phi )].
\end{equation}
 One of the main differences between SG formalism and PB formalism is that the
action of the \( exp(i\phi ) \) operator on the vacuum gives different results.
This is basically responsible for all the differences in operator relations
we have obtained in two formalisms. From equation (\ref{55qph}) it is clear
that in PB formalism the definition of phase operator\index{phase operator}
is consistent with the fact that the phase of the vacuum is random but the case
is not so in the S -G formalism.

Though in Pegg-Barnett formalism \( \phi  \) has the desirable properties of
phase but still this approach is not free from objections. In this approach
one measures the physical quantities at first in an arbitrarily large but finite
state space of \( s+1 \) dimension and then takes the limit as \( s \) tends
to \( \infty  \). These two operation do not commute so their applicability
is questionable {[}\ref{Gantstog}{]}. But Pegg and Barnett demand that their
approach is physically indistinguishable with the conventional infinite state
space model {[}\ref{PB phys rev a}{]}. Correctness of the results obtained
in different approaches may be compared only by experiments. Interesting experiments
on quantum phase\index{quantum phase} have started coming in {[}\ref{nfm,prev lett.,1991}{]}. 

People, however, use both Susskind-Glogower (SG) {[}\ref{Gerry},\ref{Fan}{]}
and Pegg- Barnett (PB) {[}\ref{Lynch}, \ref{Lynch1}{]} formalisms for the
studies of phase properties and the phase fluctuations of various physical systems.
For example, the phase fluctuations of coherent light interacting with a nonlinear
nonabsorbing medium of inversion symmetry are reported in the recent past {[}\ref{Gerry},\ref{Lynch}{]}.
It is found that both SG {[}\ref{Gerry}{]} and PB {[}\ref{Lynch}{]} formalisms
lead to same type (qualitatively) of phase fluctuations. Keeping these facts
in mind, we have chosen PB approach for the present investigation.

In chapter 1 we have seen that the interaction of a single mode of electromagnetic
field with a third order nonlinear medium can be described by the Hamiltonian
(\ref{ekpf}). \emph{}The equation of motion corresponding to the Hamiltonian
(\ref{ekpf}) is given by equation (\ref{dui}) which contains cubic nonlinearity
in field operator \( X \). The exact analytical solution of the equation (\ref{dui})
is not available. However, under RWA\index{RWA} the model~Hamiltonian reduces
to an exactly solvable form (\ref{charpf}). Gerry {[}\ref{Gerry}{]} and Lynch
{[}\ref{Lynch}{]} used this RW approximated Hamiltonian (\ref{charpf}) to
study the quantum phase fluctuations\index{quantum phase fluctuations}. On
the other hand, different techniques to obtain approximate solution of (\ref{dui})
is discussed in chapter2. Here we use first order operator solution of (\ref{dui})
obtained in chapter 2 to study the quantum phase fluctuations \index{quantum phase fluctuations}of
coherent light in third order nonlinear medium.

\section{Time evolution of the useful operators}

The purpose of the present section is to calculate the phase fluctuations\index{phase fluctuations}
of coherent light interacting with the nonlinear medium of inversion symmetry.
The corresponding Hamiltonian is given by equation (\ref{ekpf}) and the fluctuations
are measured with respect to the initial coherent state\index{coherent state}
\( |\alpha > \) which is defined as the right eigenket of the annihilation
operator \( a(0) \) corresponding to the eigenvalue equation \( a(0)|\alpha >=\alpha |\alpha > \)
{[}\ref{Louisell}{]}. The eigenvalue\index{eigenvalue} \( \alpha  \) is in
general complex and may be written as \( \alpha =|\alpha |exp(i\theta ) \),
where \( \theta  \) is the phase angle of \( \alpha  \). The modulus square
of \( \alpha  \) \( (i.e|\alpha |^{2}) \) gives the number of photons present
in the field prior to the interaction. From chapter 2 we can write the time
evolution of the annihilation operator \( a(t) \) as 
\begin{equation}
\label{choipf}
\begin{array}{lcl}
a(t) & = & X(t)+i\dot{X}(t)\\
 & = & D_{1}a(0)+D_{2}a^{\dagger }(0)\\
 & - & \left[ D_{3}a^{3}(0)+D_{4}a^{\dagger 3}(0)+D_{5}a^{\dagger 2}(0)a(0)+D_{6}a^{\dagger }(0)a^{2}(0)\right] 
\end{array}
\end{equation}
 where the parameters 
\begin{equation}
\label{satpf}
\begin{array}{lcl}
D_{1} & = & \left( 1-i\frac{3\lambda }{4}t\right) \exp (-it)\\
D_{2} & = & -i\frac{3\lambda }{4}\sin t\\
D_{3} & = & i\frac{\lambda }{4}\sin t\exp (-2it)\\
D_{4} & = & i\frac{\lambda }{8}\sin 2t\exp (it)\\
D_{5} & = & i\frac{3\lambda }{4}\sin t\\
and &  & \\
D_{6} & = & i\frac{3\lambda }{4}t\exp (-it).
\end{array}
\end{equation}
 The time evolution of the creation operator\index{creation operator} \( a^{\dagger }(t) \)
is the Hermitian conjugate of (\ref{choipf}). Thus the number operator is 
\begin{equation}
\label{noipf}
\begin{array}{lccl}
N(t) & = & a^{\dagger }(t)a(t) & =a^{\dagger }(0)a(0)+\left( D^{*}_{1}D_{2}a^{\dagger 2}(0)+h.c\right) \\
 &  &  & -\left( D^{*}_{1}D_{3}a^{\dagger }(0)a^{3}(0)+h.c\right) -\left( D^{*}_{1}D_{4}a^{\dagger 4}(0)+h.c\right) \\
 &  &  & -\left( D^{*}_{1}D_{5}a^{\dagger 3}(0)a(0)+h.c\right) -\left( D^{*}_{1}D_{6}a^{\dagger 2}(0)a^{2}(0)+h.c\right) ,
\end{array}
\end{equation}
 where \( h.c \) stands for the Hermitian conjugate. The parameters \( D^{*}_{1} \),
\( D^{*}_{2} \), \( D^{*}_{3} \), \( D^{*}_{4} \), \( D^{*}_{5} \) and \( D^{*}_{6} \)
are the complex conjugate of \( D_{1} \), \( D_{2} \), \( D_{3} \), \( D_{4} \),
\( D_{5} \) and \( D_{6} \) respectively. The equation (\ref{noipf}) is used
to obtain

\begin{equation}
\label{dashpf}
\begin{array}{lcl}
N^{2}(t) & = & \left( a^{\dagger 2}(0)a^{2}(0)+a^{\dagger }(0)a(0)\right) +\left[ 2D^{*}_{1}D_{2}\left( a^{\dagger 3}(0)a(0)+a^{\dagger 2}(0)\right) +h.c\right] \\
 & - & \left[ 2D^{*}_{1}D_{3}\left( a^{\dagger 2}(0)a^{4}(0)+2a^{\dagger }(0)a^{3}(0)\right) +h.c\right] \\
 & - & \left[ 2D^{*}_{1}D_{4}\left( a^{\dagger 5}(0)a(0)+2a^{\dagger 4}(0)\right) +h.c\right] \\
 & - & \left[ 2D^{*}_{1}D_{5}\left( a^{\dagger 4}(0)a^{2}(0)+2a^{\dagger 3}(0)a(0)\right) +h.c\right] \\
 & - & \left[ 2D^{*}_{1}D_{6}\left( a^{\dagger 3}(0)a^{3}(0)+2a^{\dagger 2}(0)a^{2}(0)\right) +h.c\right] .
\end{array}
\end{equation}
 The terms beyond the linear power of \( \lambda  \) are neglected in equation
(\ref{dashpf}).

The exponential of phase operator\index{phase operator} \( E \) and its Hermitian
conjugate \( E^{\dagger } \) under the PB formalism are given by {[}\ref{Lynch}{]}
\begin{equation}
\label{taropf}
\begin{array}{lcl}
E & = & \left( \overline{N}+\frac{1}{2}\right) ^{-1/2}a(t)\\
E^{\dagger } & = & \left( \overline{N}+\frac{1}{2}\right) ^{-1/2}a^{\dagger }(t)
\end{array}
\end{equation}
 where \( \overline{N} \) is the average number of photons present in the radiation
field after interaction. Using (\ref{noipf}) \( \bar{N} \) can be written
as

\begin{equation}
\label{chabbispf}
\overline{N}=|\alpha |^{2}\left[ 1+\frac{\lambda }{4}\left( 2(3+2|\alpha |^{2})\sin t\sin (t-2\theta )+|\alpha |^{2}\sin 2t\sin 2(t-2\theta )\right) \right] .
\end{equation}
 Interestingly, \( \overline{N} \) depends on the coupling constant \( \lambda  \),
phase angle \( \theta  \) and on the free evolution time \( t. \) Thus the
number of photons are not conserved. The result is not surprising since the
nonconserving energy terms are included in the model Hamiltonian (\ref{ekpf}).
However, in earlier studies {[}\ref{Gerry}-\ref{Lynch}{]} the photon numbers
were conserved. 

The usual cosine and sine of the phase operator\index{phase operator} are defined
in the following way 
\begin{equation}
\label{chauddopf}
\begin{array}{lcl}
C & = & \frac{1}{2}\left( E+E^{\dagger }\right) \\
S & = & -\frac{i}{2}\left( E-E^{\dagger }\right) .
\end{array}
\end{equation}
 The expectation values of the operators \( C \) and \( S \) are given by
\begin{equation}
\label{paneropf}
\begin{array}{lcl}
<C> & = & \frac{1}{2}\left( \overline{N}+\frac{1}{2}\right) ^{-1/2}\left[ (D_{1}+D^{*}_{2})\alpha \right. +(D^{*}_{1}+D_{2})\alpha ^{*}-(D_{3}+D^{*}_{4})\alpha ^{3}\\
 &  & -(D^{*}_{3}+D_{4})\alpha ^{*3}-(D_{5}+D^{*}_{6})|\alpha |^{2}\alpha ^{*}-(D^{*}_{5}+D_{6})\left. |\alpha |^{2}\alpha \right] \\
<S> & = & -\frac{i}{2}\left( \overline{N}+\frac{1}{2}\right) ^{-1/2}\left[ (D_{1}-D^{*}_{2})\alpha \right. -(D^{*}_{1}-D_{2})\alpha ^{*}-(D_{3}-D^{*}_{4})\alpha ^{3}\\
 &  & +(D^{*}_{3}-D_{4})\alpha ^{*3}-(D_{5}-D^{*}_{6})|\alpha |^{2}\alpha ^{*}+(D^{*}_{5}-D_{6})\left. |\alpha |^{2}\alpha \right] 
\end{array}
\end{equation}
 where the equations (\ref{choipf}-\ref{noipf}), and the equations (\ref{taropf}-\ref{chauddopf})
are used. Again, the square of the averages are 
\begin{equation}
\label{solopf}
\begin{array}{lcl}
<C>^{2} & = & \frac{1}{4}\left( \overline{N}+\frac{1}{2}\right) ^{-1}\left[ \left\{ (D^{2}_{1}+2D_{1}D^{*}_{2})\alpha ^{2}+c.c\right\} +2\left\{ |D_{1}|^{2}+(D_{1}D_{2}+c.c)\right\} |\alpha |^{2}\right. \\
 &  & -2\left( \left\{ (D_{1}D_{3}+D_{1}D^{*}_{4})\alpha ^{4}+c.c\right\} +\left\{ (D_{1}D^{*}_{3}+D_{1}D_{4})|\alpha |^{2}\alpha ^{*2}+c.c\right\} \right. \\
 &  & \left. \left. +\left\{ (D_{1}D_{5}+D_{1}D^{*}_{6})+c.c\right\} |\alpha |^{4}+\left\{ (D_{1}D^{*}_{5}+D_{1}D_{6})|\alpha |^{2}\alpha ^{2}+c.c\right\} \right) \right] \\
 &  & 
\end{array}
\end{equation}

\begin{equation}
\label{sateropf}
\begin{array}{lcl}
<S>^{2} & = & -\frac{1}{4}\left( \overline{N}+\frac{1}{2}\right) ^{-1}\left[ \left\{ (D^{2}_{1}-2D_{1}D^{*}_{2})\alpha ^{2}+c.c\right\} -2\left\{ |D_{1}|^{2}-(D_{1}D_{2}+c.c)\right\} |\alpha |^{2}\right. \\
 &  & -2\left( \left\{ (D_{1}D_{3}-D_{1}D^{*}_{4})\alpha ^{4}+c.c\right\} -\left\{ (D_{1}D^{*}_{3}-D_{1}D_{4})|\alpha |^{2}\alpha ^{*2}+c.c\right\} \right. \\
 &  & \left. \left. +\left\{ (D_{1}D_{5}-D_{1}D^{*}_{6})+c.c\right\} |\alpha |^{4}-\left\{ (D_{1}D^{*}_{5}-D_{1}D_{6})|\alpha |^{2}\alpha ^{2}+c.c\right\} \right) \right] \\
 &  & 
\end{array}
\end{equation}
 where \( c.c\,  \) stands for the complex conjugate. Using equations (\ref{noipf}-\ref{sateropf})
the second order variances of \( C, \) \( S \) and \( N \) can be written
as 
\begin{equation}
\label{atheropf}
\begin{array}{lcl}
(\Delta C)^{2} & = & \frac{1}{4}\left( \overline{N}+\frac{1}{2}\right) ^{-1}\left[ \left\{ |D_{1}|^{2}+(D_{1}D_{2}+c.c)\right\} -3\left\{ (D_{1}D^{*}_{3}+D_{1}D_{4})\alpha ^{*2}+c.c\right\} \right. \\
 & - & \left. 2\left\{ (D_{1}D_{5}+D_{1}D^{*}_{6})|\alpha |^{2}+c.c)\right\} -\left\{ (D_{1}D^{*}_{5}+D_{1}D_{6})\alpha ^{2}+c.c\right\} \right] ,\\
 &  & 
\end{array}
\end{equation}

\begin{equation}
\label{unishpf}
\begin{array}{lcl}
(\Delta S)^{2} & = & -\frac{1}{4}\left( \overline{N}+\frac{1}{2}\right) ^{-1}\left[ \left\{ -|D_{1}|^{2}+(D_{1}D_{2}+c.c)\right\} +3\left\{ (D_{1}D^{*}_{3}-D_{1}D_{4})\alpha ^{*2}+c.c\right\} \right. \\
 & - & \left. 2\left\{ (D_{1}D_{5}-D_{1}D^{*}_{6})|\alpha |^{2}+c.c)\right\} +\left\{ (D_{1}D^{*}_{5}-D_{1}D_{6})\alpha ^{2}+c.c\right\} \right] ,\\
 &  & 
\end{array}
\end{equation}

\begin{equation}
\label{taroapf}
\begin{array}{lcl}
(\Delta N)^{2} & = & |\alpha |^{2}\left( 1+\lambda \left[ (3+4|\alpha |^{2})\sin t\sin (t-2\theta )+|\alpha |^{2}\sin 2t\sin 2(t-2\theta )\right] \right) \\
 &  & 
\end{array}.
\end{equation}

\section{Phase fluctuations \index{phase fluctuations}}

The usual parameters for the purpose of calculation of the phase fluctuations
(for a fixed value of \( \lambda  \)) are defined as {[}\ref{Gerry}-\ref{Lynch}{]}
\begin{equation}
\label{kuripf}
U\left( \theta ,t,|\alpha |^{2}\right) =(\Delta N)^{2}\left[ (\Delta S)^{2}+(\Delta C)^{2}\right] \left/ \left[ <S>^{2}+<C>^{2}\right] \right. 
\end{equation}

\begin{equation}
\label{ekushpf}
S\left( \theta ,t,|\alpha |^{2}\right) =(\Delta N)^{2}(\Delta S)^{2}
\end{equation}
 and

\begin{equation}
\label{bishpf}
Q\left( \theta ,t,|\alpha |^{2}\right) =S\left( \theta ,t,|\alpha |^{2}\right) \left/ <C>^{2}\right. 
\end{equation}
 Now we can, in principle, analytically calculate \( U,\, S \) and \( Q \)
by using equations (\ref{chabbispf}-\ref{taroapf}). The equations (\ref{kuripf}-\ref{bishpf})
assume the following forms, 
\begin{equation}
\label{satashpf}
\begin{array}{ccc}
U\left( \theta ,t,|\alpha |^{2}\right)  & = & \frac{1}{2}\left[ 1+\frac{\lambda }{4}\left( 6(1+2|\alpha |^{2})\sin t\sin (t-\theta )+3|\alpha |^{2}\sin 2t\sin 2(t-2\theta )\right) \right] \\
 &  & 
\end{array}
\end{equation}

\begin{equation}
\label{atashpf}
\begin{array}{lcl}
S\left( \theta ,t,|\alpha |^{2}\right)  & = & \frac{|\alpha |^{2}}{4}\left( \overline{N}+\frac{1}{2}\right) ^{-1}\left[ 1+\frac{\lambda }{4}\left\{ 6(1+2|\alpha |^{2})\sin ^{2}t\right. \right. \\
 & + & 6|\alpha |^{2}t\sin 2(t-\theta )+3|\alpha |^{2}\sin 2\theta \sin 2t\\
 & + & \left. \left. 4(3+2|\alpha |^{2})\sin t\sin (t-2\theta )+4|\alpha |^{4}\sin 2t\sin 2(t-2\theta )\right\} \right] 
\end{array}
\end{equation}
 and 
\begin{equation}
\label{untrishpf}
\begin{array}{lcl}
Q(\theta ,t,|\alpha |^{2}) & = & \frac{1}{4cos^{2}(t-\theta )}\left[ 1+\frac{\lambda }{4}\left\{ 6(1+2|\alpha |^{2})\sin ^{2}t+4(3+4|\alpha |^{2})\sin t\sin (t-2\theta )\right. \right. \\
 & + & \left. |\alpha |^{2}[6t\sin 2(t-2\theta )+3\sin 2\theta \sin 2t+4\sin 2t\sin 2(t-2\theta )]\right\} \\
 & - & \frac{\lambda }{8\cos ^{2}(t-\theta )}\left\{ -6t\sin 2(t-\theta )+(6+4|\alpha |^{2})\sin t\sin (t-2\theta )\right. \\
 & - & 6(1+|\alpha |^{2})\sin ^{2}t-2|\alpha |^{2}\sin t\sin (3t-4\theta )+|\alpha |^{2}\sin 2t\sin 2(t-2\theta )\\
 & - & \left. \left. |\alpha |^{2}\sin 2t\sin 2\theta -6|\alpha |^{2}t\sin 2(t-\theta )\right\} \right] .\\
 &  & 
\end{array}
\end{equation}
 Hence the equations (\ref{satashpf}-\ref{untrishpf}) are our desired results.
In the derivation of the equation (\ref{untrishpf}), we assume \( |\alpha |^{2}\neq 0 \).
Now, \( U_{0}=\frac{1}{2},\, S_{0}=\frac{1}{4}|\alpha |^{2}\left( |\alpha |^{2}+\frac{1}{2}\right) ^{-1} \)
and \( Q_{0}=\frac{1}{4\cos ^{2}(t-\theta )} \) are the initial (i.e \( \lambda =0 \))
values of \( U, \) \( S \) and \( Q \) respectively. Thus \( U_{0} \), \( S_{0} \)
and \( Q_{0} \) signify the information about the phase of the input coherent
light. The suitable choice of \( t \) may cause the enhancement and reduction
of all the above parameters compared to their initial values. It is to be noted
that the parameters \( S \) and \( Q \) contain the secular terms proportional
to \( t. \) However, it is not a serious problem since the product \( \lambda t \)
is small {[}\ref{Mandal}{]}. The equations (\ref{satashpf}-\ref{untrishpf})
are good enough to have the flavor of analytical results. Now we would like
to discuss two interesting special cases.

\subsection{The vacuum field \index{vacuum field} effect}

The radiation field with zero photon is termed as the vacuum field\index{vacuum field}.
The interaction of vacuum field with the medium gives rise to some interesting
quantum electrodynamic effects {[}\ref{Milloni}{]}. The purpose of this subsection
is to study the effects of the vacuum field on the phase fluctuations\index{phase fluctuations}
of input coherent light. In case of vacuum field (\( |\alpha |^{2}=0 \)) the
equations (\ref{satashpf}) and (\ref{untrishpf}) reduce to 
\begin{equation}
\label{trishpf}
U(\theta ,t)=U_{0}\left( 1+\frac{3\lambda }{2}\sin t\sin (t-\theta )\right) 
\end{equation}

\begin{equation}
\label{trishbpf}
\begin{array}{lcl}
Q(\theta ,t) & = & Q_{0}\left[ 1+\frac{3\lambda }{2}\sin t\left\{ \sin t+2\sin (t-2\theta )\right\} -\frac{3\lambda }{4\cos ^{2}(t-\theta )}\right. \\
 & \times  & \left. \left\{ -t\sin 2(t-\theta )+\sin t\sin (t-2\theta )-\sin ^{2}t\right\} \right] \\
 &  & 
\end{array}
\end{equation}
\\
 where \( |\alpha |^{2}\rightarrow 0 \) is used to derive (\ref{trishbpf}).
Interestingly, the vacuum field \index{vacuum field} itself couples with the
medium and gives rise to the condition \( \lambda \neq 0. \) Thus the phase
fluctuations for vacuum field \index{vacuum field} are of purely quantum electrodynamic
in nature. Now for \( \theta =0, \) the parameters \( U \) and \( Q \) are
enhanced compared to \( U_{0} \) and \( Q_{0} \) respectively. The corresponding
maximum fluctuation of \( U \) is obtained if \( t \) is an odd multiple of
\( \pi /2. \) The value of \( Q \) is infinite if \( t \) becomes an odd
multiple of \( \pi /2. \) For \( \theta \neq 0, \) the parameters \( U \)
and \( Q \) may be reduced or enhanced by the suitable choices of \( t. \)
It is to be noted that the parameter \( U \) is 0.5 and is independent of \( \theta  \)
in the earlier occasions {[}\ref{Gerry}-\ref{Lynch}{]}. In case of the present
work, however, the parameter \( U \) depends on \( \theta  \) and \( t. \)
The parameter \( S \) is identically zero and coincides exactly with the earlier
results {[}\ref{Gerry}-\ref{Lynch}{]}.

\subsection{Phase of the input coherent light \protect\( \theta =\frac{\pi }{4}\protect \)}

The equations (\ref{satashpf}-\ref{untrishpf}) are still complicated even
for \( \theta =\pi /4. \) A further simplification is made with the choice
\( t=\pi /4. \) The equation (\ref{satashpf}) reduces to a simple form

\begin{equation}
\label{ektrishapf}
U\left( \frac{\pi }{4},\frac{\pi }{4},|\alpha |^{2}\right) =U_{0}\left( 1-\frac{3\lambda |\alpha |^{2}}{8}\right) .
\end{equation}
 A huge reduction of \( U \) is possible with the increase of the photon number
\( |\alpha |^{2}. \) However, care should be taken about the condition of the
solution during such increase. The circular nature of the trigonometric function
ensures the occurrence of similar reductions for other values of \( t \) (\( t=2m\pi +\frac{\pi }{4} \)
where \( m \) is an integer). Now for \( t=\pi /2, \) we have 
\begin{equation}
\label{ektrishbpf}
U\left( \frac{\pi }{4},\frac{\pi }{2},|\alpha |^{2}\right) =U_{0}\left( 1+\frac{3\sqrt{2}\lambda }{4}\{1+2|\alpha |^{2}\}\right) .
\end{equation}
 The equation (\ref{ektrishapf}) clearly shows the enhancement of \( U \)
parameter as the photon number \( |\alpha |^{2} \) increases. Thus we conclude
by noting that the parameter \( U \) may decrease (\ref{ektrishapf}) or increase
(\ref{ektrishbpf}) with the increase of \( |\alpha |^{2} \) by suitable choices
of \( t. \) Similarly, one can obtain the reduction and enhancement of the
remaining two parameters (\( Q \) and \( S \)) by suitable manipulations of
free evolution time.

In the earlier works {[}\ref{Gerry}-\ref{Lynch}{]}, the parameter \( Q \)
is found to decrease with the increase of photon number till a minimum is reached.
Subsequent increase of \( |\alpha |^{2} \) causes the increase of \( Q \).
However, the parameters \( U \) and \( S \) are always enhanced compared to
their initial values as \( |\alpha |^{2} \) increases. Hence, the present results
are in sharp contrast with those of the earlier studies {[}\ref{Gerry}-\ref{Lynch}{]}.
The above differences are attributed due to the fact that the nonconserving
energy terms are included in the model Hamiltonian.

\section{Conclusion}

The phase fluctuation of coherent light interacting with a nonlinear medium
of inversion symmetry is carried out by using the PB formalism. The usual parameters
for this purpose are \( U \), \( Q \) and \( S. \) These parameters (\ref{satashpf}-\ref{untrishpf})
are found to depend on \( \theta , \) \( |\alpha |^{2} \) and \( t. \) It
is interesting to note that the free evolution time \( t \) is absent in the
earlier works {[}\ref{Gerry}-\ref{Lynch}{]}. However, the presence of \( t \)
is automatic in the present calculation. It accounts for the fact that the nonconserving
energy terms are present in our model Hamiltonian.

The effect of vacuum field\index{vacuum field} on the phase fluctuation parameters
are expressed in closed analytical forms. It is found that the enhancement of
\( U \) and \( Q \) are possible when the phase angle \( \theta =0. \) However,
for \( \theta \neq 0, \) both reduction and enhancement of \( U \) and \( Q \)
are possible by suitable choices of \( t. \) The observed results are in sharp
contrast with those of the earlier studies {[}\ref{Gerry}-\ref{Lynch}{]}.
The parameter \( S \) reduces to zero for vacuum field \index{vacuum field}
and agrees exactly with those of the earlier works {[}\ref{Gerry}-\ref{Lynch}{]}.

Apart from the general expressions for \( U \), \( Q \) and \( S \), we also
made a qualitative comparison between the present results and the results obtained
earlier for the identical physical system. For \( \theta =\pi /4, \) we obtain
the reduction and enhancement of phase parameters (\( U \), \( Q \) and \( S \))
by suitable manipulation of free evolution time \( t. \) Those results are
in sharp contrast with the results already obtained for the same physical system
{[}\ref{Gerry}-\ref{Lynch}{]}. Clearly, the free evolution time \( t \) makes
the differences.

Of late the preparation of quantum states have been reported by several laboratories
{[}\ref{Smithney}-\ref{Schiller}{]}. These production of quantum states have
opened up the possibilities of experimental studies on quantum phase\index{quantum phase}
and hence the verification of the present results.

\chapter{Squeezing\index{squeezing} of coherent light coupled to a third order nonlinear
medium}

According to the Heisenberg's uncertainty principle we can't measure position
(\( X \)) and momentum (\( \dot{X} \)) of a particle with greater accuracy
than 
\begin{equation}
\label{sq1}
\triangle X\triangle \dot{X}\geq \frac{1}{2}
\end{equation}
 where uncertainty \( (\triangle C) \) in any arbitrary variable \( C \) is
defined as \( (\triangle C)=\left( <C^{2}>-<C>^{2}\right) ^{\frac{1}{2}} \).
Although, there is no position operator for photon, fundamental canonical variables
for light (component of electric field and magnetic field) satisfy an uncertainty
relation of the form (\ref{sq1}) and an electromagnetic field is said to be
electrically squeezed field if \( \triangle X \) is less than the vacuum field\index{vacuum field}
uncertainties in the quadrature (i.e. \( (\triangle X)^{2}<\frac{1}{2} \)).
Correspondingly a magnetically squeezed field is one for which \( (\triangle \dot{X})^{2}<\frac{1}{2}. \)
Here we are not going to discuss the phenomenon of squeezing\index{squeezing}
in detail. So for a review on squeezed states\index{squeezed state}, one can
see the references {[}\ref{Loudon1},\ref{Smsqueezing}{]}. In this chapter
we want to examine the possibilities of observing squeezing\index{squeezing}
in a physical situation in which a single mode of the quantized electromagnetic
field, initially prepared in the coherent state \index{coherent state} (coherent
light), is allowed to interact with a nonlinear nonabsorbing medium of inversion
symmetry.

\section{{\small Application of quantum quartic\index{quartic} oscillator: The squeezed
states}\index{squeezed state}{\small .} }

We have already shown that for the lowest order of nonlinearity (i.e cubic nonlinearity),
corresponding Hamiltonian of the system is given by the Hamiltonian of a quartic\index{quartic}
anharmonic oscillator (\ref{ekpf}). Equation of motion of the corresponding
oscillator is (\ref{dui}) and the second order operator solution of (\ref{dui})
is simply given by (\ref{15}). 

Let us now express the quadrature operator \( X \) obtained in (\ref{15})
in terms of the initial \( (i.e\, t=0) \) annihilation \index{annihilation operator}
and creation operators\index{creation operator}. It is given by 
\begin{equation}
\label{21}
\begin{array}{lcl}
X(t) & = & \left[ E_{1}a(0)+E_{2}a^{3}(0)+E_{3}a^{\dagger 2}(0)a(0)+E_{4}a^{5}(0)\right. \\
 & + & \left. E_{5}a^{\dagger }(0)a^{4}(0)+E_{6}a^{\dagger 2}(0)a^{3}(0)\right] +h.c
\end{array}
\end{equation}
where \( h.c \) stands for the Hermitian conjugate. The remaining quadrature
operator \( \dot{X}(t) \) may simply be obtained by taking the time derivative
of (\ref{21}). The parameters \( E_{i}\, (i=1,2,3,4,5 \) and \( 6) \) are
complex and are given by 
\begin{equation}
\label{21a}
\begin{array}{lcl}
E_{1} & = & \frac{1}{\sqrt{2}}\left[ \cos t-i\sin t-\frac{3\lambda }{4}\left\{ t\sin t+i(t\cos t-\sin t)\right\} \right. \\
 & + & \frac{\lambda ^{2}}{512}\left\{ (468t\sin t-63\cos t+63\cos 3t-216t^{2}\cos t)\right. \\
 & + & \left. \left. i(1188t\cos t-1053\sin t-45\sin 3t+216t^{2}\sin t)\right\} \right] 
\end{array}
\end{equation}
 
\begin{equation}
\label{21b}
\begin{array}{lcl}
E_{2} & = & \frac{1}{\sqrt{2}}\left[ -\frac{\lambda }{16}\left\{ (\cos t-\cos 3t)+i(\sin 3t-3\sin t)\right\} \right. \\
 & + & \frac{\lambda ^{2}}{512}\left\{ (156\cos t-192t\sin t-156\cos 3t-144t\sin 3t)\right. \\
 & + & \left. \left. i(156\sin 3t-324\sin t-156t\cos 3t)\right\} \right] 
\end{array}
\end{equation}
 
\begin{equation}
\label{21c}
\begin{array}{lcl}
E_{3} & = & \frac{1}{\sqrt{2}}\left[ -\frac{3\lambda }{4}\left\{ t\sin t-i(t\cos t-\sin t)\right\} \right. \\
 & + & \frac{\lambda ^{2}}{512}\left\{ (936t\sin t-126\cos t+126\cos 3t-432t^{2}\cos t)\right. \\
 & + & \left. \left. i(2376t\cos t-2106\sin t-90\sin 3t+432t^{2}\sin t)\right\} \right] 
\end{array}
\end{equation}
 
\begin{equation}
\label{21d}
\begin{array}{lcl}
E_{4} & = & \frac{\lambda ^{2}}{256\sqrt{2}}\left[ (5\cos t-12t\sin t-6\cos 3t+\cos 5t)\right. \\
 & + & \left. i(6\sin 3t-\sin t-12t\cos t-\sin 5t)\right] 
\end{array}
\end{equation}
 
\begin{equation}
\label{21e}
\begin{array}{lcl}
E_{5} & = & \frac{\lambda ^{2}}{256\sqrt{2}}\left[ (39\cos t-48t\sin t-39\cos 3t-36t\sin 3t)\right. \\
 & + & \left. i(39\sin 3t-81\sin t-36t\cos 3t)\right] 
\end{array}
\end{equation}
 and
\begin{equation}
\label{21f}
\begin{array}{lcl}
E_{6} & = & \frac{\lambda ^{2}}{256\sqrt{2}}\left[ (156t\sin t-21\cos t+21\cos 3t-72t^{2}\cos t)\right. \\
 & + & \left. i(396t\cos t-351\sin t-15\sin 3t+72t^{2}\sin t)\right] \, .
\end{array}
\end{equation}
 Initially prepared coherent state\index{coherent state} obeys the eigenvalue\index{eigenvalue}
equation \( a(0)|\alpha >=\alpha |\alpha >, \) where \( \alpha  \) is in general
complex and the parameter \( \alpha  \) is given by \( \alpha =|\alpha |\, \exp (i\theta ) \)
as before. The number of photons present in the radiation field (prior to the
interaction) is \( N_{0}=|\alpha |^{2}. \) For \( \theta =0 \) and \( \theta =\pi /2, \)
the corresponding situations are called, in phase (electromagnetic field is
in phase with the electric field) condition and out of phase condition respectively.
Using (\ref{21}) the second order variance (hereafter we shall call it as variance)
of the \( X \) quadrature is calculated with respect to the initially prepared
coherent state \index{coherent state} and is given by 
\begin{equation}
\label{22}
\begin{array}{lcl}
(\Delta X)^{2} & = & <\alpha |X^{2}|\alpha >-<\alpha |X|\alpha >^{2}\\
 & = & \left[ \left( |E_{1}|^{2}+|E_{2}|^{2}(9|\alpha |^{4}+18|\alpha |^{2}+6)+|E_{3}|^{2}(5|\alpha |^{4}+2|\alpha |^{2}\right) \right. \\
 & + & \left( 2E^{2}_{3}|\alpha |^{2}\alpha ^{*2}+2E_{1}E_{3}|\alpha |^{2}+E_{1}E^{*}_{3}\alpha ^{2}+3E_{1}E^{*}_{2}\alpha ^{*2}\right. \\
 & + & 6E_{2}E_{3}(|\alpha |^{2}\alpha ^{2}+\alpha ^{2})+3E_{2}E^{*}_{3}\alpha ^{4}+E_{1}E_{5}\alpha ^{4}+2E_{1}E_{6}|\alpha |^{2}\alpha ^{2}\\
 & + & \left. \left. 3E_{1}E^{*}_{6}|\alpha |^{4}+4E_{1}E^{*}_{5}|\alpha |^{2}\alpha ^{*2}+5E_{1}E^{*}_{4}\alpha ^{*4}+c.c\right) \right] 
\end{array}
\end{equation}
where the terms beyond \( \lambda ^{2} \) are neglected. In a similar manner,
the variance for other quadrature \( \dot{X}(t) \) may be obtained. In absence
of the interaction (i.e \( \lambda =0 \)), the variance (\ref{22}) reduces
to \( \frac{1}{2} \). The squeezing\index{squeezing} in absolute sense of
\( X \) quadrature is obtained if the value of \( (\Delta X)^{2}=0. \) We
note that the squeezing\index{squeezing} of one quadrature automatically prohibits
the squeezing\index{squeezing} of the remaining one. The variance (\ref{22})
is extremely complicated and deserves more investigation for the discussions
of squeezing\index{squeezing} effects of the input coherent light. It is not
our purpose to go through such detail discussions of the squeezing\index{squeezing}
arising out of the expression (\ref{22}). Rather, we would prefer to discuss
few special cases of physical interests. 

It is true that both the solutions (\ref{15}) and (\ref{16}) are equally good
for the purpose of getting \( (\Delta X)^{2}. \) There are good reasons for
using the solution (\ref{15}) instead of the solution (\ref{16}) though the
previous one contains secular terms. Because of the complex nature of the operators
in solution (\ref{16}), it is hard to calculate \( (\Delta X)^{2} \) . In
addition to this, the present work takes care the solution of the quartic\index{quartic}
AHO up to the second order in \( \lambda  \). Both the solutions (\ref{15})
and (\ref{16}) are identical as long as the second order solution is concerned.
The removal of the secular terms from the calculated (\( \Delta X)^{2} \),
if required, may be done by using the same Tucking-in technique.

\subsection{{\small Vacuum field} \index{vacuum field}}

In case of vacuum field\index{vacuum field} \( (i.e\, |\alpha |^{2}=0), \)
equation (\ref{22}) reduces to an extremely simple form 
\begin{equation}
\label{23}
\begin{array}{lcl}
(\Delta X)_{vac}^{2} & = & |E_{1}|^{2}+6|E_{2}|^{2}\\
 & = & \frac{1}{2}-\frac{3\lambda }{4}\sin ^{2}t+\left[ \frac{3\lambda ^{2}}{512}(201-24t^{2}-208\cos 2t+7\cos 4t-168t\sin 2t)\right] .
\end{array}
\end{equation}
 If we drop the terms involving \( \lambda ^{2} \) present in (\ref{23}) then
the corresponding variance coincides exactly with the earlier result {[}\ref{Mandalb}{]}.
Let us now discuss two special cases of particular interests. For \( t=n\pi , \)
the variance (\ref{23}) reduces to 
\begin{equation}
\label{ajebaje}
(\triangle X)_{vac}^{2}=\frac{1}{2}-\frac{9\lambda ^{2}}{64}n^{2}\pi ^{2}.
\end{equation}
Equation (\ref{ajebaje}) depends on \( \lambda ^{2} \) only and we observe
that the squeezing\index{squeezing} of the vacuum field \index{vacuum field}
is possible for \( t=n\pi  \). Note that a first order calculation can not
predict the squeezing\index{squeezing} effect for the same conditions. Hence,
a second order solution is not only useful but also essential for the present
study. For \( t=\frac{\pi }{2}, \) the variance (\ref{23}) reduces to 
\begin{equation}
\label{23a}
(\Delta X)_{vac}^{2}=\frac{1}{2}-\frac{3\lambda }{4}+\frac{3\lambda ^{2}}{256}(208-3\pi ^{2}).
\end{equation}
 For small coupling, the first order term \( (i.e\, \propto \lambda ) \) in
equation (\ref{23a}) dominates over the second order term \( (i.e\, \propto \lambda ^{2}) \)
and hence the squeezing\index{squeezing} effects are controlled accordingly.
Clearly, the monotonic increase of the squeezing\index{squeezing} effects due
to the first order term is arrested by the presence of the second order one.
Thus the second order term is responsible for the reduction of the squeezing
effect for large coupling. 

Note that the so-called tucking-in technique is also applicable to equation
(\ref{23}). Hence, the secular terms can be summed up for all orders of \( \lambda  \).
The corresponding expression for \( (\Delta X)_{vac}^{2} \) may be expressed
as 
\begin{equation}
\label{23b}
\begin{array}{lcl}
(\Delta X)_{vac}^{2} & = & \frac{1}{2}\cos 2t-\frac{3\lambda }{4}\sin ^{2}t+\frac{3\lambda ^{2}}{512}(201-208\cos 2t+7\cos 4t)\\
 & + & \frac{1}{2}(\cos \frac{3\lambda t}{4}-\cos 2t(1-\frac{63\lambda ^{2}}{64})).
\end{array}
\end{equation}

\subsection{{\small In phase with the electric field }\small }

We assume that the input radiation field is in phase with that of the electric
field. In that case \( \theta =0 \) and \( \alpha  \) is real. The corresponding
variance reduces to the following form 
\begin{equation}
\label{24}
\begin{array}{lcl}
(\Delta X)^{2} & = & (\Delta X)_{vac}^{2}+\frac{3|\alpha |^{2}}{4}\left[ -\lambda \left( 1-\cos 2t+t\sin 2t\right) \right. \\
 & + & \left. \frac{3\lambda ^{2}}{64}\left( 11+24t^{2}-32\cos 2t+21\cos 4t+71t\sin 2t+t\sin 4t-64t^{2}\cos 2t\right) \right] \\
 & + & \frac{|\alpha |^{4}\lambda ^{2}}{128}\left[ 237+72t^{2}-256\cos 2t+19\cos 4t-36t\sin 2t-36t\sin 4t-216t^{2}\cos 2t\right] .\\
 &  & 
\end{array}
\end{equation}
Equation (\ref{24}) is useful for the purpose of discussing squeezed states
\index{squeezed state} generated due to the interaction a coherent light with
a cubic nonlinear medium. For \( |\alpha |^{2}=0, \) corresponding equation
reduces to the equation (\ref{23}). If we now drop the terms involving \( \lambda ^{2} \),
the variance (\ref{24}) does have exact coincidence with the earlier reported
results {[}\ref{Mandalb}{]}. Here we are not going to present the ``Tucked-in
form'' of the equation (\ref{24}). This is because, we are interested in the
qualitative study of \( (\Delta X)^{2} \) up to the second order in \( \lambda  \)
for a weakly coupled system. If we put \( t=\frac{\pi }{2}, \) the corresponding
variance reduces to 
\begin{equation}
\label{25}
(\Delta X)^{2}=\frac{1}{2}-\frac{3\lambda }{4}+\frac{3\lambda ^{2}}{256}(208-3\pi ^{2})+\frac{3|\alpha |^{2}}{4}\left( -2\lambda +\frac{3\lambda ^{2}}{64}(64+22\pi ^{2})\right) +\frac{|\alpha |^{4}\lambda ^{2}}{16}\left( 64+9\pi ^{2}\right) .
\end{equation}
It is obvious that the equation (\ref{25}) exhibits the monotonic nature of
the squeezing\index{squeezing} as long as the first order solution is concerned.
\emph{}In most of the experiments, the percentage of squeezing gets saturated
after a certain value of the electric field and the squeezing effects are completely
destroyed if the field strength is very high. Moreover, oscillatory behaviour
in the performance of squeezing\index{squeezing} is observed in experiments.
These experimental facts are consistent with the present work (\ref{25}). These
experimental features also reveals the fact that higher order effects are inevitable
in the calculation of squeezing\index{squeezing}. Thus the present second order
calculation is more useful to predict the correct behavior o{\small f} squeezing
phenomenon observed due to the interaction of coherent light with dielectric
media.

\chapter{Photon bunching\index{bunching}, antibunching\index{antibunching} and photon
statistics}

People tried earnestly to understand the basic nature of the radiation field
and as a result of this endeavor the concept of photon came into reality {[}\ref{Planck}-\ref{Einstein}{]}.
With the due success of the photon concept, the immediate questions that come
to our mind are: How do the photons come? Are they coming together (i.e cluster)
or one after another (without cluster)? The answers to these questions have
started coming with the remarkable experiment of Hanbury-Brown and Twiss\index{HBT experiment}
{[}\ref{Hanbury}{]}. In their experiment, they obtained the intensity correlation
of an incandescent light source and they concluded that the photons come together
indeed. The phenomenon in which the photons come in a cluster is termed as photon
bunching\index{bunching}. On the other hand, the probability of detecting a
coincident pair of photons from an antibunched field\index{anibunched field}
is less than that from a fully coherent light field with a random Poisson distribution
of photons.

The explanations for photon bunching\index{bunching} may be given in terms
of the wave nature of the light in addition to the particle (quantum) nature.
However, antibunching\index{antibunching} of photon can only be explained in
terms of the particle concept of the radiation field. For this reason, the photon
antibunching\index{antibunching} is regarded as a fully quantum mechanical
phenomenon without any classical analogue. There is a natural tendency that
the photons are bunched {[}\ref{Loudon}, \ref{Hanbury}{]}. However, there
are several predictions (models) which go against this tendency and exhibit
the photon antibunching\index{antibunching}. For example, various coherent
states \index{coherent state} {[}\ref{Mahran}{]} and the degenerate parametric
amplifier {[}\ref{Stoler}{]} were predicted as possible sources of getting
antibunched light. In addition to this, the resonance fluorescence from a two
level atom {[}\ref{Carmeichael}{]}, the two photon Dick model {[}\ref{Gerry1}{]}
and the coherent light interacting with a two-photon medium {[}\ref{Gerry2}{]}
are possible sources for producing antibunched light. The fluorescence spectrum
of a single two level atom shows the antibunching\index{antibunching} and the
squeezing\index{squeezing} as well {[}\ref{Carmeichael}{]}. There are several
examples where the antibunched light are produced in the laboratory {[}\ref{Kimble1}-\ref{Mielke}{]}.
In most of these experiments {[}\ref{Died}-\ref{Basche}{]}, the resonance
fluorescence from a small number of ions {[}\ref{Died}{]}, atoms or molecules
{[}\ref{Rempe}-\ref{Basche}{]} are studied to obtain the antibunching\index{antibunching}
effects. The basic physics behind these experiments are clear. The atoms (or
molecules) emit radiation and go to a ground state from where no subsequent
radiation is possible. Hence the emitted photons are antibunched. The resonance
fluorescence field from many-atom source is not suitable for antibunching\index{antibunching}
of photons since the photons emitted are highly uncorrelated. However, a suitable
phase matching condition similar to that of four-wave mixing leads to the photon
antibunching\index{antibunching} even in the resonance fluorescence of a multi
atomic system {[}\ref{grangier}{]}. 

With the advent of different type of radiation sources, studies on the quantum
statistical properties (QSP) of the radiation field are increased considerably.
The QSP of the radiation field are usually studied with the help of second order
correlation function for zero time delay {[}\ref{Mahran}, \ref{Walls}{]}
\begin{equation}
\label{bu1}
g^{2}(0)=\frac{<a^{\dagger }a^{\dagger }aa>}{<a^{\dagger }a>^{2}}.
\end{equation}
 Equation (\ref{bu1}) can also be expressed in terms of \( <N> \) and \( (\triangle N)^{2} \)
as 
\begin{equation}
\label{bu1a}
g^{2}(0)=1+\frac{(\triangle N)^{2}-<N>}{<N>^{2}}
\end{equation}
where \( <N> \) is the average number of photons present in the radiation field
and the parameter \( (\Delta N)^{2}=<N^{2}>-<N>^{2} \) is the second order
variance in photon number. 

The conditions for bunching\index{bunching} and antibunching\index{antibunching}
of photons are given in the table 5.1 

\vspace{0.3cm}
{\centering \begin{tabular}{|c|c|}
\hline 
\( g^{2}(0)>1 \)&
photons are bunched\\
\hline 
\( g^{2}(0)=1 \)&
photons are coherent\\
\hline 
\( g^{2}(0)<1 \)&
photons are antibunched \\
\hline 
\end{tabular}\par}
\vspace{0.3cm}

{\par\centering Table 5.1\par}

\vspace{1cm}
One more useful parameter in the context of QSP of the radiation field is the
\emph{}Mandel Q-parameter {[}\ref{Mahran}, \ref{Walls}{]} which is defined
as 
\begin{equation}
\label{ab3}
Q=\frac{(\triangle N)^{2}-<N>}{<N>}=<N>\left( g^{2}(0)-1)\right) .
\end{equation}
 Statistical distribution of photons are related to the \( Q \) parameter as 

\vspace{0.3cm}
{\centering \begin{tabular}{|c|p{3.4cm}|}
\hline 
value of \( Q \)&
Photon Number  Distribution\index{photon number distribution}\\
\hline 
\hline 
\( Q<0 \)&
sub-Poissonian\index{sub-Poissonian}\\
\hline 
\( Q=0 \)&
Poissonian\index{Poissonian}\\
\hline 
\( Q>0 \)&
super-Poissonian\index{super-Poissonian}  \\
\hline 
\end{tabular}\par}
\vspace{0.3cm}

{\par\centering Table 5.2\par}
\vspace{1cm}

Now if we define a quantity 
\begin{equation}
\label{ab4}
d=\left( (\triangle N)^{2}-<N>\right) ,
\end{equation}
 then the sign of \( d \) will essentially determine the quantum statistical
properties of the radiation field. From the definition of \( d \) (\ref{ab4}),
Table 5.1 and Table 5.2 we can construct new conditions as 

\vspace{0.3cm}
{\centering \begin{tabular}{|c|p{2.4cm}|p{3.4cm}|}
\hline 
value of \( d \)&
bunching/ anti-bunching&
Photon Number  Distribution\index{photon number distribution}\\
\hline 
\hline 
 \( 0 \)&
coherent&
Poissonian\\
\hline 
\( <0 \)&
anti-bunched&
sub-Poissonian\index{sub-Poissonian}\\
\hline 
\( >0 \)&
bunched &
super-Poissonian \index{super-Poissonian}\\
\hline 
\end{tabular}\par}
\vspace{0.3cm}

{\par\centering Table 5.3\par}

The purpose of the present chapter is to study the photon number distribution
\index{photon number distribution} (PND\index{PND}) and to investigate the
possibilities of getting bunching\index{bunching} and antibunching\index{antibunching}
of photons in the interaction of a coherent light with a third order nonlinear
medium of inversion symmetry. We are specially interested to keep the off-diagonal
terms in the model Hamiltonian and to study the consequences of their inclusion
in connection with the study of nonclassical effects. The effects of input vacuum
field \index{vacuum field} on photon bunching\index{bunching} and antibunching\index{antibunching}
are also taken care of.

\section{{\small The photon bunching} \index{bunching}{\small and photon antibunching\index{antibunching}}\small }

In this chapter number of photons present in the radiation field prior to the
interaction will be denoted by nonnegative integer \( N_{0}=|\alpha |^{2}. \)
Therefore, equation (\ref{chabbispf}) may be used to write the average number
of photons present in the radiation field (initially prepared in coherent state\index{coherent state})
after interaction as 
\begin{equation}
\label{bu5}
\begin{array}{lcl}
<\alpha |a^{\dagger }a|\alpha >=<N> & = & N_{0}\left[ 1+\frac{\lambda }{4}\left\{ 2(3+2N_{0})\sin t\sin (t-2\theta )\right. \right. \\
 & + & \left. \left. N_{0}\sin 2t\sin 2(t-2\theta )\right\} \right] .
\end{array}
\end{equation}
 Thus the number of post interacted photons (i.e \( <N> \)) is not conserved
and is phase sensitive. It is not at all surprising since the Hamiltonian (\ref{ekpf})
contains the terms which are nonconserving in nature. A considerable amount
of discussions about the inclusion and impacts of the photon number nonconserving
terms in the QAHO model are available in chapter 1 and chapter 3. Now the variance
\( (\Delta N)^{2}=<\alpha |N^{2}|\alpha >-<\alpha |N|\alpha >^{2} \) is given
by (see \ref{taroapf})
\begin{equation}
\label{bu6}
\begin{array}{lcl}
(\triangle N)^{2} & = & N_{0}\left[ 1+\lambda \left( (3+4N_{0})\sin t\sin (t-2\theta )+N_{0}\sin 2t\sin 2(t-2\theta )\right) \right] .\\
 &  & 
\end{array}
\end{equation}
 Hence we have,
\begin{equation}
\label{bu7}
d=\frac{3\lambda N_{0}}{4}\left[ (2+4N_{0})\sin t\sin (t-2\theta )+N_{0}\sin 2t\sin 2(t-2\theta )\right] .
\end{equation}
 The sign of the parameter \( d \) controls the phenomena of bunching\index{bunching}
and antibunching\index{antibunching} of post-interacted photons. However, the
oscillatory nature of \( d \) ensures the revival and collapse of bunching
and antibunching effects. The revival of bunching and antibunching was obtained
in earlier occasion too {[}\ref{Gerry2}{]}. Now we investigate how the sign
of \( d \) is controlled by the phases. For \( \theta =0, \) \( d \) is always
positive and hence the photons are always bunched. Corresponding PND\index{photon number distribution}
is super-Poissonian\index{super-Poissonian}. For \( \theta =\frac{\pi }{4}(\frac{3\pi }{4}), \)
the value of \( d \) is negative (positive) if \( 0<\sin 2t<1 \) and is positive
(negative) if \( -1<\sin 2t<0. \) Thus the photons are antibunched and bunched
respectively for the former (later) and later (former) conditions. These statements
are further corroborated by other values of phases. In the equations (\ref{bu5})
and (\ref{bu6}), the terms beyond linear power of \( \lambda  \) are neglected.
This is because of two reasons. Firstly, the solution available for the equation
(\ref{dui}) is simple as long as we are interested up to the linear power in
\( \lambda  \). The second reason is that we are interested about the qualitative
studies of nonclassical properties of the radiation field and hence, the solution
linear in \( \lambda  \) is sufficient. It is tedious, however possible to
obtain the correction up to the second order in \( \lambda  \) for \( d \).
The second order correlation function (\ref{bu1a}) for zero time delay depends
on the photon number \( N_{0} \), coupling strength \( (\lambda ) \), free
evolution time \( t \) and on the phase angle \( \theta . \) Values of \( t \)
may be controlled by suitable manipulation of the length of the interaction
region. 

For \( \theta =t/2 \) we obtain, \( <N>=N_{0} \) and \( (\Delta N)^{2}=N_{0}. \)
Therefore, \( d=0 \) and hence the input coherent state\index{coherent state}
remains unchanged during its passage through the nonlinear medium. The situation
may be viewed as the propagation of coherent beam without affecting the propagating
medium. The particular value of \( t\, (=2\theta ) \) causes a destructive
interference in the absorption profile and as a result of that we observe a
phenomenon which has strong resemblance with the self induced transparency (SIT)
{[}\ref{Mc Call}{]}. Here we can also mention one more important effect which
is the outcome of the phase of the input coherent light. For \( \theta =\frac{\pi }{4} \)
and \( -1<\sin 2t<0 \) we obtain the antibunching of the emitted (post-interacted)
photons.

\section{{\normalsize The effect of vacuum field}\index{vacuum field}{\normalsize : }\normalsize }

Now, we give the flavor of analytical results for input vacuum field\index{vacuum field}.
The quantized electromagnetic field with no photons are termed as vacuum field\index{vacuum field}.
The vacuum field is responsible for various interesting physical phenomena.
For example, the vacuum field \index{vacuum field} causes Lamb shift\index{Lamb shift},
Kasimir effects and Spontaneous emission {[}\ref{Milloni}{]}. Note that all
these phenomena are purely quantum electrodynamic in nature and do not have
any classical analogue. The effects of vacuum field \index{vacuum field} interacting
with a cubic nonlinear medium are studied to obtain the squeezing\index{squeezing}
{[}\ref{Mandalb}{]} and phase fluctuations\index{phase fluctuations} {[}\ref{anirban}{]}
of input coherent light. It was found that these nonlinear effects are the outcome
of inclusion of photon number nonconserving terms. Therefore, it is quite natural
to explore the effects of vacuum field \index{vacuum field} on the bunching\index{bunching}
and antibunching\index{antibunching} of photons. In case of input vacuum field
\index{vacuum field} \( (i.e\, N_{0}=0), \) \( <N> \) and \( (\triangle N)^{2} \)
are identically zero and hence the numerator of the equation (\ref{bu1a}) is
zero \( (i.e.\, d=0). \) Hence, \( g^{2}(0) \) is indeterminate. However,
the denominator \( <N>^{2}\rightarrow 0 \) faster compare to the numerator.
Thus the value of \( g^{2}(0)\gg 1, \) if the sign of \( d\, (i.e\, d\rightarrow 0^{+}) \)
is positive. Therefore the photons are strongly bunched for vacuum field\index{vacuum field}.
It is already established that the vacuum field produces the squeezing\index{squeezing}
effect {[}\ref{Mandalb}{]}. Hence the retainment of the non conserving energy
terms in the model Hamiltonian (\ref{ekpf}) leads to photon bunching\index{bunching}
and squeezing\index{squeezing} effects of the vacuum fields\index{vacuum field}.
Of course, the two photon coherent state \index{coherent state} with zero photon
(i.e squeezed vacuum state) gives rise to the photon bunching along with the
super-Poissonian\index{super-Poissonian} photon statistics {[}\ref{Mahran},\ref{Walls}{]}.
However the explanation of photon bunching for vacuum field \index{vacuum field}
were not given in these earlier publications. The situation is interesting,
since, the photons are bunched although no photons are present in the radiation
field! The result does not have classical analogue and is a consequence of pure
quantum electrodynamic effect. The vacuum field interacts with the medium and
produces photons which are bunched\emph{.} The creation of photons through nonlinear
interactions is ensured with the incorporation of the photon number nonconserving
energy terms in the model Hamiltonian (\ref{ekpf}). 

Interestingly, the suitable choice of phases (for example \( \theta =\pi /4 \)
and \( 0<t<\pi /2 \)) leads to negative \( d \) \( (i.e\, d\rightarrow 0^{-}) \)
and hence the photon antibunching\index{antibunching}. Thus, the antibunched
photons may be produced through the nonlinear interaction between the medium
and the vacuum field\index{vacuum field}. 

For \( N_{0}=0 \) the \( Q \) parameter (\ref{ab3}) has indeterminate form
\( (\frac{0}{0}) \). However, in the limit \( N_{0}\rightarrow 0 \), the corresponding
value of \( Q \) is found to be 
\begin{equation}
\label{bu8}
Q=\frac{\frac{3\lambda }{2}\sin t\sin (t-2\theta )}{1+\frac{3\lambda }{2}\sin t\sin (t-2\theta )}.
\end{equation}
 Now for \( \theta =0 \) and \( t\neq n\pi , \) corresponding PND\index{photon number distribution}
is super-Poissonian\index{super-Poissonian}. Interestingly, for \( \theta =\pi /4, \)
the corresponding PND is sub-Poissonian if \( 0<t<\pi /2 \) and super-Poissonian
\index{super-Poissonian} if \( \pi /2<t<\pi  \) respectively. Hence, the sub-
and super-Poissonian\index{super-Poissonian} distribution of the photons are
obtained periodically. For \( \theta =\pi /2, \) the corresponding PND is sub-Poissonian\index{sub-Poissonian}
for all \( t. \)

\section{{\small Conclusion:}\small }

We obtain the simultaneous appearance of photon bunching\index{bunching} (classical
phenomenon) and the sub-Poissonian\index{sub-Poissonian} photon statistics
(nonclassical phenomenon). In other words classical phenomena may appear simultaneously
with the nonclassical phenomena.  These results are consistent with the previous
observations {[}\ref{Died},\ref{Mandel1}{]}. The effects of the phase of the
input coherent light on \( g^{2}(0) \) are pointed out clearly. We obtain the
bunching\index{bunching} and antibunching\index{antibunching} of the vacuum
field \index{vacuum field} for \( \theta \ne 0. \) It is found that the vacuum
field \index{vacuum field} generates photons which may be bunched or antibunched. 

We conclude the present chapter with the following observations. 

\begin{enumerate}
\item As a result of the interaction between a single mode of coherent light and a
nonlinear medium classical and nonclassical phenomena may appear simultaneously.
There is no reason to believe that the appearance of one of the classical (nonclassical)
phenomena warrants the presence of the others. 
\item We report the sub-Poissonian\index{sub-Poissonian} PND \index{photon number distribution}and
the antibunching\index{antibunching} of photons for input coherent light and
vacuum field\index{vacuum field}. 
\item The vacuum field \index{vacuum field} interacting with the nonlinear medium
produces photons through nonlinear interaction. 
\end{enumerate}

\chapter{Application of the \protect\( m\protect \)-th anharmonic oscillator: Interaction
of coherent light with an \protect\( (m-1)\protect \)-th order nonlinear medium}

The concept of phase plays a very crucial role in the understanding of basic
physics. In fact, phase is responsible for all the interference phenomena we
observe in classical and quantum physics. So people have studied the properties
of the quantum phase\index{quantum phase} from different contexts since the
beginning of quantum mechanics {[}\ref{perinova}{]}. But the interest increased
considerably in the recent past when Berry published his famous paper {[}\ref{Berry}{]}.
He showed that the phase acquired by the quantum system during a cyclic evolution
under the action of an adiabatically varying Hamiltonian is a sum of two parts.
The first part is dynamical in nature and the second is geometric. Later on
Aharonov and Anandan generalized the idea of the Berry phase and they defined
a geometric phase factor for any cyclic evolution of quantum system {[}\ref{Aharonov}{]}.
The existence of geometric phase\index{geometric phase} is found in many areas
of physics. The examples are from the quantum Hall effect to the Jahn-Teller\index{Jahn-Teller effect}
effect and from the spin orbit interaction to quantum computation {[}\ref{Joshi-Pati},\ref{perinova},\ref{blais}{]}. 

In the present work, interaction of electromagnetic fields with matter is modeled
by AHO. These facts provoked us to study the possibilities of observing nonadiabatic
geometric phase\index{nonadiabatic geometric phase} or Aharonov-Anandan phase
\index{Aharonov Anandan phase}{[}\ref{Aharonov},\ref{Moore}{]} for anharmonic
oscillators in general. In the present chapter we will find an analytic expression
in closed form for the Aharonov-Anandan phase for a generalized anharmonic oscillator.
The generalized expression is then used to study a particular case of physical
interest in which an intense laser\index{laser} beam interacts with a third
order nonlinear nonabsorbing medium having inversion symmetry. In this chapter
results of chapters 4 and 5 are extended to the case of \( (m-1) \)-th order
nonlinear medium in general to study the possibilities of generating nonclassical
states in an \( (m-1) \)-th order nonlinear medium.

\section{Aharonov-Anandan phase\index{Aharonov Anandan phase}}

In quantum mechanics of any system, one has a complex Hilbert space\index{Hilbert space} $\cal{H}$
whose nonzero vectors represent states of the system. A vector \( |\psi > \)
in $\cal{H}$ and any complex multiple of it (i.e \( \lambda |\psi > \)) represent
the same state in the Hilbert space $\cal{H}$. This arbitrariness in the representation
of states can be reduced by imposing the normalization condition 
\begin{equation}
\label{ek}
<\psi |\psi >=1.
\end{equation}
 There still remains an arbitrariness of a phase factor because a normalized
vector \( |\psi > \) and the vectors \( \alpha |\psi > \) represents the same
state in the Hilbert space\index{Hilbert space} provided that the modulus of
\( \alpha  \) being unity (i.e \( |\alpha |=1 \)). The collection of all vectors
\( \alpha  \)\( |\psi > \) with \( |\psi > \), a fixed normalized vector
and \( \alpha  \) taking all possible complex values of modulus one is called
a unit ray\index{unit ray} and is denoted by \( |\widetilde{\psi }> \) {[}\ref{Tulshidas}{]}.
More general rays are consist of collection of all vectors of the form \( \lambda  \)\( |\psi > \).
So rays represent vectors without any arbitrariness of phase. 

To understand the geometric phase let us start from a simple situation in which
a state vector \( |\psi > \) evolves cyclically in the Hilbert space\index{Hilbert space} $\cal{H}$
and after a complete period \( T \) returns to the same state vector with a
different phase which may be written as \( \alpha |\psi >=\exp (i\phi )|\psi > \).
Therefore, 
\begin{equation}
\label{tin}
|\psi (T)>=\exp \left( i\phi \right) |\psi (0)>
\end{equation}
 where \( \phi  \) is the total phase which has two parts. The first part is
the dynamical phase and the other part is the nonadiabatic geometric phase\index{nonadiabatic geometric phase}
or the Aharonov Anandan phase (AA phase)\index{Aharonov Anandan phase}. We
are interested in the later type of phase which can be found out by subtracting
the dynamical phase part from the total phase. 

The cyclic evolution of the state vector may be described by a curve \( C \)
in $\cal{H}$ because \( |\psi > \) and \( \alpha |\psi > \) are two different
points in $\cal{H}$. Since there are infinitely many possible values of \( \alpha  \)
so there are infinitely many curves \( C \) in $\cal{H}$ which describes the
same cyclic evolution. In the projective Hilbert space\index{projective Hilbert space}
of rays \( P \) or in the ray representative space\index{ray representative space}
all these points are represented by the same ray (i.e the same point). So we
will have a single closed curve \( \widehat{C} \) in the ray representative
space\index{ray representative space} \( P \) corresponding to the infinitely
many possible curves in the usual Hilbert space\index{Hilbert space}. As there
are no arbitrariness of phase in \( P \) so in the interval \( [0,T] \) we
must have 
\begin{equation}
\label{four}
|\widetilde{\psi }(T)>=|\widetilde{\psi }(0)>.
\end{equation}
 Aharonov and Anandan exploited this property of the single valuedness of the
unit ray\index{unit ray} in the ray representative space\index{ray representative space}
\( P \) to establish the presence of a geometric phase \index{geometric phase}for
all cyclic evolutions. They started their work by supposing that the normalized
state \( |\psi (t)>\in  \)$\cal{H}$ evolves according to the Schr\"{o}dinger
equation\index{Schrodinger equation} 
\begin{equation}
\label{dui6}
H(t)|\psi (t)>=i\hbar \left( \frac{d}{dt}|\psi (t)>\right) .
\end{equation}
 We can always define a state vector \( |\widetilde{\psi }(t)> \) in \( P \)
which is equivalent to \( |\psi (t)> \) as 
\begin{equation}
\label{char}
|\widetilde{\psi }(t)>=\exp (-if(t))|\psi (t)>.
\end{equation}
 Now substituting (\ref{tin}) and (\ref{char}) in (\ref{four}) we obtain
\begin{equation}
\label{aa5}
f(T)-f(0)=\phi .
\end{equation}

If we assume that the total wave function rotates by \( 2\pi  \) radian for
this \( T \) which results in a cyclic motion of every state vector of the
Hilbert space\index{Hilbert space} $\cal{H}$, then we obtain another condition
as
\begin{equation}
\label{cond1}
f(T)-f(0)=2\pi .
\end{equation}
 Mere satisfaction of condition (\ref{cond1}) will ensure the single valuedness
of the wave function in the usual Hilbert space\index{Hilbert space} $\cal{H}$\emph{.}
Again from (\ref{dui6}) and (\ref{char}) we have 
\begin{equation}
\label{choi.1}
H|\psi (t)>=i\hbar \left( \frac{d}{dt}\exp (if(t)|\widetilde{\psi }(t)>\right) 
\end{equation}
 or, 
\begin{equation}
\label{choi.2}
H|\psi (t)>=-\hbar \frac{df}{dt}\exp (if(t)|\widetilde{\psi }(t)>+i\hbar \exp (if(t)\frac{d}{dt}|\widetilde{\psi }(t)>.
\end{equation}
 Therefore,
\begin{equation}
\label{choi.3}
<\psi (t)|H|\psi (t)>=-\hbar \frac{df}{dt}<\psi (t)|\exp (if(t)|\widetilde{\psi }(t)>+i\hbar <\psi (t)|\exp (if(t)\frac{d}{dt}|\widetilde{\psi }(t)>
\end{equation}
 or,
\begin{equation}
\label{choi.4}
-\frac{df}{dt}<\psi (t)|\psi (t)>=\frac{1}{\hbar }<\psi (t)|H|\psi (t)>-<\widetilde{\psi }(t)|i\frac{d}{dt}|\widetilde{\psi }(t)>.
\end{equation}
 Since \( \psi (t) \) is normalized so we have 
\begin{equation}
\label{choi}
-\frac{df}{dt}=\frac{1}{\hbar }<\psi (t)|H|\psi (t)>-<\widetilde{\psi }(t)|i\frac{d}{dt}|\widetilde{\psi }(t)>.
\end{equation}
 Now integrating equation (\ref{choi}) from \( 0 \) to \( T \) and using
(\ref{aa5}) we have 
\begin{equation}
\label{aa7}
\phi =-\frac{1}{\hbar }\int ^{T}_{0}<\psi (t)|H|\psi (t)>dt+\int ^{T}_{0}<\widetilde{\psi }(t)|i\frac{d}{dt}|\widetilde{\psi }(t)>dt=\gamma +\beta 
\end{equation}
 where 
\begin{equation}
\label{aa8}
\gamma =-\frac{1}{\hbar }\int ^{T}_{0}<\psi (t)|H|\psi (t)>dt
\end{equation}
 is proportional to the time integral of the expectation value of the Hamiltonian,
it is dynamical in origin and hence called dynamical phase. Now if we subtract
the dynamical phase part \( \gamma  \) from the total phase \( \phi  \) then
we get the geometric phase \index{geometric phase} or the Aharonov Anandan
phase\index{Aharonov Anandan phase}
\begin{equation}
\label{9}
\beta =\int ^{T}_{0}<\widetilde{\psi }(t)|i\frac{d}{dt}|\widetilde{\psi }(t)>dt.
\end{equation}
 It is clear that the same \( |\widetilde{\psi }(t)> \) can be chosen for infinitely
many possible curves \( C \) by appropriate choice of \( f(t) \). This phase
is geometrical in the sense that it does not depend on the Hamiltonian responsible
for the evolution. Moreover it is independent of the parameterization and redefinition
of phase of \( |\psi (t)> \). Hence, \( \beta  \) depends only on the evolution
of the shadow of \( |\psi (t)> \) in projective Hilbert space\index{projective Hilbert space}
\( P \) and is a pure geometrical entity.

\subsection{Anharmonic oscillator:}

The Hamiltonian of a generalized anharmonic oscillator can be written as 
\begin{equation}
\label{ten}
\begin{array}{lcl}
H & = & H_{0}+H_{I}\\
 & = & a^{\dagger }a+\lambda ^{\prime }F(a^{\dagger },a)
\end{array}.
\end{equation}
 Now the total eigenstate of the unperturbed Hamiltonian \( H_{0} \) is 
\begin{equation}
\label{chnew1}
|\psi (0)>=\sum _{n}C_{n}|n>
\end{equation}
 where the coefficients \( C_{n} \)'s depend on the initial conditions. The
total wave function of the Hamiltonian \( H \) satisfying the Schr\"{o}dinger
equation \index{Schrodinger equation}(\ref{dui6}) is 
\begin{equation}
\label{label2}
|\psi (t)>=\sum _{n}C_{n}\exp \left( -iHt\right) |n>=\sum _{n}C_{n}\exp \left( -ia^{\dagger }at-i\lambda ^{\prime }F(a^{\dagger },a)t\right) |n>.
\end{equation}
 For the calculation of the geometric phase, we have to chose the single-valued
state \( |\widetilde{\psi }(t)> \). This can be easily done by choosing a \( f(t) \)
which simultaneously satisfies equation (\ref{four}) and (\ref{cond1}). The
simplest choice is \( f(t)=\lambda ^{\prime }t \) and with that choice we have
\begin{equation}
\label{new7}
\begin{array}{lcl}
|\widetilde{\psi }(t)> & = & \exp (-if(t))|\psi (t)>\\
 & = & \sum _{n}C_{n}\exp \left( -ia^{\dagger }at-i\lambda ^{\prime }t\left( 1+F(a^{\dagger },a)\right) \right) |n>
\end{array}
\end{equation}
 and 
\begin{equation}
\label{new7.1}
f(T)=\lambda ^{\prime }T=2\pi .
\end{equation}
Now from equations (\ref{9}, \ref{new7} and \ref{new7.1}) a compact expression
for the Aharonov Anandan phase of a generalized anharmonic oscillator is obtained
as 
\begin{equation}
\label{new8}
\begin{array}{lcl}
\beta  & = & \int ^{T}_{0}<\widetilde{\psi }(t)|i\frac{d}{dt}|\widetilde{\psi }(t)>dt\\
 & = & \int ^{T}_{0}\sum _{k}\sum _{n}<k|C_{k}^{*}C_{n}\exp \left( ia^{\dagger }at+i\lambda ^{\prime }t\left( 1+F(a^{\dagger },a)\right) \right) \\
 & \times  & \left[ a^{\dagger }a+\lambda ^{\prime }\left( 1+F(a^{\dagger },a)\right) \right] \exp \left( -a^{\dagger }at-i\lambda ^{\prime }t\left( 1+F(a^{\dagger },a)\right) \right) |n>dt\\
 & = & \int ^{T}_{0}\sum _{k}\sum _{n}<k|C_{k}^{*}C_{n}\left[ a^{\dagger }a+\lambda ^{\prime }\left( 1+F(a^{\dagger },a)\right) \right] |n>dt\\
 & = & \sum _{k}\sum _{n}<k|C_{k}^{*}C_{n}\left[ 2\pi \left( 1+\frac{a^{\dagger }a}{\lambda ^{\prime }}+F(a^{\dagger },a)\right) \right] |n>\\
 & = & 2\pi +\sum _{n}|C_{n}|^{2}\left( \frac{2\pi n}{\lambda ^{\prime }}\right) +2\pi \sum _{k}\sum _{n}<k|C_{k}^{*}C_{n}\left( F(a^{\dagger },a)\right) |n>.
\end{array}
\end{equation}

The quantum nature of the geometric phase is manifested through the discrete
sum appearing in the expression (\ref{new8}) and the state dependence of the
phase appears through the coefficients \( C_{k}^{*}C_{n} \). Now let us think
of some special situations of physical interest.

\subsubsection{Anharmonic part is a polynomial of the number operator:}

When the anharmonic part can be written as a polynomial \( P \) of number operator
\( a^{\dagger }a \), i.e when \( F(a^{\dagger },a)=P(a^{\dagger }a) \) then
from (\ref{new8}) we obtain 
\begin{equation}
\label{new8.1}
\beta =\sum _{n}|C_{n}|^{2}\left[ 2\pi \left( 1+\frac{n}{\lambda ^{\prime }}+P(n)\right) \right] .
\end{equation}
 One of the simplest possible polynomial of number operator is 
\[
P(a^{\dagger }a)=a^{\dagger ^{2}}a^{2}=(a^{\dagger }a)^{2}-a^{\dagger }a.\]
 This kind of interaction is very common in Physics and Joshi \emph{et al} {[}\ref{Joshi-Pati}{]}
have shown that a geometric phase exists for such a Hamiltonian. Actually they
considered the propagation of electromagnetic radiation in a dispersive medium
and under certain approximation they obtained the approximated Hamiltonian.
Now from (\ref{new8.1}) we obtain the geometric phase\index{geometric phase}
for the approximated Hamiltonian of Joshi \emph{et al} (\ref{charpf}) as 
\begin{equation}
\label{new8.2}
\beta =\sum _{n}|C_{n}|^{2}\left[ 2\pi \left( 1+\frac{n}{\lambda ^{\prime }}+(n^{2}-n)\right) \right] .
\end{equation}
 This expression for geometric phase\index{geometric phase} does not coincide
exactly with that of Joshi et al {[}\ref{Joshi-Pati}{]}. The term \( \sum _{n}|C_{n}|^{2}\left[ 2\pi \frac{n}{\lambda ^{\prime }}\right]  \)
is absent in their calculation. A conceptual oversight is responsible for that.
In their paper they have worked in the interaction picture and found out the
correct and exact wave function (in the interaction picture) but they have not
considered the time evolution of the differential operator \( \frac{d}{dt} \)
present in equation (\ref{9}). The time evolution of the operator \( \frac{d}{dt} \)
in the interaction picture is \( \exp (iH_{0}t)\frac{d}{dt}\exp (-iH_{0}t) \)
and this is the operator one should use in equation (\ref{9}) while working
in the interaction picture. Use of the corrected operator will yield the same
result.

\subsubsection{Interaction with an (m-1)-th order nonlinear medium}

Now let us think of a physical situation in which an intense electromagnetic
field interacts with an \( (m-1) \)-th order nonlinear nonabsorbing medium.
If the symmetry of the medium is chosen in such a way that the only nonlinear
interaction in the medium appears due to the presence of \( (m-1) \)-th order
susceptibility then the interaction Hamiltonian (\ref{ten.1}) of the system
may be written in an alternative form as 
\begin{equation}
\label{ten.1aa}
\begin{array}{lcl}
H & = & \frac{x^{2}}{2}+\frac{\dot{x}^{2}}{2}+\frac{\lambda }{m}x^{m}\\
 & = & a^{\dagger }a+\lambda ^{\prime }(a^{\dagger }+a)^{m}
\end{array}
\end{equation}
where \( \lambda ^{\prime }=\frac{\lambda }{m(2)^{\frac{m}{2}}} \). The above
Hamiltonian (\ref{ten.1aa}) represent a generalized anharmonic oscillator for
which \( F(a^{\dagger },a)=(a^{\dagger }+a)^{m} \) and the geometric phase
is 
\begin{equation}
\label{beta1}
\begin{array}{lcl}
\beta  & = & 2\pi +\sum _{n}|C_{n}|^{2}\left( \frac{2\pi n}{\lambda ^{\prime }}\right) +2\pi \sum _{k}\sum _{n}C_{k}^{*}C_{n}<k|(a^{\dagger }+a)^{m}|n>\\
 & = & 2\pi +\sum _{n}|C_{n}|^{2}\left( \frac{2\pi n}{\lambda ^{\prime }}\right) +2\pi \sum _{n}\sum ^{\frac{m}{2}}_{r=0}\sum ^{m-2r}_{p=0}t_{2r}\, ^{m}C_{2r}\, ^{m-2r}C_{p}\\
 & \times  & C_{n-m+2r+2p}^{*}C_{n}\frac{\left[ n!(n-m+2r+2p)!\right] ^{\frac{1}{2}}}{(n-m+2r+p)!}
\end{array}
\end{equation}
 where 
\begin{equation}
\label{beta2}
t_{r}=\frac{(r)!}{2^{(\frac{r}{2})}(\frac{r}{2})!}.
\end{equation}
 Normal ordering \index{normal order} theorems 1 and 2 are used here to derive
(\ref{beta1}). Now we have a closed form analytic expression (\ref{beta1})
for the geometric phase \( \beta  \) for a generalized anharmonic oscillator
which depends on the photon statistics of the input radiation field. So an observation
of the geometric phase \index{geometric phase} may help us to conclude the
nature of the input electromagnetic field. We use this generalized expression
(\ref{beta1}) to study a very special case.

\subsubsection{An intense laser\index{laser} beam interacts with a third order nonlinear
medium}

Let us now consider the case where an intense electromagnetic field having Poissonian
statistics (\( C_{n}=\exp (-\frac{|\alpha |^{2}}{2})\frac{\alpha ^{n}}{\sqrt{n!}} \))
interacts with a third order nonlinear nonabsorbing medium. If the symmetries
of the medium is chosen in such a way that the only nonlinear interaction in
the medium appears due to the presence of third order susceptibility, then the
Hamiltonian of the system is 
\begin{equation}
\label{cond2}
H=\frac{x^{2}}{2}+\frac{\dot{x}^{2}}{2}+\frac{\lambda }{4}x^{4}=H_{0}+\lambda ^{\prime }(a^{\dagger }+a)^{4}
\end{equation}
where \( \lambda  \) is the coupling constant and \( \lambda ^{\prime }=\frac{\lambda }{16} \).
With the help of equations (\ref{beta1}) the geometric phase \index{geometric phase}\( \beta  \)
for this physical system can be written as 
\begin{equation}
\label{cond3}
\begin{array}{lcl}
\beta  & = & 2\pi +\sum _{n}|C_{n}|^{2}\left( \frac{2\pi n}{\lambda ^{\prime }}\right) +2\pi \sum _{n}\sum ^{2}_{r=0}\sum ^{4-2r}_{p=0}t_{2r}\, ^{4}C_{2r}\, ^{4-2r}C_{p}C_{n-4+2r+2p}^{*}C_{n}\frac{\left[ n!(n-4+2r+2p)!\right] ^{\frac{1}{2}}}{(n-4+2r+p)!}\\
 & = & 8\pi +2\pi \exp (-|\alpha |^{2})\left\{ \sum _{n}\frac{|\alpha |^{2n}}{n!}\right. \left[ \left( 6n^{2}+\left( 6+\frac{1}{\lambda ^{\prime }}\right) n\right) \right. \\
 & = & \left. \left. 4|\alpha |^{2}\left( 2n+3\right) \cos (2\theta )+2|\alpha |^{2}\cos (4\theta )\right] \right\} \\
 & = & 8\pi +2\pi \left[ \left( 6|\alpha |^{4}+12|\alpha |^{2}+\frac{|\alpha |^{2}}{\lambda ^{\prime }}\right) +4|\alpha |^{2}\left( 2|\alpha |^{2}+3\right) \cos (2\theta )+2|\alpha |^{2}\cos (4\theta )\right] .
\end{array}
\end{equation}
 Here we observe that the geometric phase \index{geometric phase} depends on
the phase of the initial laser\index{laser} beam. The presence of the off-diagonal
terms in the interaction Hamiltonian is manifested through the presence of \( \theta  \)
dependent terms in the expression for the geometric phase \( \beta  \).

\subsection{Remarks on the results:}

The geometric phase \index{geometric phase} for the physical systems considered
by us can be experimentally observed with the help of interferometric experiments.
To be more precise we can think of an optical analogue of the Suter-Muller-Pines
experiment {[}\ref{Suter}{]}. A gedanken experiment for this purpose is also
proposed by Joshi et al. {[}\ref{Joshi-Pati}{]}.  

We have seen that the geometric phase arising due to the interaction of intense
electromagnetic field with a nonlinear medium is experimentally observable and
the \( C_{n} \)'s depend on the photon statistics of the input field (see Table
6.1). So an observation of the geometric phase can give us the photon statistics
of the input electromagnetic field. Earlier we have shown that the photon statistics
in a third order nonlinear nonabsorbing medium of inversion symmetry changes
with the interaction time and in this chapter we will show that this fact is
true for \( (m-1)-th \) order nonlinear medium also. Using the output of such
an interaction as the input of a geometric phase experiment our predictions
on photon statistics can be verified. 

\begin{table}
{\centering \begin{tabular}{|l|c|}
\hline 
{\small Photon Statistics}&
{\small \( \, \, \, \, \, \, \, \, \, |C_{n}|^{2} \)}\\
\hline 
{\small Sub-Poissonian}\index{sub-Poissonian}&
{\small \( \frac{N!}{(N-n)!n!}p^{n}(1-p)^{N-n}\, \, (0\leq p\leq 1) \)}\\
\hline 
{\small Poissonian}\index{Poissonian}&
\( \exp (-|\alpha |^{2})\frac{\alpha ^{n}}{\sqrt{n!}} \)\\
\hline 
{\small Super-Poissonian}\index{super-Poissonian}&
{\small \( \frac{(n+W-1)!}{(W-1)!n!}q^{n}(1-q)^{W}\, \, (0\leq q\leq 1) \)}\\
\hline 
\end{tabular}\small \par}
{\par\centering Table 6.1\par}
\end{table}

The general Hamiltonian of the form (\ref{ten.1aa}) and its special case (\ref{cond2})
are very common in quantum optics and we often use rotating wave approximation\index{rotating wave approximation}
(RWA\index{RWA}) to deal with these Hamiltonians. Under RWA calculations off-diagonal
terms are neglected. But in the present case the presence of the off-diagonal
terms in the interaction Hamiltonian is manifested through the presence of \( \theta  \)
dependent terms in the expression for the geometric phase \( \beta  \)\index{geometric phase}.
From these facts we can conclude two things firstly, it is not right to use
RWA\index{RWA} for the calculation of the geometric phase and, secondly, geometric
phase\index{geometric phase} can be tuned because it depends strongly on the
phase of the input field which can be tuned.

\section{Higher harmonic generation}

If an input of frequency \( \omega  \) generates a output frequency \( n\omega  \)
due to the nonlinear frequency mixing in a nonlinear medium, then we say that
the \( n-th \) harmonic generation is taken place in the medium. It is well
known that the \( n-th \) harmonic generation is observed in an \( n-th \)
order nonlinear medium. Now, a close look at the term, \( \exp \left( i\Omega t(2p-m+2r-1)\right)  \)
in (\ref{ev11}), will show that the expansion of the summation present in (\ref{ev11})
will give us terms with frequency \( 3\Omega ,5\Omega ,7\Omega ,....,(m-1)\Omega  \)
when \( m \) is even and \( 2\Omega ,4\Omega ,6\Omega ,......(m-1)\Omega  \)
when \( m \) is odd. Thus we can conclude from the above discussion and (\ref{ev11})
that the frequency mixing occurred in such a way that all the odd / even higher
harmonics up to the order of the nonlinearity of the medium will be present
in the output.

\section{Bunching\index{bunching}, antibunching\index{antibunching} and statistical
distribution of the photons}

In chapter 5 we have seen that the quantum statistical properties of radiation
field can be studied with the knowledge of sign of \( d=\left( (\triangle N)^{2}-<N>\right)  \).
To study the QSP of the output of an interaction of coherent light with an \( (m-1) \)-th
order nonlinear medium we can construct a closed form expression for \( d \)
by using (\ref{ev6}) and (\ref{ab4}) as 
\begin{equation}
\label{ab5}
\begin{array}{lcl}
d & = & \frac{4\lambda }{m(2)^{\frac{m}{2}}}\left( \sum ^{\frac{m}{2}}_{r}t_{2r}\, ^{m}C_{2r}\sum ^{(m-2r)}_{p\neq \frac{m-2r}{2}}\, ^{(m-2r)}C_{p}\frac{p(p-1)}{2p-m+2r}|\alpha |^{m-2r}\sin \left[ \frac{\left( 2p-m+2r\right) }{2}(t-2\theta )\right] \right. \\
 & \times  & \left. \sin \left[ \frac{\left( 2p-m+2r\right) }{2}t\right] \right) 
\end{array}
\end{equation}
 where we have taken the expectation value with respect to the initial coherent
state\index{coherent state} \( |\alpha >=\sum ^{\infty }_{n=0}\exp \left( \frac{|\alpha |^{2}}{2}\right) \frac{\alpha ^{n}}{\sqrt{n!}}|n> \)
with \( \alpha =|\alpha |\exp (i\theta ) \), having Poissonian statistics.
From this expression we can see that for \( m=4 \), i.e for a third order nonlinear
medium having inversion symmetry, we have 
\begin{equation}
\label{ab6}
d=\frac{3\lambda |\alpha |^{2}}{4}\left[ 2\left( 2|\alpha |^{2}+1\right) \sin (t-2\theta )\sin (t)+|\alpha |^{2}\sin \left( 2(t-2\theta )\right) \sin (2t)\right] .
\end{equation}
 This is in exact accordance with our earlier result. From (\ref{ab5}) we have
following observations, 

i) When \( t=2\theta  \) then \( d=0 \). Therefore, input coherent state\index{coherent state}
remains coherent in the output. 

ii) When \( \theta =0 \) or \( \theta =n\pi  \), i.e. input is real then \( d \)
is a sum of square terms only. So \( d \) is always positive and we have bunched
photons having super-Poissonian\index{super-Poissonian} statistics in output.

iii) For other values of phase \( \theta  \) of input radiation field, value
of \( d \) oscillates from positive to negative, so we can observe bunched,
antibunched or coherent output depending upon the interaction time \( t \).

\section{Squeezing\index{squeezing}}

Here we will study the possibilities of electrically squeezed field. Possibilities
of magnetically squeezed field may also be studied in the similar process. Now
the uncertainty in electric field is
\begin{equation}
\label{sq2}
\begin{array}{lcl}
(\triangle X)^{2} & = & \frac{1}{2}\left[ 1+\frac{2\lambda t}{2^{\frac{m}{2}}m}\sum ^{\frac{m}{2}}_{r=0}t_{2r}\, ^{m}C_{2r}\, ^{m-2r}C_{\frac{m-2r}{2}}(\frac{m-2r}{2})(\frac{m-2r}{2}-1)|\alpha |^{m-2r-2}\sin \left( 2(\theta -t)\right) \right. \\
 & - & \frac{4\lambda t}{2^{\frac{m}{2}}m}\sum ^{\frac{m}{2}}_{r}t_{2r}\, ^{m}C_{2r}\sum ^{(m-2r)}_{p\neq \frac{m-2r}{2}}\, ^{(m-2r)}C_{p}\frac{p|\alpha |^{m-2r-2}}{(2p-m+2r)}\sin \left( \frac{(m-2r-2p)}{2}t\right) \\
 & \times  & \left. \left[ (p-1)\sin \left( \frac{(m-2r-2p+2)}{2}(2\theta -t)-t\right) +(m-2r-2p)\sin \left( \frac{(m-2r-2p)}{2}(2\theta -t)\right) \right] \right] .\\
 &  & \\
 &  & 
\end{array}
\end{equation}
 Here we can note that the squeezed state \index{squeezed state} may be generated
due to the interaction of an intense laser\index{laser} beam with the \( (m-1) \)-th
order nonlinear medium for particular values of \( \theta  \) and \( t \).
For example, for \( \theta =0 \) and \( t<\frac{2\pi }{m} \) we will get electrically
squeezed field in the output. To compare the above expression with the existing
results we put \( m=4 \) and \( \theta =0 \), which corresponds to the case
in which an intense laser\index{laser} beam having zero input phase interacts
with a third order nonlinear non absorbing medium of inversion symmetry and
obtain, 
\begin{equation}
\label{sq3}
(\triangle X)^{2}=\frac{1}{2}-\frac{3\lambda }{4}\sin ^{2}(t)-\frac{3\lambda |\alpha |^{2}}{4}\left( t\sin (2t)+2\sin ^{2}(t)\right) .
\end{equation}
 This is in exact accordance with our earlier result {[}\ref{PM01}{]}. From
(\ref{sq3}) we can see that the squeezing\index{squeezing} in electric field
quadrature will be observed for \( t<\frac{\pi }{2} \) for any arbitrary allowed
values of other parameters. So a nonclassical phenomenon (squeezing\index{squeezing})
is observed for \( m=4 \) and \( \theta =0 \) but for the same conditions
photons of the output radiation field are always bunched. Thus we can conclude,
although the nonclassical phenomenon of antibunching\index{antibunching} and
squeezing\index{squeezing} usually appears together but it is not essential
that the two nonclassical phenomena have to appear together.

\chapter{Summary and concluding remarks}

In the present work coherent light interacting with a nonlinear medium of inversion
symmetry is modeled by a general quantum anharmonic oscillator. The quartic-anharmonic-oscillator\index{quartic}
model emerges if only the lowest order of nonlinearity (i.e third order nonlinear
susceptibility) is to be present in the medium. The quartic-oscillator model
has considerable importance in the study of nonlinear and quantum optical effects
that arise in a nonlinear medium of inversion symmetry. For example, silica
crystals {\large }constitute an inversion symmetric third order nonlinear medium
and these crystals are used to construct the optical fibers. In the optical
communication electromagnetic beam passes through the optical fiber and the
interaction of the electromagnetic field (single mode) with the fiber can be
described by the one dimensional quartic\index{quartic} anharmonic oscillator
Hamiltonian. Depending on the nature of nonlinearity in a physical problem the
treatment of higher anharmonic oscillators also assumes significance. But anharmonic
oscillator models are not exactly solvable in a closed analytic form. On the
other hand, we need operator solutions of the equations of motion corresponding
to these models in order to study the quantum fluctuations of coherent light
in nonlinear media. So we have two alternatives, either we can use an approximate
Hamiltonian which is exactly solvable or we can use an approximate operator
solution. Approximate operator solutions of anharmonic oscillator problems (solutions
in the Heisenberg approach) were not available even in the recent past presumably
because the existing methods tend to introduce inordinate mathematical complications
in a detailed study. Due to the unavailability of the operator solution people
would use RW approximated Hamiltonians to study the quantum fluctuations of
coherent light in nonlinear media. But the situation is now improved and many
proposals to obtain approximate operator solutions of anharmonic oscillators
have appeared. Some of these solutions are obtained as a part of the present
work. For example, we have constructed a second order operator solution for
quartic\index{quartic} oscillator and have generalized the first order operator
solutions available for the quartic\index{quartic} oscillator to the \( m \)-th
anharmonic oscillator. From the generalized solutions we have observed that
an apparent discrepancy is present between the solutions obtained by different
techniques. Then the question arise: Which solution should be used for physical
applications? Therefore, we have compared different solutions and have concluded
that all the correct solutions are equivalent and the apparent discrepancy whatsoever
is due to the use of different ordering of the operators. These solutions are
then exploited to study the possibilities of observing different nonlinear optical
phenomena in a nonlinear dielectric medium. To be precise, we have studied quantum
phase fluctuations\index{quantum phase fluctuations} of coherent light in third
order inversion symmetric nonlinear medium in chapter 3. Fluctuations in phase
space quadrature for the same system is studied in chapter 4 and the possibility
of generating squeezed state \index{squeezed state} is reported. In chapter
5 fluctuations in photon number is studied and the nonclassical phenomenon of
antibunching\index{antibunching} is predicted. In chapter 6 we have generalized
the results obtained for third order nonlinear medium and have studied the interaction
of an intense laser\index{laser} beam with an \( (m-1) \)-th order nonlinear
medium in general. Aharonov Anandan nonadiabatic geometric phase\index{nonadiabatic geometric phase}
is also discussed in the context of \( (m-1) \)-th order nonlinear medium in
general. 

We have observed different interesting facts regarding quantum fluctuations
of coherent light interacting with nonlinear media and now we can conclude the
present work with the following observations. 

\begin{enumerate}
\item In a third order nonlinear medium reduction and enhancement of phase fluctuation
parameters (\( U\left( \theta ,t,|\alpha |^{2}\right)  \), \( Q\left( \theta ,t,|\alpha |^{2}\right)  \)
and \( S\left( \theta ,t,|\alpha |^{2}\right)  \)) are possible for suitable
choice of free evolution time \( t. \) This result is in sharp contrast with
the earlier results {[}\ref{Gerry}-\ref{Lynch}{]}. 
\item Electrically squeezed electromagnetic field can be generated due to the interaction
of an intense beam of coherent light with an \( (m-1) \)-th order nonlinear
medium. 
\item Antibunching\index{antibunching} of photons can be observed due to the interaction
of an intense beam of coherent light with an \( (m-1) \)-th order nonlinear
medium in general.
\item Simultaneous appearance of photon bunching\index{bunching} (classical phenomenon)
and squeezing\index{squeezing} (nonclassical phenomenon) is observed. This
establishes the fact that there is no reason to believe that the appearance
of one of the nonclassical phenomena warrants the presence of the others. 
\item The vacuum field \index{vacuum field} interacting with the nonlinear medium
produces photons through nonlinear interaction. We report sub-Poissonian\index{sub-Poissonian}
PND\index{photon number distribution} and antibunching\index{antibunching}
of these photons. 
\item A nonvanishing noadiabatic geometric phase\index{nonadiabatic geometric phase}
appears due to the interaction coherent light with an \( (m-1) \)-th order
nonlinear medium. The geometric phase for the physical systems considered by
us can be experimentally observed with the help of interferometric experiments. 
\item Under RWA\index{RWA} calculations off-diagonal terms are neglected. But in
the present case the presence of the off-diagonal terms in the interaction Hamiltonian
is manifested through the presence of \( \theta  \) dependent terms in the
expressions for the \( \beta ,\, (\triangle X)^{2} \) etcetera. Therefore special
care should be taken before using RWA for the calculations related to the matter-field
interaction. 
\item Percentage of squeezing\index{squeezing} and geometric phase \index{nonadiabatic geometric phase}can
be tuned because they strongly depend on the phase of the input coherent field
which can be tuned. 
\end{enumerate}

\section{{\large Limitations and scope for future works}\large }

In most of the studies envisaged in the present work we have used either first
order or second order solutions. But higher order terms might have important
effects in comparatively strongly coupled systems. Higher order operator solutions
are now appearing in the literature such that it is now technically possible
to study the strongly coupled systems using the methods used in the present
work.

In some particular cases we have used secular solutions to study the quantum
fluctuations. This is valid for short interaction time. In order to have corresponding
expressions valid for all times we have to remove the secular terms from the
expressions. Tucking in technique\index{tucking in technique} may be used for
the purpose.

We started working on the quantum anharmonic oscillator problem in 1998. At
that time only two techniques were available to provide operator solution of
the quantum quartic\index{quartic} anharmonic oscillator. The situation has
considerably improved during the last four years. Now we have sixth order solution
of quartic oscillator {[}\ref{6th order}{]}\footnote{
In fact we can construct solution of anharmonic oscillators (for any particular
\( m \)) up to arbitrary order. The procedure is reported in our work {[}\ref{APFM01}{]}
but it is not included in the present thesis.
}. So we can now address the question of convergence of the operator series.
We can also extend our works on one dimensional anharmonic oscillators to two
or higher dimensions. We also expect that the present work will serve as the
back bone of the nonlinear dynamical studies (in Heisenberg approach) of more
complex systems. For example, we can study a system of coupled oscillators,
toda lattices etcetera. This kind of study may provide many new information
in near future. 

The other thing is that almost all the effects discussed in the present thesis
are experimentally realizable and all of them have potential applications in
optical communications and other areas of physics which are related to public
life. We finish the present work with an expectation that it will be useful
for the future development of theoretical and experimental studies of nonlinear
dynamical systems.

\printindex

\end{document}